\documentclass[11pt,a4paper]{article}
\usepackage[utf8]{inputenc}
\usepackage{jheppub}
\usepackage{amsthm,amsbsy,amsfonts,mathrsfs,enumerate,float,wrapfig,amsmath}
\usepackage{subfigure}
\usepackage{tikz}
\usepackage{textcomp}
\usepackage{ytableau}
\usepackage{mathtools} 
\usepackage{extarrows}
\usepackage{mleftright}
\usepackage{float}
\usepackage{mathtools}

\newcommand{\nn}{\nonumber}
\def\a{{\alpha}}

\def\b{{\beta}}

\def\0{{\emptyset}}
\def\u{{\mu}}
\def\v{{\nu}}

\def\p{{\rho}}

\def\sqqt {{ \sqrt{\frac{q}{t}}}}
\def\sqtq {{ \sqrt{\frac{t}{q}}}}

\def\inf{{\infty}}

\def\half{\text{half}}

\def\sinh{{\text{sinh}}}

\def\>{{  \succcurlyeq} }
\newcommand{\bsmall}{\begin{small}   }
	\newcommand{\esmall}{ \end{small}   }

\def\qbrane{{q\text{-brane}}}
\def\qbarbrane{{\bar{q}\text{-brane}}}
\def\tbrane{{t\text{-brane}}}
\def\tbarbrane{{\bar{t}\text{-brane}}}

\def\PE{\mathrm{PE}}

\def\F{\mathbf{F}}
\def\AF{\mathbf{AF}}
\def\N{\mathcal{N}}

\def\tm{\tilde{m}}

\def\eff{\text{eff}}
\def\CS{{\text{Chern-Simons }}}

\def\B{{\mathcal{B}}}

\def\W{{\mathcal{W}}}

\def\D{\text{D}}
\def\NS{\text{N}}
\def\half{\frac{1}{2}}
\def\aa{\mathfrak{a}}
\def\tW{\widetilde{\W}}
\def\Li{\text{Li}  }
\def\dd{\text{d}}

\makeatletter
\newcommand\mathcircled[1]{%
	\mathpalette\@mathcircled{#1}%
}
\newcommand\@mathcircled[2]{%
	\tikz[baseline=(math.base)] \node[draw,circle,inner sep=1pt] (math) {$\m@th#1#2$};%
}
\makeatother

\title{
3d  $\mathcal{N}=2$ Brane Webs and Quiver Matrices
}

\author[a]{Shi Cheng,}

\affiliation[a]{Faculty of Physics, University of Warsaw, ul. Pasteura 5, 02-093 Warsaw, Poland}

\emailAdd{scheng@fuw.edu.pl} 

\abstract{
	
We discuss 3d brane webs and effective Chern-Simons levels for 3d $\mathcal{N}=2$ gauge theories. We find that turning on real masses for chiral multiplets leads to various equivalent brane webs that are related by flipping positions of D5-branes. We interpret flips as $ST$-transformations for chiral multiplets.  $ST$-transformations could turn abelian theories into dual theories with mixed Chern-Simons levels that are interpreted as quiver matrices $C_{ij}$ encoding DT-invariants. We notice that each brane web corresponds to a  quiver matrix. $ST$-transformations of holomorphic blocks are discussed to verify results.
We also discuss the movement of flavor D5-branes, which leads to double-layer brane webs and manifests fiber-base duality. In the second part, we compute refined vortex partition functions of nonabelian theories with the gauge group $U\left(N\right)$ and find corresponding quiver matrices. The computation shows that on Higgs branch nonabelian groups are broken to abelian groups. 

	\rule{0pt}{0pt}
	\\
	\rule{0pt}{0pt}
	\\
	\\
	\\

}

\begin{document}

\maketitle
	

\section{Introduction}

3d $\N=2$ gauge theories relate many aspects of string theories and geometry. Some 3d theories can be constructed using brane systems \cite{Boer:1997ts} and enjoy various dualities \cite{Aharony:1997aa,Intriligator:2013lca,Aharony_1997,Giveon_2009,Boer:1997ts}. Geometrically, in the context of 	3d/3d correspondence \cite{Dimofte:2011ju,Terashima:2011qi}, 3d $\N=2$ gauge theories can be constructed by compactifing 6d $\left(2,0\right)$ superconformal field theories on three manifolds. In addition, some 3d $\N=2$ theories can also be viewed as surface defect theories by Higgsing 5d $\N=1$ gauge theories \cite{Dimofte:2010tz,Alday:2009fs}, and the later can be engineered by Calabi-Yau three-manifolds in M-theory or 5-brane webs in type IIB string theory \cite{Aharony:1997bh,Aharony:1997uv,Morrison:1996xf,Douglas:1996xp,Ganor:1996pc,Intriligator:1997pq}. In this work, we further develop this Higgsing construction by studying the 3d brane webs in type IIB string theory. 

There are some physical quantities that could characterize 3d $\N=2$ theories, such as gauge groups, representations of chiral multiplets, Chern-Simons levels, etc. These physical quantities are supposed to be encoded in 3d brane webs. For instance, the relative angle $\theta$ between NS5-brane and NS5'-brane is related to the Chern-Simons level \cite{Bergman:1999na,Kitao:1999aa}. Turning on real mass parameters separate overlapped D5-branes. Decoupling matters changes Chern-Simons levels. Following the story in 5d $\N=1$ theories in e.g.\cite{Intriligator:1997pq,Aharony:1997bh,Aharony:1997uv}, one should be able to read off these information from 3d brane webs. Fortunately, we can compute the vortex partition functions using topological string methods, in particular topological vertex. Comparing  terms and factors in vortex partition functions of various brane webs, one could see how 3d brane webs encode these physical quantities.
We implement refined topological vertex to compute open topological string amplitudes that are interpreted as 3d vortex partition functions, in particular the cases that can be obtained by Higgsing closed topological string amplitudes \cite{Dimofte:2010tz,Cheng:2021aa}. For the recently development of topological vertex, see e.g. \cite{Kim:2017jqn,Hayashi:2020aa,Nawata:2021uu,Kimura:2019wi,Hayashi:2018bkd,Hayashi:2021pcj}. 

We firstly discuss abelian theories with matters. We find that there are many equivalent brane webs by turning on real mass parameters, which compose different chambers of the 3d $\N=2$ theories. A subset of these brane webs composes Higgs branch. An observation is that 3d vortex partition functions of these 3d brane webs are related by flipping the signs of real mass parameters, which is equivalent to flipping the positions of D5-branes. We propose that this flip can be interpreted as the $ST$-transformation from the perspective of their $ST$-dual theories. It is the operator $ST \in SL(2, \mathbb{Z})$ that plays a crucial role in 3d-3d correspondence.  Moreover,
$ST$-transformation can also be interpreted as the functional Fourier transformation on partition functions \cite{Kapustin:1999ha}. This property has been implemented to construct mirror dual theories in \cite{Benvenuti:2016wet,Cheng:2020aa}. 
It is found in \cite{Cheng:2020aa} that abelian theories can be turned into $ST$-dual theories by performing $ST$-transformations. We note that each $ST$-dual theory has one associated mixed Chern-Simons level matrix which is also called quiver matrix in the context of Donaldson-Thomas invariants and knots-quivers correspondence \cite{Kontsevich:2010px,Kucharski:2017poe,Kucharski:2017ogk}. We argue that each equivalent brane web corresponds to such a quiver matrix. One useful tool to read off \CS levels is the effective superpotential \cite{Shadchin_2007,Hori:2013ika}, which can be computed by taking the classical limit of partition functions.

 The above discussion on 3d theories is not complete, because if we put 3d theories on the noncompact spacetime $\mathbb{R}^2 \times S^1$, then there are gauge and flavor anomalies, which should be canceled by add 2d (0,2) theories on the boundary. Therefore, we discuss the holomophic blocks following \cite{Beem:2012mb} to fill this potential loophole. We discuss the interplay between $ST$-transformations and mixed \CS levels using holomorphic blocks. The results we obtain match with the conclusions from analyzing sphere partition functions \cite{Cheng:2020aa} and 3d brane webs.	 

After the discussion on abelian theories, we discuss nonabelian theories with gauge group $U\left(N\right)$ and  find the above conclusion also applies to nonabelian theories. By comparing 3d vortex partition functions obtained by localization \cite{Hwang:2012jh,Benini:2014aa}, we notice that on Higgs branch 3d brane webs of nonabelian theories are abelianized and some internal lines on toric diagrams (brane webs) need to be assigned with empty Young tableaux. We find there are also quiver matrices for nonabelian theories. Besides, we can move flavor D5-branes in the brane systems of 3d $\N=2$ theories, which leads to different kinds of equivalent brane webs, and even brane webs satisfying fiber-base duality; see also \cite{Nieri:2018vc,Liu:2021ui}. 

The organization of the paper is as follows. In section \ref{sec2}, we review 3d vortex partition functions for abelian theories, and discuss quiver matrices of their $ST$-dual theories.   
In section \ref{holoblocks}, we mainly discuss $ST$-transformations and mixed \CS levels using holomorphic blocks.
In section \ref{seceffCS}, we discuss the relations between relative theta angles and \CS levels.
In section \ref{secmassdef}, we focus on mass deformations that lead to equivalent brane webs and quiver matrices for abelian theories. We also discuss the movement of D5-branes.
In section \ref{secnonabelian} we compute refined 3d vortex partition functions of nonabelian theories and read of quiver matrices.

\section{Abelian theories }\label{sec2}

In this section, we discuss the abelian theory $U\left(1\right)_k+N_{f} \F+N_{a}\AF$. We firstly review 3d vortex partition functions obtained by topological vertex. Then we discuss  that using $ST$-transformations, abelian theories can be turned into $ST$-dual theories that have mixed Chern-Simons levels.

\subsection{Vortex partition function}\label{geotrans}
\begin{figure}[h!]
	\centering
	\includegraphics[width=6in]{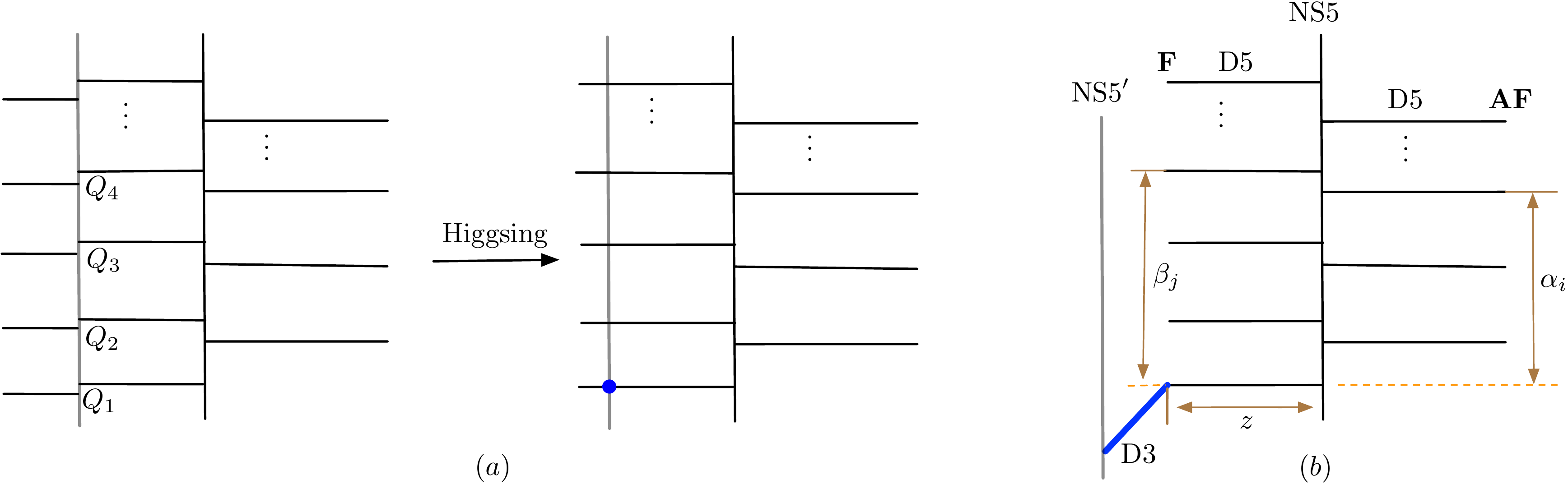}
	\caption{ 
		In the diagram $\left(a\right)$, we show the Higgsing process. We prefer to create a D3-brane at the bottom line. After Higgsing local conifolds $Q_1,\,Q_2\,,\dots$, the NS5'-brane is lifted along the perpendicular direction and a D3-brane is created. 
		The more precise 3d brane web is the diagram $\left(b\right)$, where the D3-brane is represented by a blue line. D5-branes and the NS5-brane are on the same plane, while D3-brane is perpendicular to this plane. Note that there could be a relative angle $\theta$ between  the NS5-brane and the 	NS5'-brane. In the diagram $(b)$, we draw NS5-brane and NS5'-brane in parallel for simplicity.}
	\label{fig:strippara}
\end{figure}
The 3d vertex partition function on a particular vacuum can be obtained by Higgsing closed topological string amplitudes \cite{Dimofte:2010tz,Koz_az_2010,Pasquetti:2011fj,Cheng:2020aa,Zenkevich:2017ylb}. Geometrically, Higgsing is the geometric transition that acts on the local conifold. Depending on different expectation values of K\"ahler parameters, local conifolds lead to different types of D3-branes. More explicitly, if $Q_i =\sqrt{\frac{t}{q}}$, then no D3-brane is created, and if $Q_i=t\sqrt{\frac{t}{q}}$, a D3-brane called $\tbrane$ is created, and if $Q_i=\frac{1}{q}\sqrt{\frac{t}{q}}$, a D3-brane called $\qbarbrane$ is created \cite{Aganagic:2012hs,Cheng:2021aa}.

The procedure of computation  is the following: firstly, we compute the closed topological string amplitudes using topological vertex formula, and then Higgsing some local conifolds $\{Q_i\}$ to produce open topological string amplitudes that are vortex partition functions of 3d theories. Namely,
\begin{align}
	Z^{ \text{closed}}  \xlongrightarrow{ \text{Higgsing }  }Z^{ \text{open}} =Z^{\text{3d vortex}} \,.
	\end{align}

For example, the brane web $\left(b\right)$ in Figure \ref{fig:strippara} is produced by giving values $Q_1=t\sqtq \left( \text{or}~ \frac{1}{q}\sqtq \right)$ and $ Q_{i} =0$ for $ \forall \,i > 1$.
This 3d brane web constructs the theory $U\left(1\right)_k+ N_{f} \, \mathbf{F}+N_{a} \,\mathbf{AF}$. The associated vortex partition function takes the following form in unrefined limit $q=t$ \cite{Dimofte:2017tpi,Cheng:2020aa},
\begin{align}\label{vortexNCNAC}
Z_{U\left(1\right)_k+ N_{f} \, \mathbf{F}+N_{a} \,\mathbf{AF}}^{\text{vortex}}\left(z,\a,\b \right)
=\sum\limits_{n=0}^{\inf} {  \left(-\sqrt{q}\right)^{ k^{\eff} \, n^2} z^n} 
\cdot \frac{
	\left(\a_1 ;q \right)_n \left(\a_2;q\right)_n  \cdots \left(\a_{N_{a}};q \right)_n 
}
{ 	\left(\b_1 ;q \right)_n \left(\b_2  ;q \right)_n \cdots \left(\b_{N_{f}} ;q \right)_n 	
}   \,, 
\end{align}
where $\b_1=q$ and the effective \CS level $k^{\eff}=k+N_{f}/2-N_{a}/2$ receives corrections from chiral multiplets. 
The mass parameters for fundamental chiral multiplets $\F$ are associated to $\b_i= e^{  -\tilde{m}_i}$, and mass parameters for antifundamental chiral multiplets $\AF$ are associated to $\a_i = e^{  -m_i}$\,. 
The Fayet–Iliopoulos (FI) parameter is $z = e^{\xi}$. We illustrate the assignment of these parameters in Figure \ref{fig:strippara}.

(Anti)fundamental chiral multiplets ($\AF$) $\F$ come from open strings connecting the D3-brane and (anti)fundamental D5-branes that are semi-infinite horizontal lines on the (right) left hand side of the NS5-brane.
The contributions of chiral multiplets $\F$ and $\AF$ to vortex partition functions are the following terms
\begin{align}\label{identifyFAF}
	\AF \rightarrow \left(\a;q \right)_n\,,\quad \F \rightarrow  \frac{1}{\left( \b ;q \right)_n}   \,.
\end{align}
In particular, the term $\left(q;q \right)_n^{-1}$ in the vortex partition function is contributed by the massless $\F$ coming from the massless string locating at the D3-D5-brane interestion.	

The one-loop part cannot be obtained directly using topological vertex but can be borrowed from the localization computation in e.g.\cite{Benini:2014aa}:
\begin{align}
	Z_{U\left(1\right)_k+ N_{f} \, \mathbf{F}+N_{a} \,\mathbf{AF}}^{\text{one-loop}}\left(z,\a,\b \right) = \frac	{ 	\left(\b_1 ;q \right)_\inf \left(\b_2  ;q \right)_\inf \cdots \left(\b_{N_{f}} ;q \right)_\inf	
	} {
		\left(\a_1 ;q \right)_\inf \left(\a_2;q\right)_\inf  \cdots \left(\a_{N_{a}};q \right)_\inf
	} \,.
	\end{align}

Actually, the above vortex partition function is only the  component  on a vacuum of Higgs branch.
The Higgs branch consists of some discrete point $\mathcal{M}_H=\{  \aa_i\,, i=1,\,2, \cdots,\,N_{f} \}$ where $\aa_i$ denote positions of D3-branes, which can end on any fundamental D5-branes. The vacuum of \eqref{vortexNCNAC} is  $\aa_1$ if we denote D5-branes from bottom to top. We will show more examples in section \ref{secmassdef}. The total vortex partition function is the summation of all vacua
\begin{align}
	Z^{\text{total}} = \sum_{ \aa \in \mathcal{M}_H} Z_{\aa  }^{\text{3d}}\,.	
\end{align}
The 3d partition function on a vacuum $\aa$ of the Higgs branch contains three parts:
\begin{align}\label{vortexalpha}
Z_{\aa }^{\text{3d}}=Z^{\text{class}}_\aa \cdot Z^{\text{one-loop}}_\aa \cdot Z^{\text{vortex}}_\aa \,,
	\end{align}
where $Z^{\text{class}}$ is the classical part\footnote{We do not discuss this classical part in this note, since it is a constant factor.}.	
In this note, we mainly discuss the component $Z_{\aa}^{\text{3d}}$, as we will shown latter that vortex partition function components are equivalent to each other. For discussions or examples on other aspects, see e.g. \cite{Beem:2012mb,Zenkevich:2017ylb,Nieri:2015yia,Aprile:2018oau}.


\subsection{Quiver matrices}
Following the recent development in knot theory \cite{Kucharski:2017poe,Kucharski:2017ogk,Ekholm:2018eee,Ekholm:2019lmb, Cheng:2021aa}, we define the quiver as the symmetric matrix $C_{ij}$ that involves the generating function:
\begin{align} \label{quivergenfun}
P_{C_{ij}}\left({q; x_1, \cdots, x_N} \right):=\sum_{  d_1,...,d_N=0}^{\inf}  
\left(-\sqrt{q} \right)^{ \sum\limits_{i,j=1}^N C_{ij}d_i d_j} 
\frac{
	x_1^{d_1}	x_2^{d_2}\cdots x_N^{d_N}
} 
{ 
	\left(q;q\right)_{d_1}	\left(q;q \right)_{d_2}\cdots \left(q;q\right)_{d_N} }
\,.
\end{align}
This matrix $C_{ij}$ is called quiver as it can be represented as the quiver graph that contains arrows and nodes.
This quiver generating function \eqref{quivergenfun} has a product decomposition form:
\begin{align}\label{DTgen}
P_{C_{ij}}\left(q;x_1,x_2,\cdots,x_m \right) =\prod_{d_1,...,d_m=0}\prod_{ J \in \mathbb{Z}} \prod_{n=0}^{\inf} \left( 1- q^{n+\frac{J-1}{2} }  x_1^{d_1}\cdots x_N^{d_N}  \right) ^{\left(-1\right)^{J+1} \Omega_{d_1,...,d_N;J}   }  \,,
\end{align}
where $\Omega_{d_1,...,d_N;n} $ are integer motivic Donaldson-Thomas (DT) invariants \cite{Kontsevich:2010px} which are interpreted as BPS invariants in many physical contexts. 

In \cite{Cheng:2020aa}, these quiver matrices are interpreted as effective  mixed \CS levels of $ST$-dual theories.


\subsection{$ST$-dual theories}\label{stdualtheory}
The \CS terms in the Lagrangian of 3d $\N=2$ theories enjoy the $ST$-transformation which satisfies $\left(ST\right)^3=1$, where $S$ and $T$ are generators in  ${SL}\left(2 \,, \mathbb{Z}\right)$\footnote{Please do not confuse it with the $SL(2,\mathbb{Z})$ symmetry of brane webs in the type-IIB string theory.}. The $T^{k}$-operator shifts the \CS level by $k$, and the $S$-operator gauges one flavor symmetry and hence introduces a $U\left(1\right)$ gauge group \cite{Witten:2003ya,Dimofte:2011ju}. Since $ST$-transformations preserve partition functions, we refer to  these dual theories obtained by this transformation as $ST$-dual theories.

One well known example is the mirror pair: 
\begin{align}\label{mirrorpair}
1\, \F~\xlongleftrightarrow{{ST}} ~ U\left(1\right) + 1\F  \,.
\end{align}
which is the duality between a free chiral multiplet and a theory with the gauge group $U(1)$ and one chiral multiplet \cite{Boer:1997ts}.
The action of $ST$-transformation on the chiral multiplet is viewed as gauging the flavor symmetry.
For this mirror pair, the $ST$-transformation is the mirror symmetry that exchanges Higgs branch and Coulomb branch. 

The $ST$-transformations can also be applied on chiral multiplets that are coupled to gauge nodes. These chiral multiplets can be in fundamental representation $\F$, antifundamental representation $\AF$, bifundamental representation, and so on. In this case, $ST$-transformations lead to mixed CS terms between gauge groups. For instance, the theory $U\left(1\right)_k+N_{f}\F+N_{a} \AF$ can be transformed into abelian theories with  gauge group $U\left(1\right)^{ N_{f}+N_{a}}$ after integrating out the original gauge group $U\left(1\right)$. One can continue performing $ST$-transformations on each chiral multiplet and obtain a group of $ST$-dual theories that consist of many building blocks $U(1)+1\F$ coupled together by  mixed \CS levels $k_{ij}$:
\begin{align}\label{striptheory}
	U\left(1\right)_k+N_{f} \F+N_{a}\AF  ~~ \xrightarrow{~ST~} ~~ \Big\{ ~\left(U\left(1\right) + 1\F\right)_{k_{ij}}^{N_{f}+N_{a}}  \Big\}  \,.
\end{align}
Note that these $ST$-dual theories have different  mixed \CS levels $k_{ij}$ and each gauge node has a FI parameter. Since they are obtained by equivalent transformations ($ST$-transformations), these theories are dual to each other, and their FI parameters are related to match partition functions.

Computation shows vortex partition functions of $ST$-dual theories take form \eqref{quivergenfun} and $C_{ij}$ are effective mixed CS levels
$
	 C_{ij} = k^{\eff}_{ij}
$.
More explicitly, each vortex partition function \eqref{vortexalpha} can be written as the form \eqref{quivergenfun} if we ignore the classical part
\cite{Panfil:2018faz, Cheng:2021aa} and use properties of $q$-Pochhammer products \eqref{quiverfrompoch1} and \eqref{quiverfrompoch2}:
\begin{align}\label{vortextoPc}
	Z_{ \aa }^{\text{3d}} (q;z,\a,\b)=  P_{k_{ij}^{\eff}}\left(q;{x_1, \cdots, x_N} \right)  \,,
\end{align}
where FI parameters $x_i=\xi_i^{\eff} = e^{\pm m_i} ~( \text{or}~e^{\pm \tilde{m}_i})$.
The right hand side is the 3d vortex partition function of the $ST$-dual theory. Note that $ST$-dual theories are massless after absorbing mass parameters into FI parameters. The one-loop parts of $ST$-dual theories are trivial.
 For more discussions on these theories, see \cite{Dorey:1999rb,Cheng:2020aa}.

$ST$-dual theories  in \eqref{striptheory}  are not unique. They have different effective CS levels, but their vortex partition functions are equivalent:
\begin{align}
	P_{k_{ij}^{\eff}}\left(q;{ \cdots, x_l, \cdots} \right)   ~\xlongequal{ST}~ P_{k_{ij}^{'\eff}}\left(q;{ \cdots, x_l^{-1}, \cdots} \right) ~\xlongequal{ST}~  \cdots \,.
\end{align}
where we get a group of $\{ k_{ij}^{\eff} \}$, which are related to each other by flipping FI parameters and any $x_l$ can be flipped \cite{Cheng:2020aa}. 
We define the flip as
\begin{align}\label{flipmass}
	m_i \rightarrow - m_i \,~\text{or}\,~\tilde{m}_j \rightarrow - \tilde{m}_j, ~~\forall \,i \in 1\,,\dots\,,N_{f}\,,~~\forall\, j \in 1\,,\dots\,,N_{a} \,.
\end{align}

Let us show explicitly the mixed CS levels of the theory $U(1)_k +N_f \F+N_a \AF$ to illustrate. The effective CS levels of its $ST$-dual theories are
\begin{align}\label{STdualpattern}
	\begin{split}
\includegraphics[width=5in]{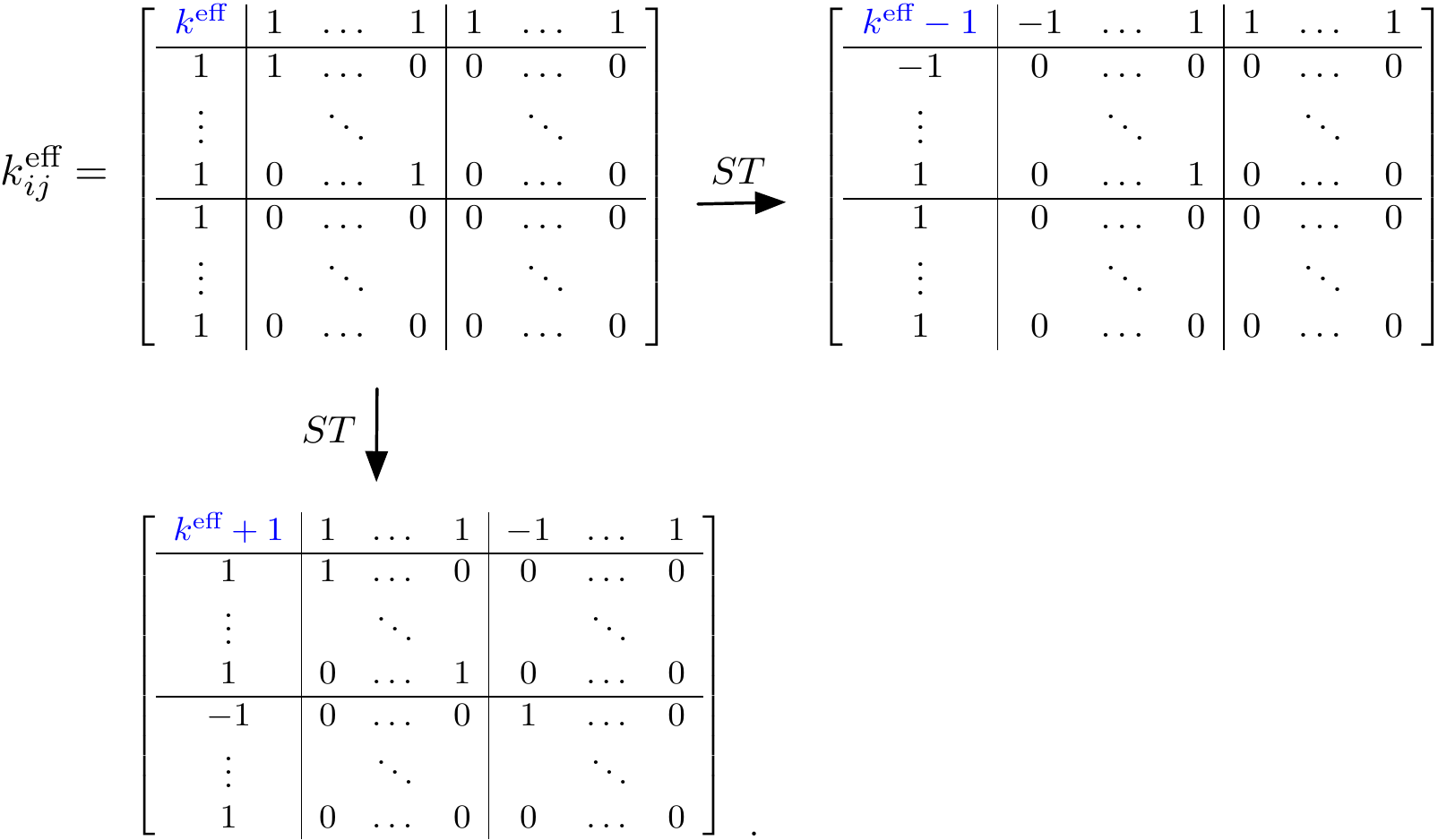} 
\end{split}
\end{align}
in which different mixed CS levels are related by $ST$-transformations.
Note that the effective mixed CS levels of $ST$-dual theories satisfy $k^{\eff}_{ij}=k_{ij}+ \delta_{ij}/2 \in \mathbb{Z} $, which is the constraint imposed by parity anomaly. The $1/2$ in second term is contributed by $\F$. In \eqref{STdualpattern}, $k^{\eff}$ is the effective CS level of the original theory $U(1)_k+N_f\F+N_a\AF$ and $k^{\text{eff}} = k +\frac{N_{f} - N_{a} }{2}  \in \mathbb{Z} $.



We can separate the $k_{ij}^{\eff}$ into matrix components, and each of them is contributed by a chiral multiplet. Flipping  the sign of the chial multiplet changes the mixed CS level component:
\begin{align}\label{FtoC}
	& \F: \qquad ~~ 
	\left[
	\begin{array}{cc}
		~	\color{blue}{0} &~1	  \\
		~1& ~1 \\
	\end{array}
	\right]  ~~ 
	\xrightarrow{ m_i \rightarrow -m_i }
	~~
	\left[
	\begin{array}{cc}
			\color{blue}{-1} &-1	 \\
		-1&~0 \\
	\end{array}
	\right]  \,, \\
	& \AF:\qquad
	\left[
	\begin{array}{cc}
		~	\color{blue}{0}&~1	  \\
		~1& ~0 \\
	\end{array}
	\right] 
	~~ 
	\xrightarrow{ \tilde{m}_i \rightarrow -\tilde{m}_i }
	~~
	\left[
	\begin{array}{cc}
		~\color{blue}{1} &-1	  \\
		-1&~ 1 \\
	\end{array}
	\right]  \,. \label{AFtoC}
\end{align}
One can derive this by expanding contributions of chiral multiplets \eqref{identifyFAF}, using \eqref{quiverfrompoch1} and \eqref{quiverfrompoch2} \footnote{Note that the first element marked in blue is special, since it is associated to the original FI parameter of $U(1)_k+N_{f}\F+N_a\AF$.  }.
The blue element stands for the first element $k_{0\,0}^{\eff} $. If we flips the real mass parameter of any chiral multiplet, then mixed CS levels change accordingly. More explicitly, if we flips a $\tilde{m}_j$, then the first element $k^{\eff}_{0\,0}$ increases by one, namely $k^{\eff}+1$, and if we flips a ${m}_j$, then the first element  minus  one, namely $k^{\eff} -1$. 


 In section \ref{secSTtrans}, we will show that flip is interpreted as $ST$-transformation for chiral multiplets, using holomorphic blocks.
In section \ref{secmassdef}, we will show that the flip of mass parameters can be represented in terms of 3d brane webs, since positions of D5-branes are associated with mass parameters.  Hence flipping mass parameters can be viewed as flipping D5-branes.
One may wonder if there is a problem with the vacuum, since the vacuum $\aa \in \mathcal{M}_H$ has many choices. Our answer is that any vacuum gives rise to the same group of $ST$-dual theories. Changing the vacuum $\aa$ only permutes these $ST$-dual theories. Different vacua are exchanged by flipping D5-branes. We also observe the correspondence between mixed CS levels of $ST$-theories and 3d brane webs. 

\section{Holomorphic blocks }\label{holoblocks}

If the spacetime of 3d theories is not compact, one needs to add a 2d theory on boundaries \cite{Honda:2013uca,Yoshida:2014ssa,Longhi:2019hdh,Beem:2012mb}. In this case one should consider holomorphic blocks of 2d-3d coupled theories
 \cite{Beem:2012mb}, which are lifted versions of 3d vortex partition functions.
After doing this, anomalies of 3d $\N=2 $ theories cancels with that of 2d $\left(0,2\right)$ theories on the boundary of the spacetime $\mathbb{R}^2 \times S^{1}$.

Contributions from 2d boundary theories are some theta functions, which  in  \cite{Beem:2012mb} are shown to be the lifted versions of 3d \CS terms $e^{\frac{1}{2}k_{ij} X_i X_j}$ after taking into account fluxes, where $X_i:= \log\, x_i$ is the variable for the gauge group $U\left(1\right)_i$. Namely,
\begin{align}
e^{-\half X^2} = e^{ - \frac{1}{2} \left( \log x\right)^2 }  \rightarrow ~\theta\left(x;q\right) \,.	
	\end{align}
 The theta function is defined as
\begin{align}
	\theta\left(x;q\right):=\left( - q^{\half} x;q\right)_\inf \left(- q^{\half} x^{-1};q \right)_\inf  = \left(q;q\right)^{-1}_\inf \sum_{n \in \mathbb{Z}}  \left(-\sqrt{q}\right)^{ {n^2} } \left(-x\right)^n \,,
\end{align}
which satisfies 
\begin{align}\label{thetaproperty}
	\theta\left(- q^{-\half} x;q \right)= \left(x;q\right)_\inf \left(q x^{-1};q \right)_\inf \,,~~\theta\left(x;q\right)=\theta\left(x^{-1};q\right) \,,~~ \frac{\theta\left(q^nx;q\right)}{\theta\left(x;q\right)} = q^{\frac{n^2}{2}} x^{-n} \,.	
\end{align}
We often use shorthand $\theta\left( z\right):=\theta\left( z;q\right)$ in this note.

In \cite{Yoshida:2014ssa}, it is shown that these theta functions are contributions of Fermi multiplets and chiral multiplets of 2d $\left(0,2\right)$ theories. Holomorphic blocks are defined as  partition functions of 2d-3d coupled theories \cite{Beem:2012mb}.

The 3d chiral multiplets can be given Dirichlet or Neumann boundary conditions \cite{Yoshida:2014ssa}.
The one-loop contributions of 3d chiral multiplets are
\begin{align}
	Z^{\text{chiral},~ \D}_{\text{1-loop} ,\F} &= \left(q z^{-1};q\right)_\inf^{\D} \,, ~~~~Z^{\text{chiral}, ~\NS}_{\text{1-loop}, \F} =\frac{1} {\left( z;q\right)_\inf^{\NS}} \,, \\
	Z^{\text{chiral},~ \D}_{\text{1-loop} ,\AF} &= \left(q z;q\right)_\inf^{\D} \,,	~~~~~~Z^{\text{chiral}, ~\NS}_{\text{1-loop}, \AF} =\frac{1} {\left( z^{-1};q\right)_\inf^{\NS}} \,,
\end{align}
where $\F$ denote fundamental chiral multiplets and $\AF$ denote antifundamental chiral multiplets, and $z = e^{\rho(\sigma) + F_l m_l}$ where $\rho(\sigma)$ is the fundamental weight of the scalar in the vector multiplet, and $F_l$ are charges for global symmetries \cite{Yoshida:2014ssa}. We use $\D$ to stand for  Dirichlet boundary conditions and N to stand for Neumann boundary conditions.

 \subsection{Abelian theories }
 In this section, we compute the holomorphic block of the theory $U\left(1\right) + N_{f} \F+N_a \AF$, which is the main example that we will discuss later.

The 3d partition function on a vacuum \eqref{vortexalpha} is lifted to the holomorphic block
\begin{align}
Z_{\aa}^{\text{3d}}~~	 \rightarrow ~~\B_\aa = \int_{\Gamma_\aa} \frac{d s}{s}\Upsilon \,,
	\end{align}
where $\Upsilon$ contains one-loop contributions. Each convergent contour $\Gamma_\aa$ gives rise to a component of the 2d-3d partition function $\B_\aa $, and $\aa$ is the vacuum on Higgs branch, which corresponds to the flat connection on the three manifold in 3d-3d correspondence, see e.g. \cite{Beem:2012mb,Chung:2016aa}.

For the theory  $U\left(1\right)_k + N_{f} \F+N_a \AF$, we have
\begin{align}\label{blocks}
&\B_\aa = \int_{\Gamma_\aa} \frac{d s}{2 \pi i\,s}\Upsilon =\int_{\Gamma_\aa} \frac{d s}{2 \pi i\,s}\, Z^{\text{2d}} \,
\cdot Z^{\text{3d}}
 \,,\\
 &Z^{\text{2d}} =  \frac{1}{\theta\left( - q^{-\frac{1}{2}}s\right)^{k^{\eff}}}  \cdot
 \frac{ \theta\left(z\right)\,\theta\left( -q^{-\half} s \right)    }{  \theta \left(-q^{-\half} s z\right) } \cdot 
  \prod_{j=1}^{N_a}   \frac{ \theta\left(\a_i\right)\,\theta\left( -q^{-\half} s \right)     }{  \theta \left(-q^{-\half} s \a_i^{-1}\right) } 
\cdot 
  \prod_{j=1}^{N_{f}}   \frac{ \theta\left(\b_j\right)\,\theta\left( -q^{-\half} s \right)     }{  \theta \left(-q^{-\half} s \b_j^{-1}\right) } \,,    \label{2dblocks} \\
      &Z^{\text{3d}}= \frac{ \prod_{j=1}^{N_{f}}\left( q \b_j s^{-1};q \right)_{\inf}^{\D}   }{ \prod_{i=1}^{N_a}  \left(  \a_i s^{-1}  ;q\right)_\inf^{\NS}  } \,,  \label{3dpartblock}
\end{align}
where the first term of the 2d boundary theory comes from the effective \CS level
\begin{align}
	\exp\left( \frac{1}{2} k^{\eff}  s^2 \right) ~~\rightarrow~~ \frac{1}{\theta\left( s;q\right)^{k^{\eff}}} \,.
\end{align}
The rest terms of $Z^{2\dd}$ come from 3d FI parameters. We follow \cite{Beem:2012mb} to write down this holomorphic block to match with \eqref{vortexNCNAC}. Notice that both gauge anomalies and gauge-flavor anomalies need to be canceled \cite{Yoshida:2014ssa}.
The vacua are Higgs branch $\mathcal{M}_H = \{s = s_\aa \,, \aa=1,2,\cdots, N_{f} \} $\footnote{In this note, we use $s_\aa$ or its index $\aa$ interchangeable to denote vacua for the lack of symbols. } and correspond to poles of contributions of 3d fundamental chiral multiplets $\left(q \b_j s^{-1};q\right)_\inf$, whose poles are $s_\aa^{}=\b_\aa q^{-n}$ with $n= 0, 1, 2, \cdots$ and $\aa=1,2,\cdots, N_{f}$.

After taking the values of poles into each term of \eqref{blocks} and  using identities
\begin{align}
	\left(Q q^n;q\right)_\inf = \frac{\left( Q;q\right)_\inf  }{ \left( Q;q \right)_n } \,, \quad \left(  Q q^{-n} ;q\right)_\inf  = \frac{  \left(Q q^{-1} ;q^{-1}\right)_n }{ \left(  Q q^{-1} ; q^{-1} \right)_\inf  }  \,,
\end{align}
one can get the residue of each term
\begin{align}
	&\theta\left(-q^{-\half}s_\aa y\right) =  \theta\left( -q^{-\half} \b_\aa y\right) \,\left(-\sqrt{q}\right)^{- n^2}  \left( \frac{\b_\aa y}{\sqrt{q}}\right)^n \,, \\
	& \left(q \b_j s^{-1}_{\aa} ;q \right)_\inf = \frac{ \left(q \b_j/\b_{\aa} ;q \right)_\inf  }{ \left( q \b_j/\b_{\aa} ;q  \right)_n  } \,, \quad \frac{1 } { \left( \a_i s_{\aa}^{-1} ;q  \right)_\inf  }  =\frac{ \left( \a_i/\b_{\aa} ;q \right)_\inf  }{ \left(  \a_i/\b_{\aa} ;q  \right)_n  } \,.
	\end{align}
Finally, the holomorphic block takes the form 
\begin{align}
&	\mathcal{B}_{\aa} =Z^{\text{2d}}_\aa
\cdot  Z^{3 \dd~\text{one-loop}}_\aa \cdot
Z^{3\dd~	\text{vortex}}_\aa  \,,\\
&Z^{\text{2d}}_\aa =\frac{ \theta\left(z\right)  \theta\left(-q^{ -\half}  \b_\aa \right)^{1+N_{f}+N_a-k^{\eff}}  } {  \theta\left( -q^{-\half} \b_\aa z \right)  }
\cdot
 \prod_{i=1}^{N_a}   \frac{ \theta\left(\a_i\right)   }{  \theta \left(-q^{-\half} \b_\aa/\a_i\right) } \cdot
\prod_{j=1}^{N_{f}}   \frac{ \theta\left(\b_j\right)  }{  \theta \left(-q^{-\half} \b_\aa/\b_j\right) } \,,
 \\
& Z^{3\dd~\text{one-loop}}_\aa	  = \frac{ \prod_{j=1}^{N_{f}} \left(q \b_j/\b_{\aa} ;q\right)_\inf} {\prod_{i=1}^{N_a} \left( \a_i/\b_{\aa} ;q\right)_\inf}    \,,  \label{3doneloop}\\
& Z^{3\dd~\text{vortex}}_\aa =\sum_{n=0}^{\inf}  
\left(-\sqrt{q}\right)^{k^{\eff} n^2}
\left(  \left( \frac{\b_\aa}{\sqrt{q}}  \right)^{-k^\eff} z^{-1} \prod_{j=1}^{N_{f}} \b_j^{} \prod_{i=1}^{N_a} \a_i^{} \right)^n \cdot \frac{\prod_{i=1}^{N_a} \left( \a_i/\b_{\aa} ;q\right)_n}{\prod_{j=1}^{N_{f}} \left(q \b_j/\b_{\aa} ;q\right)_n}  \,. \label{vortexcomplete}
	\end{align}
In order to match \eqref{vortexcomplete} with \eqref{vortexNCNAC}, one needs to shift 
\begin{align}
	\b_j \rightarrow \b_j \b_\aa q^{-1} \,,~~	\a_i \rightarrow \a_i \b_\aa q^{-1} \,,
	\end{align}
which is not necessary for discussing physics, so we do not shift parameters for simplicity in this section.

Note that if we choose the chamber $m_{N_{f}} > \cdots >m_2 > m_1$, then $\log\left(\b_j/\b_a\right)=m_j-m_\aa \geqslant0$ if $j\geqslant \aa$, and $\log\left(\b_j/\b_a\right)= m_j-m_{\aa}<0$ if $j<\aa$. This is consistent with the flips of K\"ahler parameters in open topological amplitudes and the assignment of mass parameters for 3d brane webs that will be discussed in section \ref{phasesabelian}. For example, we have the equivalence \eqref{shiftKahler} for the theory $U(1)_k+2\F$. On the vacuum $A$, the K\"ahler parameter is $e^{-(m_2-m_\aa)} = e^{-(m_2-m_1)}  = Q_1$, while on the vacuum $B$,  it is $e^{-(m_1-m_\aa)} = e^{-(m_1-m_2)}  = Q_1^{-1}$.

Taking the limit $\hbar \rightarrow 0$, then $q \rightarrow 1$, one can get the leading term
\begin{align}
	\Upsilon \xrightarrow{\hbar \rightarrow 0}    \exp \left( \frac{1}{\hbar} \widetilde{\W} \right) \,.
	\end{align}
At this limit $\left(x;q\right)_\inf \sim e^{ \frac{1}{\hbar}  \text{Li}_2\left(x\right)}$. The leading terms is the effective superpotential. For abelian theories, we get\footnote{where we use the identity
$ \Li_2 \left(x\right) + \Li_2\left(x^{-1}\right) = -\frac{\pi^2}{6} -\frac{1}{2} \left(\log\left(-x\right)\right)^2  $. }
\begin{align}\label{Weff}
	\tW  = & \sum_{j=1}^{N_{f}} \Li_2\left( \frac{\b_j}{s} \right)  
	-\sum_{i=1}^{N_a} \Li_2\left( \frac{\a_i}{s} \right) 
	+ \frac{ k^{\eff}}{2} \left(\log\left(-s\right)\right)^2 + 
	\log\left(-s\right) \left( \log\,z-\sum_{j=1}^{N_{f}}\log\, \b_j   - \sum_{i=1}^{N_a}\log\, \a_i  \right) 
	 \,,
\end{align}
where $\b_j$ and $\a_i$ are associated with mass parameters through $\b_j= e^{-m_j}$ and $\a_i = e^{-\widetilde{m}_i}$\footnote{Here the mass parameters $m_i$ and $\tm_j$ contain signs that can be positive or negative, while in the following sections, signs of mass parameters are shown on the brane webs, for instance, the brane web in Figure \ref{fig:turnonmass}.}. The superpotential \eqref{Weff} matches with examples in \cite{Gadde_2014}. Notice that the 3d theory only contributes the first and second terms in \eqref{Weff}, which the 2d boundary theory contributes the rest terms which are CS terms and the FI term. We turns on the FI term in \eqref{2dblocks} to match effective superpotentials obtained in other ways, such as sphere partition functions and open topological string amplitudes.

\subsection{Exchange boundary conditions}

As it is discussed in \cite{Dimofte:2017tpi}, boundary conditions $\D$ and $\NS$ can be exchanged by a theta function
\begin{align}\label{NDexchange}
	\left(q \b_j s^{-1};q \right)_\inf^\D  = \frac{ \theta\left(-q^{-\half} \b_j^{-1} s \right)}{ \left( \b_j^{-1} s;q\right)_\inf^{\NS}  } \,,
	\end{align}
which is interpreted as the $T$-transformation in \cite{Gadde_2014}, which is the operator $T \in SL(2 , \mathbb{Z})$
After taking the poles $s_\aa=\b_\aa q^{-n}$ with $n \in \mathbb{N}$, one can find both sides of \eqref{NDexchange} give rise to the same result
\begin{align}
\frac{\left(q \b_j/\b_\aa;q\right)_\inf}{ \left(q \b_j/\b_\aa; q\right)_n  } \,.
	\end{align}
Therefore 3d partition functions \eqref{3doneloop} and \eqref{vortexcomplete} are not sensitive to boundary conditions. The 3d part \eqref{3dpartblock} in holomorphic blocks can be given boundary conditions randomly, and then the 2d part \eqref{2dblocks} changes accordingly.


In terms of effective superpotentials,  exchanging boundary conditions is equivalent to
\begin{align}
\Li_2\left(e^{-Z}\right) ~=~ -\Li_2\left(e^Z\right) - \frac{Z^2}{2} \,,
\end{align} which is an identity.
Explicitly, exchanging boundary conditions leads to the following change to effective superpotential \eqref{Weff}  \begin{align}
 	\Li_2\left( \frac{\b_j}{s}\right) ~=~	-\Li_2\left( \frac{s}{\b_j} \right) - \frac{1}{2} \log\left(-s\right)^2 -\frac{1}{2 } \log\left(\b_j\right)^2 + \log \left(-s\right) \, \log\left(\b_j\right) \,.
 \end{align}

\subsection{$SL(2, \mathbb{Z})$-transformation}\label{secSTtrans}

The mirror pair \eqref{mirrorpair} gives rise to an identity for their holomorphic blocks
\begin{align}\label{STtransf}
	ST:\quad	\left(q;q\right)_\inf \left(q x^{-1};q \right)_\inf^{\D} = \int \frac{ d z}{z} \frac{ \theta\left( -q^{-\half}x;q\right)}{ \theta \left( -q^{-\half} x z;q\right) } \cdot \left( q z^{-1};q\right)_\inf^{\D} \,,
\end{align}
which is the $ST$-transformation in terms of holomorphic blocks \cite{Beem:2012mb,Gadde_2014}. We can ignore the normalization factor $\left(q;q\right)_\inf$.
Note that \eqref{STtransf} can be written as the following 
\begin{align}\label{STtrans2}
	\frac{1}{\left(x;q\right)^{\NS}_\inf} = \int \frac{d z}{z} \frac{1}{ \theta\left( -q^{-\half} xz ;q\right)} \cdot \left(q z^{-1} ;q \right)_\inf^\D \,.
\end{align}
This suggests that this form of $ST$-transformation not only gauges the flavor symmetry but also changes the boundary conditions and representations of chiral multiplets.


The Fourier transformation of \eqref{STtrans2} is
\begin{align}\label{STtransf3}
	{\left(q x^{-1};q\right)^{\D}_\inf} = \int \frac{d z}{z} { \theta\left( -q^{-\half} x z  ;q\right)} \cdot \frac{ 1}{\left(z ;q \right)_\inf^{\NS}} \,,
\end{align}
which should be the $\left(ST\right)^2$-transformation, since $(ST)^3=1$.

In \cite{Gadde_2014,Yoshida:2014ssa}, the $S$-transformation for the chiral multiplet with Neumann boundary condition is found to be
\begin{align}
	\frac{1}{\left(x;q\right)_{\inf}^{\NS}} = \int \frac{d z}{z} \frac{\theta\left(-q^{-\half} z;q \right) \theta\left(-q^{-\half} x;q  \right) }{ \theta\left( -q^{-\half} x z;q\right) } \cdot \frac{ 1}{\left(z;q\right)^{\NS}_{\inf}  } \,,
\end{align}
which can be written as follows using \eqref{thetaproperty}
\begin{align}
	\frac{1}{\left(x;q\right)_{\inf}^{\NS}} = \int \frac{d z}{z} \frac{  
		\theta\left(-q^{-\half} x;q  \right) }{ \theta\left( -q^{-\half} x z;q\right) } \cdot  \left(q z^{-1};q\right)^{\D}_{\inf}  \,,
\end{align}
which matches with \eqref{STtransf} after exchanging the boundary condition of the chiral multiplet. This verifies again that exchanging boundary conditions \eqref{NDexchange} is $T$-transformation.

\vspace{5mm}
Let us summarize transformations in terms of holomorphic blocks. Then we discuss how these $SL(2, \mathbb{Z})$ transformations change CS levels.
	
\subsection*{$T$-transformation}
$T$-transformation exchanges boundary conditions:
\begin{align}
	\left(q  z^{-1};q \right)_\inf^\D  = \frac{ \theta\left(-q^{-\half} z;q \right)}{ \left(  z;q\right)_\inf^{\NS}  } \,,
\end{align}
which inverse the variable $z$. Note that  this is not the flip that we mentioned in \eqref{flipmass}. The story is a bit complicated. The flips of mass parameters from the perspectives of $ST$-dual theories should be $ST$-transformations.

\subsection*{$ST$-transformation}
$ST$-transformation introduces  not only a theta function, but also an additional gauge group. $ST$-transformation also leads to mixed \CS levels that come from the 2d boundary theory.

The $ST$-transformation is the mirror symmetry for the chiral singlet, and takes different forms:
\begin{align}
\left(q x^{-1};q \right)_\inf^{\D} &= \int \frac{ d z}{z} \frac{ \theta\left( -q^{-\half}x\right)}{ \theta \left( -q^{-\half} x z\right) } \cdot \left( q z^{-1};q\right)_\inf^{\D} \,, \label{ST1} \\
\frac{1}{\left( x;q \right)_\inf^{\NS}} &= \int \frac{ d z}{z} \frac{1}{ \theta \left( -q^{-\half} x z\right) } \cdot \left( q z^{-1};q\right)_\inf^{\D} \,, \label{ST2}\\
	\frac{1}{\left( x^{};q\right)^{\NS}_\inf} &= \int \frac{d z}{z} \,\frac{  \theta\left( -q^{-\half} z \right)}{ \theta\left( -q^{-\half} x z \right)  } \cdot \frac{ 1}{\left(z ;q \right)_\inf^{\NS}}  \,, \label{ST3} 	\\
	\left(q x^{-1};q \right)_\inf^{\D} &= \int \frac{ d z}{z} \frac{ \theta\left( -q^{-\half}x\right) \theta\left(-q^{-\half} z  \right)}{ \theta \left( -q^{-\half} x z\right) }  \cdot \frac{ 1}{\left(z ;q \right)_\inf^{\NS}}  \,, \label{ST4}
\end{align}
which can be used to turn the theory $U\left(1\right)_k +N_{f} \F$ into its $ST$-dual theories  $\left(U\left(1\right) + 1\, \F\right)_{k_{ij}}^{\otimes N_{f}} $ with mixed \CS levels between gauge nodes. 

\subsection*{$(ST)^2$-transformation}
Different forms of the $(ST)^2$-transformation are Fourier transformations of (\ref{ST1}-\ref{ST4}) respectively:
\begin{align}
	\frac{1}{\left(x;q \right)_\inf^{\NS}} &= \int \frac{ d z}{z} \frac{ \theta \left( -q^{-\half} x z\right) }{ \theta\left( -q^{-\half}x\right)} \cdot 	\frac{1}{\left(z;q \right)_\inf^{\NS}} \,,\label{STST1} \\
	{\left( q x^{-1};q \right)_\inf^{\D}} &= \int \frac{ d z}{z} { \theta \left( -q^{-\half} x z\right) } \cdot \frac{1}{\left( z^{};q\right)_\inf^{\NS} }\,, \label{STST2} \\
	{\left(q x^{-1} ;q \right)_\inf^{\D}} &= \int \frac{d z}{z} \,\frac{ \theta\left( -q^{-\half} x z \right)  }{  \theta\left( -q^{-\half} z \right)}\cdot   	{\left(q z^{-1} ;q \right)_\inf^{\D}}  \,, \label{STST3}	\\
\frac{ 1}{\left(x ;q \right)_\inf^{\NS}} 	 &= \int \frac{ d z}{z} \frac{ \theta \left( -q^{-\half} x z\right) } { \theta\left( -q^{-\half}x\right) \theta\left(-q^{-\half} z  \right)} \cdot \left(q z^{-1};q \right)_\inf^{\D}  \,, \label{STST4}
\end{align}
\subsection*{$S$-transformation}
For the antifundamental chiral multiplet, we have the $S$-transformation:
\begin{align}\label{STforN1}
	\frac{1}{\left(x;q\right)_{\inf}^{\NS}} &= \int \frac{d z}{z} \frac{\theta\left(-q^{-\half} z\right) \theta\left(-q^{-\half} x  \right) }{ \theta\left( -q^{-\half} x z\right) } \cdot \frac{ 1}{\left(z;q\right)^{\NS}_{\inf}  } \,,   
\end{align}
which can be written as
\begin{align}\label{STforN2}
	\frac{1}{\left(x;q\right)^{\NS}_\inf} &= \int \frac{d z}{z} \frac{\theta\left( -q^{-\half} x\right)}{ \theta\left( -q^{-\half} xz\right)} \cdot \left(q z^{-1} ;q \right)_\inf^\D \,,
\end{align}
Similarly, one can also write down its Fourier transformation, which only differs to $ST$-transformations by a theta function.
We do not further discuss this $S$-transformation in this note. 

\vspace{5mm}
In the following, we analyze $ST$- and $(ST)^2$-transformations of chiral multiplets and show how these operations lead to mixed CS levels, using the fact that $\theta\left(xz;q\right)$ leads to the mixed term 
\begin{align}
	\theta\left(xz;q\right)^{\pm}~~ \rightarrow~~ 	\exp \left(  \mp\frac{1}{2 } \left( \log \left(x z\right)\right)^2  \right) =~ \exp \left( \mp\frac{1}{2} X^2  \mp \frac{1}{2} Z^2 \mp X Z \right) \,,
\end{align}
where $\log\,x := X$ and $\log\, z:=Z$.

\subsubsection*{Fundamental chiral multiplets}
Using \eqref{ST1} or \eqref{ST4}, one can find the $ST$-transformation leads to a mixed term in the effective superpotenital:
\begin{align}\label{FD}
	\Li_2\left( e^{-X}\right)  \rightarrow  \Li_2\left( e^{-Z}\right) +\frac{1}{2} (-Z)^2 + (-X) (-Z) \,,
\end{align}
which contributes to effective  mixed \CS levels:
\begin{align}\label{STofF1}
		\left[
	\begin{array}{cc}
		 k_{(-X)(-X)}&~ k_{(-X)(-Z)}	  \\
		 k_{(-Z)(-X)}& ~k_{(-Z)(-Z)} \\
	\end{array}
	\right] 
=
	\left[
	\begin{array}{cc}
		~0&~1	  \\
		~1& ~1 \\
	\end{array}
	\right]  \,.
	\end{align}

Using \eqref{STST2} or \eqref{STST3}, one can find $\left(ST\right)^2$-transformation leads to
a mixed term:
\begin{align}\label{FN}
	\Li_2\left( e^{-X}\right)  \rightarrow \Li_2\left( e^{-Z}\right) -\frac{1}{2} (-X)^2  - (-X) (-Z)  
	   \,,
\end{align}
which contributes to effective mixed \CS levels:
\begin{align}\label{STofF2}
		\left[
\begin{array}{cc}
	 k_{(-X)(-X)}&~ k_{(-X)(-Z)}	  \\
k_{(-Z)(-X)}& ~k_{(-Z)(-Z)} \\
\end{array}
\right] 
=
	\left[
	\begin{array}{cc}
		-1&-1	  \\
		-1& ~0 \\
	\end{array}
	\right]  \,.
\end{align}

Note that the mass parameter becomes the effective FI parameter   $X = \xi^{\eff}$ after $ST$-transformation. The mixed CS terms \eqref{STofF1} and \eqref{STofF2} match with \eqref{FtoC}. 
The right hand side of \eqref{ST1} and \eqref{STST3} only differ by some theta functions, and this is caused by another $ST$-transformation on the 2d boundary theory as $(ST)^2= ST \, ST$. Therefore, this verifies that the flip in \eqref{flipmass} is caused by the $ST$-transformation.

\subsubsection*{Antifundamental chiral multiplets}
Using \eqref{ST2} and \eqref{ST3}, one can find the $ST$-transformation leads to some mixed terms in the effective superpotenital: 	
\begin{align}
	-\Li_2\left( e^X\right)  \rightarrow  \Li_2\left( e^{-Z}\right) +\frac{X^2}{2}+ \frac{Z^2}{2}  - X \left(-Z\right) \,,
\end{align}
which contributes to effective mixed \CS levels:
\begin{align}\label{STofAF1}
		\left[
\begin{array}{cc}
	k_{XX}&~ k_{X(-Z)}	  \\
	k_{(-Z)X}& ~k_{(-Z)(-Z)} \\
\end{array}
\right] 
=
	\left[
	\begin{array}{cc}
		~1&-1	  \\
		-1& ~1 \\
	\end{array}
	\right]  \,.
\end{align}

Using \eqref{STST1} and \eqref{STST4}, one can find $\left(ST\right)^2$-transformation leads to a mixed term:
\begin{align}
	-\Li_2\left( e^X\right)  \rightarrow  \Li_2\left( e^{-Z}\right) 	 + X \left(-Z\right) \,,
\end{align}
which gives rise to mixed \CS levels
\begin{align}\label{STofAF2}
		\left[
\begin{array}{cc}
	k_{XX}&~ k_{X(-Z)}	  \\
k_{(-Z)X}& ~k_{(-Z)(-Z)} \\
\end{array}
\right] 
=
	\left[
	\begin{array}{cc}
		~0&~1	  \\
		~1&~0 \\
	\end{array}
	\right] 
\end{align}
The mixed CS terms \eqref{STofAF1} and \eqref{STofAF2} match with \eqref{AFtoC}.


\subsection*{Elliptic stable envelope and further}
$ST$-transformation is the mirror symmetry that exchanges Coulumb branch and Higgs branch of the free chiral multiplet, but for generic theories it is not true. What is interesting is that $ST$-transformation relates theories from one chamber to other chambers of Higgs branchs by flipping the signs of mass parameters, which is also discussed in section \ref{secSTtrans}.
The 2d-3d vertex functions on different chambers are related by R-transition matrices that can be expressed in term of elliptic stable envelops, see e.g. \cite{Aganagic:2016jmx,Rimanyi:2019zyi}, which are  Stokes jumps discussed in \cite{Beem:2012mb,Nieri:2015yia,Ashok:2019gee,Jain:2021bjf}. It is straightforward to compute the R-matrices caused by $ST$-transformations, which only consist of theta functions.

Moreover, $ST$-transformation plays an important role in 3d-3d correspondence \cite{Dimofte:2011ju} by connecting mixed \CS levels to  three-manifolds, which is beyond the topics in this note, so we prefer to further discuss $ST$-transformations for abelian theries with mixed CS levels  in \cite{plumb22}, in particular the abelian theory $(U(1) + 1 \,\F)_{k_{ij}}^{N}$.

\section{Effective Chern-Simons levels}\label{seceffCS}
In this section, we discuss how to read off effective \CS levels from 3d brane webs. Decoupling massive chiral multiplets could change \CS levels, which shows some interesting properties.

\subsection{Relative angles}
We assume the relative angle between NS5-brane and NS5'-brane is $\theta$.  It was found in \cite{Bergman:1999na,Kitao:1999aa} that  this relative angle relates to the CS level
$
  k= \tan\theta 
$.
For generic 3d $\N=2$ brane webs, we note that this angel $\theta$ should relate to the effective Chern-Simons level, since there are one-loop corrections from chiral multiplets \cite{Aharony:1997aa,Intriligator:2013lca}. In this note, we mainly discuss abelian theories with the gauge group $U\left(1\right)$, but the conclusion also applies to nonabelian theories with the gauge group $U\left(N\right)$ in section \ref{secnonabelian}.


\begin{figure}[H]
	\centering
	\includegraphics[width=2.2in]{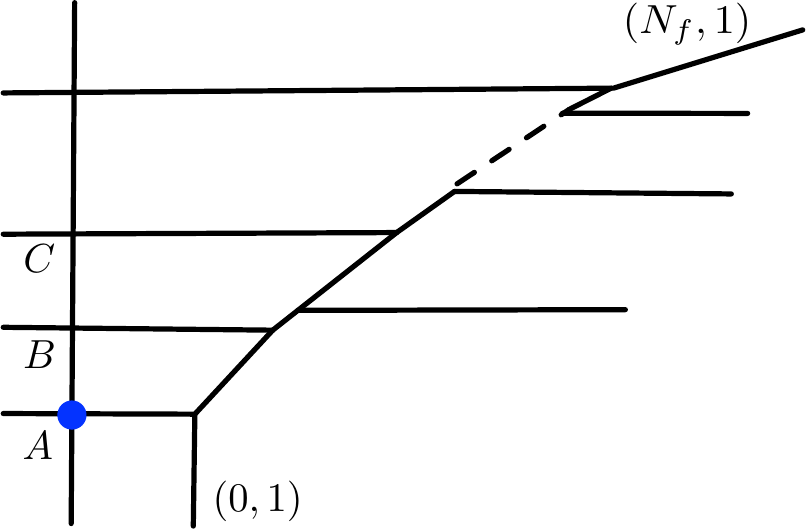}
	\caption{The blue node denotes the D3-brane. We can also put the D3-brane on other intersection points $\{A\,,B\,, C\,,\cdots\}$. Intersection points denote vacua of the theory, which compose the Higgs branch. Each intersection point can be regarded as a local conifold singularity on the dual toric diagram through M-theory/type-IIB duality \cite{Leung:1997tw}.  }
	\label{fig:CSslope}
\end{figure}
For the theory $U\left(1\right)_k + N_{f} \textbf{F} + N_{a} \textbf{AF}$, we illustrate its 3d brane web in Figure \ref{fig:CSslope}. Its effective Chern-Simons level is
\begin{align}\label{keff_k}
k^{\eff}= k+ \frac{N_{f}}{2} - \frac{N_{a}}{2} =\tan \theta\,.
\end{align}
In particular, $k^{\eff}=0$ for all cases that D3-brane locates at $\{ A\,,B\,, C\,, \cdots\}$ in Figure \ref{fig:CSslope}, because the relative angle $\theta=0$. Then the bare CS level is
$k= \frac{N_{a}-N_{f}}{2} $. 

\subsection{$N_{f}\geqslant N_a$}
\begin{figure}[h!]
	\centering
	\includegraphics[width=2.3in]{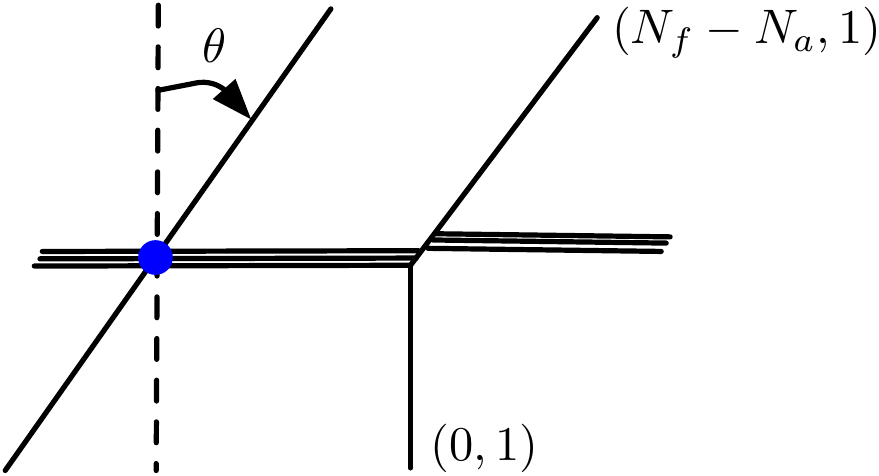}
	\caption{ The angle $\theta$ is between the dash line (parallel to the (0,1)-brane) and the $(N_f-N_a,1)$-brane. The blue node denotes the D3-brane that is along the direction perpendicular to 5-branes webs.
	}
	\label{fig:k_range}
\end{figure}

3d brane web have nonzero effective CS levels if one rotates the NS5'-brane to finite angles, as it is illustrated in Figure \ref{fig:k_range}.
If we do not want any intersection between the NS5'-brane and the NS5-brane, then there is the bound
\begin{align}\label{keff1}
k^{\eff}= \tan \theta \in [\, 0\,, N_{f}-N_a \,]  \,.
\end{align}
Using \eqref{keff_k}, we can get the bound for the bare CS level:
\begin{align}\label{keff2}
k =\tan\theta -\frac{N_f- N_{a}}{2}  \,\in\, \bigg[\, -\frac{N_{f}- N_a}{2} \, , \frac{N_{f}- N_a}{2}    \, \bigg] \,. 
\end{align}
Particularly, when $N_{f}=N_a$, we have $k^{\eff}=k=0$.

\begin{figure}[H]
	\centering
	\includegraphics[width=3.5in]{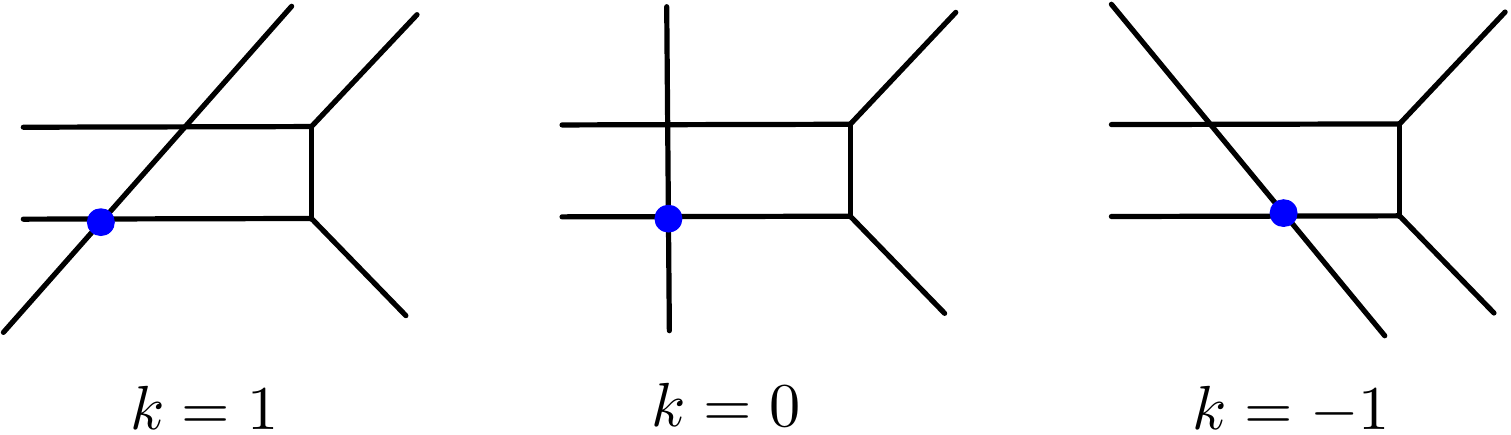}
	\caption{ Note that the $SL\left(2, \mathbb{Z}\right)$ symmetry in type IIB string theory preserves in 3d brane webs. This symmetry only changes slopes of 5-branes but does not change the relative angle and the effective CS level. }
	\label{fig:2F_k}
\end{figure}

Using \eqref{keff2}, we can determine that the bare CS level can only be $k=\pm \frac{1}{2}$ for the theory $U\left(1\right)_k +1\, \F$. For the theory $U\left(1\right)_k +2\, \F$, 
we illustrate all possible values that satisfy this bound in Figure \ref{fig:2F_k}. 



\subsection{$N_{f} < N_{a}$ and decoupling }
In this case, the NS5-brane bends to the left and always intersects with the NS5'-brane. Note that intersections are allowed in 3d brane webs,
as this kind of intersections can be regarded as local conifold singularities that can be resolved by blowing up. Intersections can also be avoided by introducing additional fundamental  D5-branes. Finally, one can send these additional D5-branes to infinity to return to the original theory.
\begin{figure}[h!]
	\centering
	\includegraphics[width=1.6in]{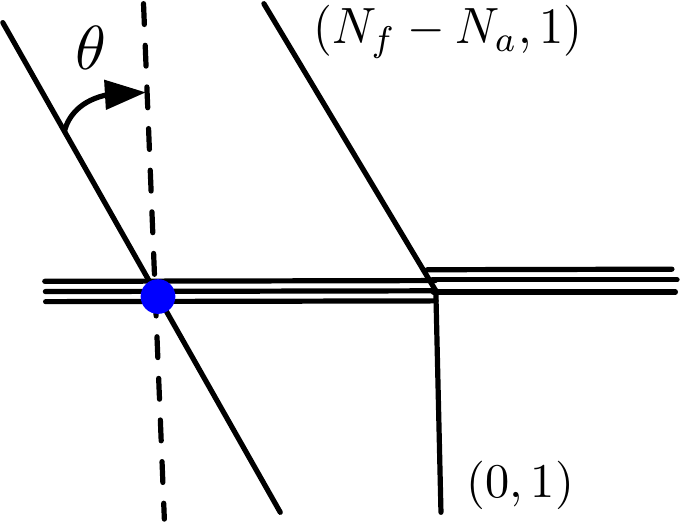}
	\caption{ When $N_{f} < N_{a}$, there is always an intersection between NS5'-brane and NS5-brane.  }
	\label{fig:krange2}
\end{figure}
\begin{figure}[h!]
	\centering
	\includegraphics[width=3.5in]{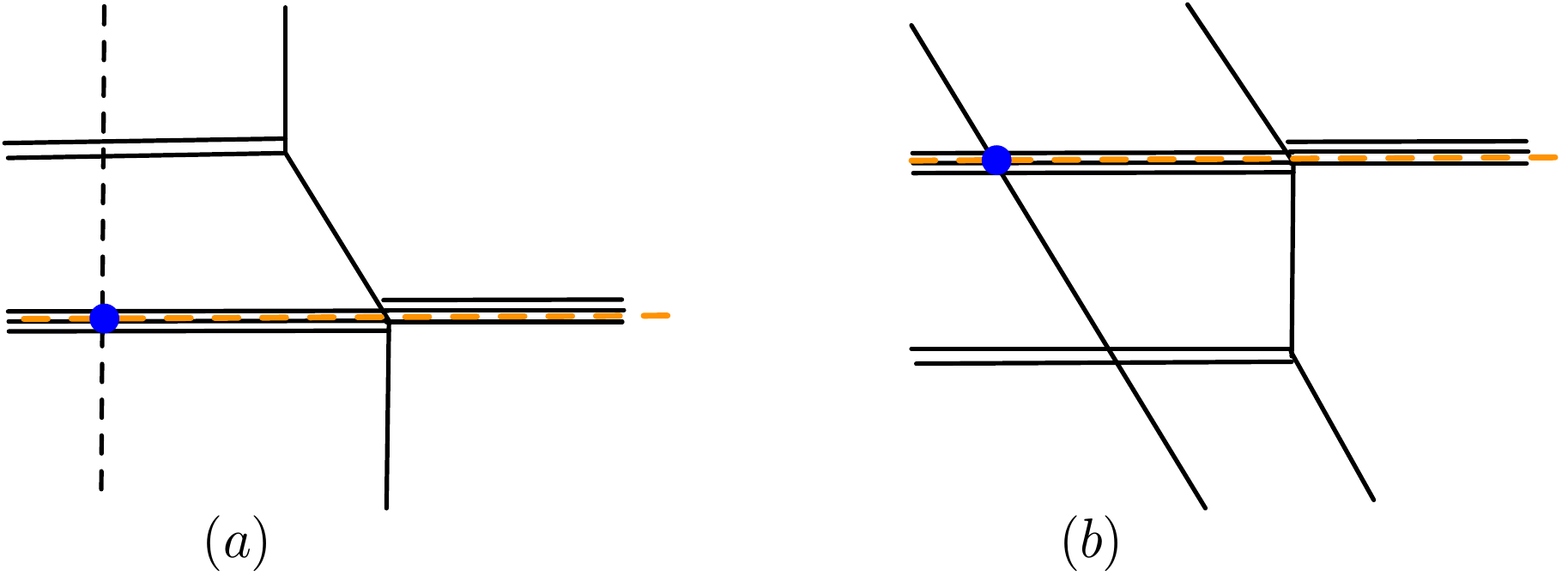}
	\caption{ The original line is denoted by the dashed orange line. In the case $(a)$, we add $(N_{a}-N_{f})\F$ above the original line  to avoid the intersection, and in the case $(b)$, we add the same number of $\F$ below the original line to avoid the intersection. Case $(a)$ and case $(b)$ are  maximal and minimal relative angles respectively for the 3d brane web in Figure \ref{fig:krange2}. }
	\label{fig:depNAFNF}
\end{figure}

We prefer to firstly introduce some fundamental multiplets $\F$ and finally decouple them. Decoupling means absolute values of mass parameters are sent to infinity. For simplicity, we add a particular number of fundamental chiral multiplets $\F$ such that the new effective Chern-Simons level vanishes. We note that there are two different cases, as illustrated in Figure \ref{fig:depNAFNF}. In the case $(a)$ we introduce $n \F$ above the original line, while in the second case $(b)$ we introduce $n \F$ below it, where $n = N_{a}-N_{f}$. In both cases effective \CS levels vanish $
k^{\eff} =0
$.

In the case $(a)$ decoupling $n\F$ does not change the effective \CS level, because if we decouple a fundamental multiplet $\F$ by sending the real mass parameter $m$ to infinity, and hence $\b = e^{-m}\rightarrow 0$, then its contribution becomes trivial as $\left(\b , q \right)_n^{-1} \rightarrow 1$, using the limit that 
\begin{align}
	(Q;q)_n ~ \xrightarrow{Q \,\rightarrow\,  0}  ~ 1 \,.
	\end{align}
As decoupling $\F$ does not change the \CS level, we have the maximal value  $
k^{\eff}_{\rm max} =0
$ for the 3d brane web in Figure \ref{fig:krange2}.

Let us discuss how to decouple a fundamental multiplet $\F$ below the original line, which is the case $(b)$. The K\"ahler parameter in this case should be $\b =e^{-(-m)} =e^m$, because the sign of mass parameter is negative\footnote{See the assignment rule that will be discussed in section \ref{phasesabelian}.}. Decoupling it means the large mass limit $m \rightarrow +\inf$,  so $\b$ is not the proper K\"ahler parameter as it is divergent. We should inverse this K\"ahler parameter and  send $\b^{-1} =e^{-m} \rightarrow 0$. Then the contribution from $\F$ becomes
\begin{equation}
	\left(\b ; q \right)_n^{-1}  = \left(\b^{-1}; q^{-1} \right)_n^{-1}  ( -\sqrt{q})^{-n^2} \left( \b^{-1} {\sqrt{q}} \right)^n  \,, \end{equation}
where the last term $(\b^{-1}\sqrt{q})^n$ cancels with the term $\b$ in the FI parameter (see \eqref{vortexcomplete}) and hence does not join the decoupling. Then sending send $\b^{-1}\rightarrow 0$ only leaves a term $(-\sqrt{q})^{-n^2}$, which reduces the effective Chern-Simons level by one, namely $k^{\eff} \rightarrow k^{\eff} -1$. 

Similarly, if we decouple an antifundamental chiral multiplet $\AF$ below the original line, then the effective CS level increases by one, namely $k^{\eff} \rightarrow k^{\eff} +1$, because its contribution is 
\begin{equation}( \a;q )_n = (\a^{-1};q^{-1})_n (-\sqrt{q})^{n^2} (\a/\sqrt{q})^n  \,, \end{equation}
 and we should absorb $(\a/\sqrt{q})^n$ into the FI parameter term $z^n$ in \eqref{vortexNCNAC} and then send $\a^{-1} =e^{- \tilde{m}} \rightarrow 0$.

We summarize different cases of decoupling in the table:
\begin{align}
\begin{tabular}{ c|c|c|c } 
	\hline
\text{decouple}&\text{position}&	mass  &~~ \text{CS level }$k^{\eff}$ ~~  \\  \hline
$\F$&\text{above} &	\text{positive} &  $k^{\eff}$   \\  
$\AF$&\text{above}&	\text{positive}  &  $k^{\eff}$  \\    \hline
$\F$&\text{below} &	\text{negative} &  $k^{\eff}-1$   \\  
$\AF$&\text{below}&	\text{negative} &  $k^{\eff}+1$  \\  
	\hline
\end{tabular}
\end{align}
where the position means if the chiral multiplet is above or below the original line which is also the location of the D3-brane. 

Let us return to the case $(b)$ in Figure  \ref{fig:depNAFNF}, for which the effective CS level is zero.  
We can at most decouple $N_{a}-N_{f}$ number of $\F$ and get a minimum for the effective CS level for the brane web in Figure \ref{fig:krange2}:
\begin{align}
	k^{\eff}_{\rm min}= 0-(N_{a}-N_{f})=N_{f}-N_{a} \,.
\end{align}  
Now we get  the range of the effective CS level for the brane web in Figure \ref{fig:krange2}, which is between minimum and maximum:
\begin{align}
	k^{\eff} \in [N_{f}-N_{a},0] \,.
\end{align}
Then the bare \CS level is
\begin{align}
	k =\tan\theta -\frac{N_{f}- N_{a}}{2}  \,\in\, \Bigg[\, -\frac{N_{a}- N_{f}}{2} \, , \frac{N_{a}- N_{f}}{2}    \, \Bigg]  \,.
\end{align}

We note that for both cases $N_{f}\geqslant N_{a}$ and $N_{f} < N_{a}$, the Chern-Simons level falls in the bound 
\begin{align}
	k =\tan\theta-\frac{N_{f} -N_{a}}{2}  \in \Bigg[ - \frac{|N_{f}-N_{a}|}{2} , \frac{|N_{f}-N_{a}|}{2}  \Bigg] \,,
\end{align}
which agrees with the bound found using localization methods \cite{Benini:2014aa,Benini:2011aa}.  This bound is also the constraint on Chern-Simons levels for Aharony duality \cite{Aharony_1997}.

\section{Equivalent brane webs}\label{secmassdef}
In this section, we discuss equivalent 3d brane webs that are given by real mass deformations, and show that their 3d partition functions are equivalent. 
\subsection{Real mass deformations}\label{phasesabelian}
\begin{figure}[H]
	\centering
	\includegraphics[width=4.5in]{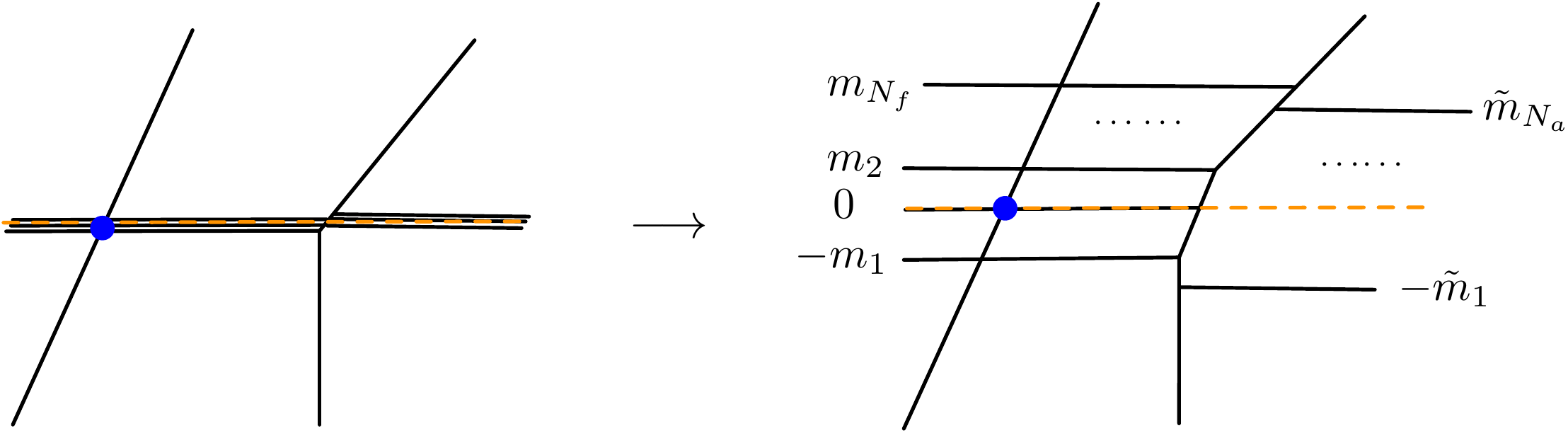}
	\caption{In the right brane web, $m_i$ are the real mass parameters for $\F$, and $\tilde{m}_i$ are real mass parameters for $\AF$. 
		The assignment of mass parameters are shown on the 3d brane web. $m_i$ and $\tm_j$ are absolute values of mass parameters, and we assume they are positive and real.
		The D3-brane (blue node) can locate on any fundamental D5-branes. The original line (dash orange line) is drawn along the fundamental D5-brane where the D3-brane locates. }
	\label{fig:turnonmass}
\end{figure}

The procedure of turning on real mass parameters is illustrated in Figure \ref{fig:turnonmass}.  Firstly, we need to pick up a fundamental flavor D5-brane and locate the D3-brane on it. Then we separate other D5-branes by turning on real masses parameters.  These are various configurations to separate D5-branes, since some mass parameters are larger than others. We could get many equivalent 3d brane webs. In particular, possible locations of D3-brane compose the Higgs branch $\mathcal{M}_H$ defined by $\{ s_\aa=m_\aa\,, \forall \aa=1,2, \cdots,N_f \}$. See \cite{Dorey:1999rb} for discussions on Higgs branch. 

These equivalent brane webs are physically equivalent, so their 3d partition functions should be equivalent. These different brane webs are different phases (chambers) of 3d theories. Notice that some mass parameters between two chambers have opposite signs, which is similar to the interface discussed in \cite{Aganagic:2016jmx,Dedushenko:2021mds,Bullimore:2021rnr}.

Through comparing 3d vortex partition functions and 3d brane webs, we note that the assignment rule for real mass parameters: D5-branes below the original line should be assigned with negative masses, and D5-branes above the original line should be assigned with positive masses. The mass parameter for the D5-brane (original line) at where D3-brane is located, remains zero. If we locate the D3-brane on other fundamental D5-branes or change the positions of antifundamental D5-branes, then mass parameters will be changed accordingly but will still follow the same assignment rule. Note that different brane webs obtained by different mass deformations correspond to the same theory and  hence have the same effective CS level, and the relative angle $\theta$ between NS5-brane and NS5'-brane is independent of real mass deformations.

\begin{figure}[h!]
	\centering
	\includegraphics[width=4.5in]{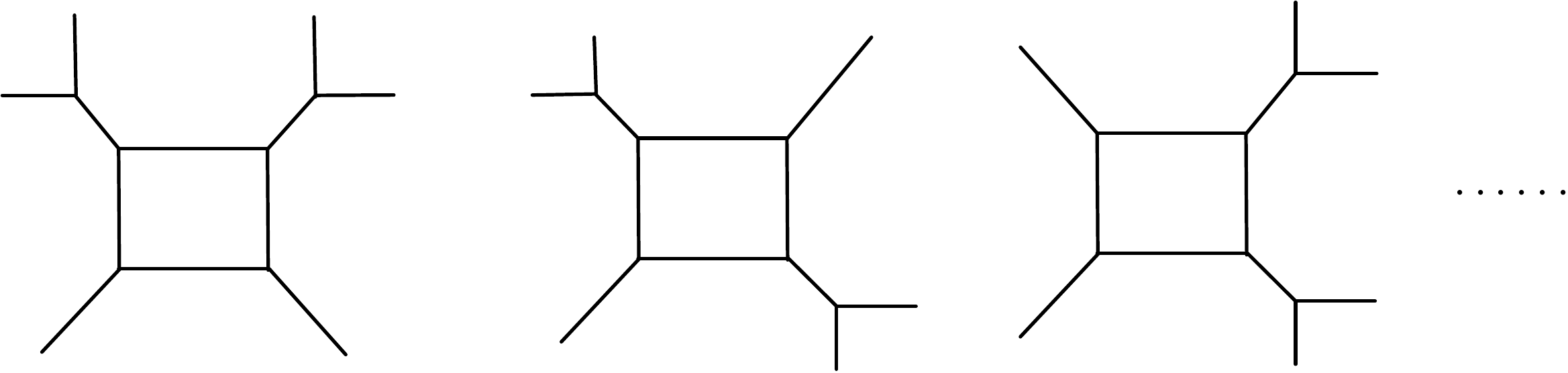}
	\caption{Some equivalent 5d brane webs for the 5d $\N=1$ theory with  the gauge group $SU(2)$ and two fundamental hypermultiplets.}
	\label{fig:5dsu2web}
\end{figure}

The real mass deformations of overlapped branes is similar to 5d $\N=1$ gauge theories, for which fundamental hypermultiplets are  given by semi-infinite D5-branes, and the distances between these flavor D5-branes and the original line are mass parameters. 
Mass parameters are negative below the original line and positive above the original line. 
As it is discussed in \cite{Taki:2014pba}, there are various brane webs for 5d $\N=1$ theories, depending on the positions of flavor D5-branes. Through Hanany-Witten (HW) transitions \cite{Hanany:1996ie} and flop transitions, these different 5d brane webs can be transformed to each other, hence even through they seem different but correspond to the same theory, and associated 5d Nekrasov partition functions are equal. For example, the 5d theory with a gauge group $SU(2)$ and two fundamental hypermultiplets has some equivalent 5d brane webs  in Figure \ref{fig:5dsu2web}. These 5d brane webs are related by HW transitions.

One would expect that there may be some similar operations to relate equivalent 3d brane webs given by various real mass deformations. However, it turns out that HW transitions cannot relate these equivalent 3d brane webs, even for the very simple theory $U(1)_k+2 \F$ shown in Figure \ref{fig:2fwebcs}. 

Note that in 3d brane webs we should be very careful with Hanany-Witten transitions, since 7-brane may cross 5-branes that attach to D3-branes, which often cause crossing complication. Fortunately, there are special cases without this crossing, namely performing HW transitions vertically. This kind of HW transitions gives rise to equivalent non-toric 3d brane webs with the same vortex partition function upon some extra open strings\footnote{These open strings correspond to chiral singlets}; see \cite{Cheng:2021aa} for more details. We will discuss the HW transitions along the horizontal direction in section \ref{movingD5}, which do not lead to extra open strings.



\subsection{Examples}\label{secexample}
In the following subsection, we compute the 3d partition functions of some brane webs to verify the assignment rule for real mass parameters and check the equivalence of various brane webs. 
We use refined topological vertex found in \cite{Iqbal:2007ii} and Higgsing method to produce D3-branes as surface defects; see e.g.\cite{Dimofte:2010tz,Cheng:2021aa,Koz_az_2010,Pasquetti:2011fj,Cheng:2020aa,Zenkevich:2017ylb,Kim:2020npz,Aganagic:2012hs,Kimura:2021ngu}. We have discussed the procedure in section \ref{geotrans}.
\subsubsection*{$U\left(1\right)_k+ 2\, \mathbf{F}$}
\begin{figure}[h!]
	\centering
	\includegraphics[width=2.8in]{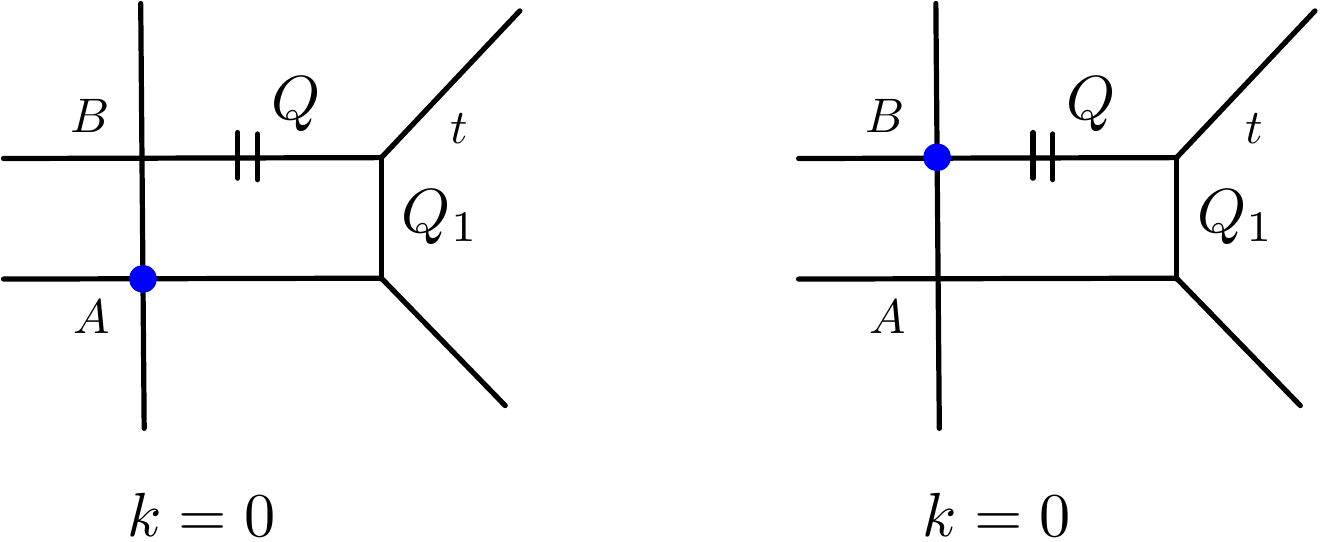}
	\caption{The bare CS level for this web is zero. The assignment of refined parameter $t$ is shown in brane webs. The short double line $||$ is a notation in topological vertex method, denoting that the preferred direction is along horizontal lines. }
	\label{fig:P1}
\end{figure}

The Higgs branch of this theory contains two discrete points $A$ and $B$ as it is shown in Figure \ref{fig:P1}.
If we only introduce a D3-brane (denoted by the blue node) at vacuum $A$,  its $\qbarbrane$ version of refined 3d partition functions is
\begin{align}\label{u12FA}
Z^{A}_{\qbarbrane}\left(Q,Q_1\right) =\left(Q_1 \frac{t}{q};t\right)_\inf  \cdot \sum_{n=0}^{\inf} \frac{ \left( -\sqrt{t}\right)^{n^2} \left( \frac{\sqrt{t}}{q} Q\right)^n  }{ \left(t;t\right)_n \left(Q_1 \frac{t}{q};t\right)_n   } \,.
\end{align}
Similarly, if we only introduce a D3-brane at vacuum $B$, then the 3d partition function is
\begin{align}\label{u12FB}
Z^{B}_{\qbarbrane}\left(Q,Q_1\right)  = \left(Q_1^{-1} \frac{t}{q};t \right)_\inf  \cdot \sum_{n=0}^{\inf} \frac{ \left( -\sqrt{t}\right)^{n^2} \left( \sqrt{t} Q_1^{-1} Q\right)^n  }{ \left(t;t\right)_n \left(Q_1^{-1} \frac{t}{q};t \right)_n   }  \,.
\end{align}
One can find that upon the identification of parameters 
\begin{align}\label{shiftKahler}
&Z^{A}_{\qbarbrane} \left(Q,Q_1\right)= Z^{B}_{\qbarbrane} \left(  Q Q_1^{-1} q^{-1},Q_1^{-1} \right)  \,.
\end{align}
 
 For $Z^{A}_{\qbarbrane} $, the real mass parameter is $Q_1 =e^{-m}$, while for  $Z^{B}_{\qbarbrane} $, mass parameter is  $Q_1^{-1} =e^{-m}$. This verifies the argument that mass parameters above the original line are positive, and mass parameters below it are negative. 
In addition,
effective CS levels for both $A$ and $B$ are the same  $k^{\eff} = 0 +\frac{2}{2} =1$, using the formula \eqref{keff1}.

\subsubsection*{$U\left(1\right)_k+ 2\, \mathbf{F}+1 \,\mathbf{AF}$}

\begin{figure}[h!]
	\centering
	\includegraphics[width=5.5in]{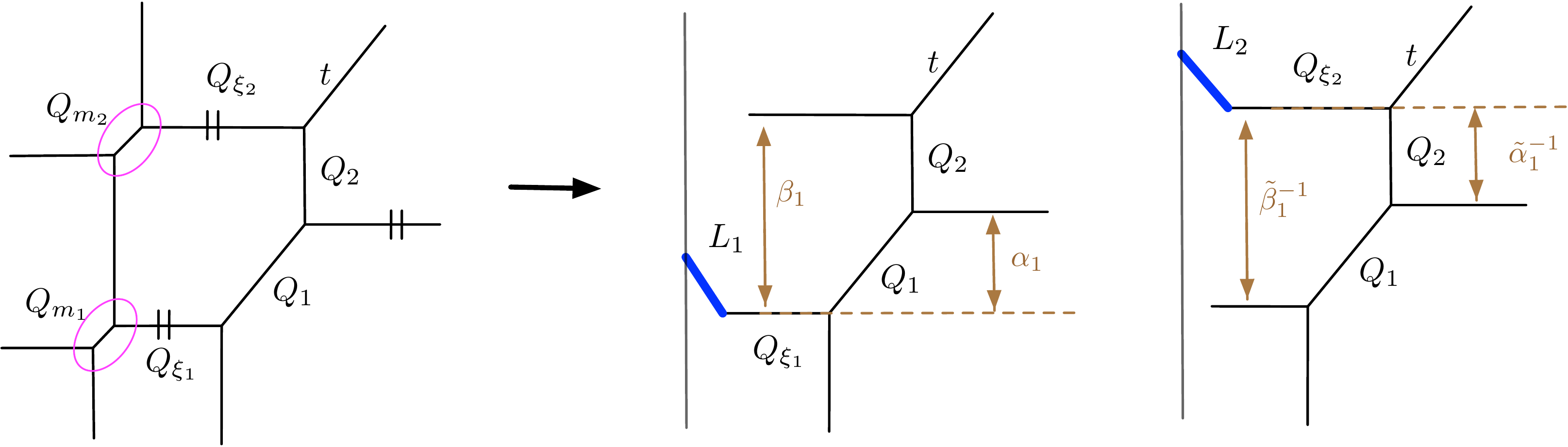}
	\caption{We denote the D3-brane introduced by Higgsing $Q_{m_1}$ as $L_1$, and the one  by Higgsing $Q_{m_2}$ as $L_2$. The K\"ahler parameters corresponding to mass parameters have been assigned on brane webs.}
	\label{fig:L1L2}
\end{figure}
We draw the brane web for this theory in Figure \ref{fig:L1L2}. The D3-brane is introduced by Higgsing $Q_{m_1}$ and $Q_{m_2}$.
Setting  $Q_{m_1}=\frac{1}{q}\sqtq $ and $Q_{m_2} =\sqtq$ leads to  the $\qbarbrane$ whose refined 3d partition function is 
\begin{align}
Z^{L_1}_{\qbarbrane} (Q_{\xi_1} , Q_1, Q_2)=\frac{ \left(Q_1 Q_2 t q^{-1};t\right)_\inf}{ \left(Q_1 \sqtq;t \right)_\inf   } \cdot
	 \sum_{n=0}^{\inf}  \frac{
	\left(Q_{\xi_1} \sqtq \right)^n  \left(Q_1 \sqtq;t \right)_n
}{  \left(t;t\right)_n \left(Q_1Q_2 \frac{t}{q};t\right)_n }    \,.
	\end{align}
In this case, relations between gauge theory parameters and K\"ahler parameters are
\begin{align}	
\a_1 =Q_1 \,, ~~ \b_1=Q_1 Q_2   \,,~~ z= Q_{\xi_1}  \,.
\end{align}
This 3d partition function can be written as 
\begin{align}\label{L1vortex}
Z^{L_1}_{\qbarbrane}  \left( z, \a,\b\right)=\frac{\left(\b_1 \frac{t}{q};t\right)_\inf }{  \left(\a_1 \sqtq;t \right)_\inf  }   \cdot
\sum_{n=0}^{\inf}  \frac{
	\left(z \sqtq \right)^n  \left(\a_1 \sqtq;t \right)_n
}{  \left(t;t\right)_n \left(\b_1 \frac{t}{q};t\right)_n }    \,.
\end{align}

The $\qbarbrane$ partition function for the D3-brane $L_2$ is given by setting  $Q_{m_1}=\sqtq $ and $Q_{m_2} =\frac{1}{q}\sqtq$. We find it is
\begin{align}\label{L2vortex}
	Z^{L_2}_{\qbarbrane} (Q_{\xi_1} , Q_1, Q_2)	&= 
	\frac{ \left(Q_1^{-1}Q_2^{-1} \frac{t}{q};t \right)_\inf    }{ \left(Q_2^{-1} \sqtq ;  t \right)_\inf   }
 \cdot \sum_{n=0}^{\inf}  \frac{
	\left(\frac{Q_{\xi_2}}{Q_1} \frac{t}{q} \right)^n  \left(Q_2^{-1} \sqtq ;  t \right)_n
}{  \left(t;t\right)_n \left(Q_1^{-1}Q_2^{-1} \frac{t}{q};t \right)_n }   \,.
\end{align}
In this case, relations between gauge theory parameters and K\"ahler parameters are
\begin{align}	
\tilde{\a}_1 =Q_2^{-1} \,, ~~ \tilde{\b}_1=Q_1^{-1} Q_2^{-1}   \,,~~ \tilde{z}= \frac{Q_{\xi_2}}{Q_1} \sqtq \,.
\end{align}
This partition function can be  written as
\begin{align}\label{L2dvortex}
	Z^{L_2}_{\qbarbrane}  \left( \tilde{z}, \tilde{\a},\tilde{\b}\right)=\frac{\left(\tilde{\b}_1 \frac{t}{q};t\right)_\inf }{  \left(\tilde{\a}_1 \sqtq;t \right)_\inf  }   \cdot
	\sum_{n=0}^{\inf}  \frac{
		\left(\tilde{z} \sqtq \right)^n  \left(\tilde{\a}_1 \sqtq;t \right)_n
	}{  \left(t;t\right)_n \left(\tilde{\b}_1 \frac{t}{q};t\right)_n }    \,.
\end{align}

Note that \eqref{L1vortex} and \eqref{L2dvortex} take the same form,
which implies that D3-brane configurations $L_1$ and $L_2$ in Figure  \ref{fig:L1L2} describe the same theory, but the maps between mass parameters and K\"ahler parameters are different. In this example, effective CS levels for both 3d brane webs are zero, which can be known by comparing \eqref{L1vortex} with the generic form \eqref{vortexNCNAC}.

In addition, the refined Ooguri-Vafa (OV) formula that we will discuss in section \ref{OVformula} can be used to extract refind OV invariants from \eqref{L2vortex}. 
These refined OV invariants should be positive integers as they are degeneracy numbers of vortex particles in the representation $\left(r,s\right)$ of the rotation symmetry and R-symmetry $SO\left(2\right) \times U\left(1\right)_R$.  However, if one straightforwardly applies refined OV formula \eqref{openGV} to \eqref{L2vortex} and expand partition functions in terms of K\"ahler parameter $Q_1$ and $Q_1Q_2$, one would get some negative refined OV invariants, which implies that expansion parameters are not chosen properly and proper expansion parameters should be $\tilde{\a}$ and $\tilde{\b}$.

Based on the above two examples, we can confirm the assignment rule for mass parameters. Namely, mass parameters below the original line are negative and above the original line are positive.
One can expect that the same argument also applies for more generic theories $U(1)_k+N_{f}\F+N_{a}\AF$. In section \ref{secnonabelian}, vortex partition functions show that nonabelian theories also follow this rule.


\subsection{Brane webs and quiver matrices  }\label{quiverwebs}

In this section, we discuss the correspondence between brane webs and quiver matrices that are connected by $ST$-transformations. We observe this correspondence based on several examples.

The examples in last section show that 3d partition functions for 3d brane webs in Figure \ref{fig:L1L2}   are related by an operation
\begin{align}
	Z^{L_1}_{\qbarbrane}  (Q_i) \xrightarrow{ Q_1 \rightarrow Q_1^{-1}\,,~ Q_2 \rightarrow Q_2^{-1}} Z^{L_2}_{\qbarbrane}  (Q_i) \,.
\end{align}
The operation
\begin{align}\label{examplemir}
	\{ Q_1 \rightarrow Q_1^{-1}\,,~ Q_2 \rightarrow Q_2^{-1}\}
\end{align}
relates 3d brane webs $L_1$ and $L_2$. If one expresses K\"ahler  parameters \eqref{examplemir} in terms of real mass parameters, then this operation becomes the flips of mass parameters:
\begin{align}
	\{	\tilde{m}_1 \rightarrow -\tilde{m}_1 \,,~~ 	m_1 \rightarrow -m_1 \}\,
\end{align}
Flipping mass parameters is usually a phenomenon of wall crossings between chambers; see e.g.\cite{Aganagic:2016jmx,Rimanyi:2019zyi}.


 More examples show that flips relate equivalent 3d brane webs. 
For the abelian theory $U(1)+N_{f} \F+N_{a}\AF$, there are many mass parameters, and one can flip any of them. This is represented as reflecting positions of D5-branes, because positions of D5-branes stand for real mass parameters. By flipping mass parameters in a sequence, one can obtain all equivalent 3d brane webs that are obtained by real mass deformations.
Moreover, we has discussed in section \ref {stdualtheory} and  section \ref{secSTtrans} that $ST$-transformations for chiral multiplets also flip the signs of mass parameters, so we conjecture that equivalent brane webs obtained by mass deformations are related by $ST$-transformations.  Hence, there is the correspondence:
\begin{align} \label{STflip}
&ST\text{-transformations } ~~\xleftrightarrow{~~~~}~~\text{flip D5-branes} ~~\xleftrightarrow{~~~~}~~ \text{3d brane webs} 
\end{align}
As we review in section \ref {stdualtheory}, for $ST$-dual theories, flipping the sign of mass parameters is along with the change of mixed CS levels of $ST$-dual theories. Putting these considerations together, we propose that quiver matrices $C_{ij}$ can be used to characterize 3d brane webs:
\begin{align}
\text{3d brane webs} ~~\xleftrightarrow{~~~~}~~ \text{quiver matrices}~ C_{ij} 	 \,.
\end{align}
which can also be viewed as a correspondence.

Let us look at brane webs for some abelian theories and show their quiver matrices. The first example is $U\left(1\right)_k+ 1 \F$, which only has one brane web, so it does enjoy any flip and its quiver matrix is just the effective CS level $C_{ij} =k^{\eff}=k+1/2$.
\begin{figure}[h!]
	\centering
	\includegraphics[width=3in]{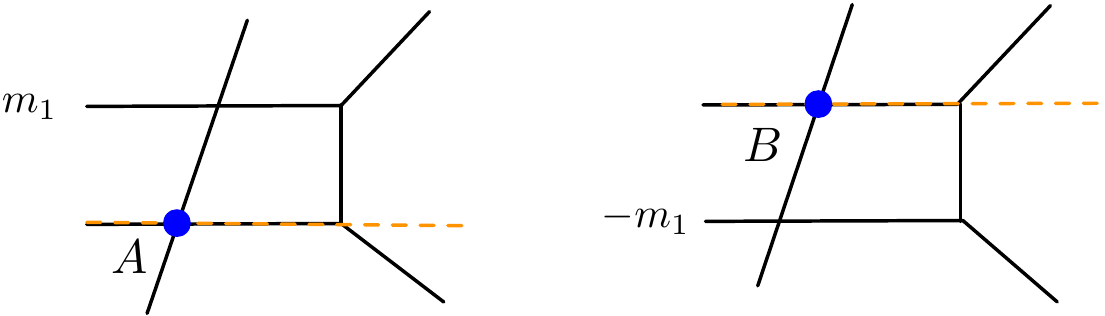}
	\caption{ There are two points $A$ and $B$ on its Higgs branch. Flipping (reflecting) the massive D5-brane relates these two vacua. Note that these two 3d brane webs cannot be related by HW transitions or flop transitions.}
	\label{fig:2fwebcs}
\end{figure}

The first non-trivial example is $U\left(1\right)_k+2\F$, which has two equivalent brane webs as illustrated in Figure \ref{fig:2fwebcs}, corresponding to different vacua on Higgs branch. 
These two brane webs are related by flips:
\begin{align}
	\begin{split}
		&\text{flip 0 times:} ~~\{m_1 \} \,,\\ &\text{flip 1 times:}~~ \{-m_1 \} \,.
	\end{split}
\end{align}
The associated quiver matrices (mixed CS levels) are:
\begin{align}
C_{ij}^{A} =
\left[
\begin{array}{cc}
k+1 &1	  \\
1& 1 \\
\end{array}
\right]  \,, \quad 
C_{ij}^{B}=
\left[
\begin{array}{cc}
k &-1	  \\
-1& 0 \\
\end{array}
\right]  \,.
\end{align}

Similarly, there are also two quiver matrices for $U(1)_k+1\F+1\AF$, for which we can flip the $\AF$, but the $\F$ cannot be flipped as it is connected to the D3-brane.
\begin{figure}[h!]
	\centering
	\includegraphics[width=5.5in]{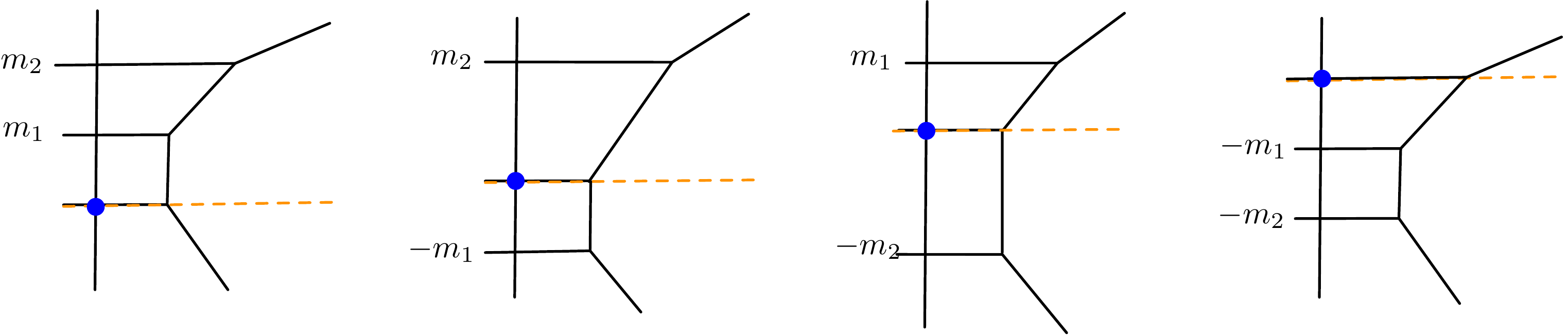}
	\caption{In this figure, brane webs for the theory $U\left(1\right)_{-1/2}+3\F$ have the same effective CS level $k^{\eff}=1$.	 }
	\label{fig:3Fexample}
\end{figure}

The next example is $U\left(1\right)_k+ 3 \F$, whose brane webs are shown in Figure \ref{fig:3Fexample}. There are in total four brane webs that are related by flipping D5-branes:
\begin{align}
	\begin{split}
&\text{flip 0 times:} ~~\{m_1, m_2 \} \,,\\ &\text{flip 1 times:}~~ \{-m_1, m_2  \} \,,~ \{m_1, -m_2\}\,,\\&\text{flip 2 times:}~~ \{-m_1, -m_2\} \,.
\end{split}
\end{align}
The corresponding quiver matrices are:
\begin{align}
\left[
\begin{array}{ccc}
k+\frac{3}{2} &1	  &1 \\
1& 1&0 \\
1 &0  &1 \\
\end{array}
\right]  \,,~~
\left[
\begin{array}{ccc}
k+\frac{1}{2} &-1	  &1 \\
-1& 0&0 \\
1 &0  &1 \\
\end{array}
\right]  \,,~~
\left[
\begin{array}{ccc}
k+\frac{1}{2} &1	  &-1 \\
1& 1&0 \\
-1 &0  &0 \\
\end{array}
\right]  \,,~~
\left[
\begin{array}{ccc}
k-\frac{1}{2} &-1	  &-1 \\
-1& 0&0 \\
-1 &0  &0 \\
\end{array}
\right]  \,.
\end{align}

These quiver matrices can be read off from effective superpotentials, which are the leading terms of vortex partition functions by taking the classical limit $q \rightarrow1$. For more details on computation, see \cite{Cheng:2020aa}. We notice that the first element $C_{0,0}$ is the effective CS level that corresponds to the D5-brane where D3-brane locates, and each row and each column corresponds to other D5-brane. 
We summarize quiver components associated to D5-branes in the following
\begin{align}
	\begin{tabular}{ c|c|c|c } 
		\hline
		\text{matter}&\text{mass}&	\text{K\"ahler parameter}  &~~ \text{quiver component} ~~  \\  \hline
			&&&\\
		$\F$& $0$ &	$\b=e^{-m}$&  $ \left[
	\begin{array}{c}
	~k^{\eff}~ 
\end{array}	\right]$   \\  
		&&&\\\hline
		&&&\\
		$\F$& $m$ &	$\b=e^{-m}$&  $ \left[
		\begin{array}{cc}
			~0~ &~1	~  \\
			~1~& ~1~ \\
		\end{array}
		\right]$   \\  
				&&&\\
		$\F$&$-m$&	$\b=e^{-(-m)}$ &  $\left[
		\begin{array}{cc}
			-1 &-1	 \\
			-1&~0 \\
		\end{array}
		\right] $  \\    
			&&&\\\hline
			&&&\\
		$\AF$& $\tilde{m}$ &	$\a=e^{-\tilde{m}}$&  $ \left[
	\begin{array}{cc}
		~0~ &~1	~  \\
		~1~& ~0~ \\
	\end{array}
	\right]$   \\  
	&&&\\
	$\AF$&$-\tilde{m}$&	$\a=e^{-(-\tilde{m})}$ &  $\left[
	\begin{array}{cc}
		~1 &-1	 \\
		-1&~1 \\
	\end{array}
	\right] $  \\  	&&&\\  \hline
	\end{tabular}
\end{align}
which match with \eqref{FtoC} and \eqref{AFtoC}. Using this table, one can easily write down the quiver matrices for given brane webs.

\begin{figure}[h!]
	\centering
	\includegraphics[width=5in]{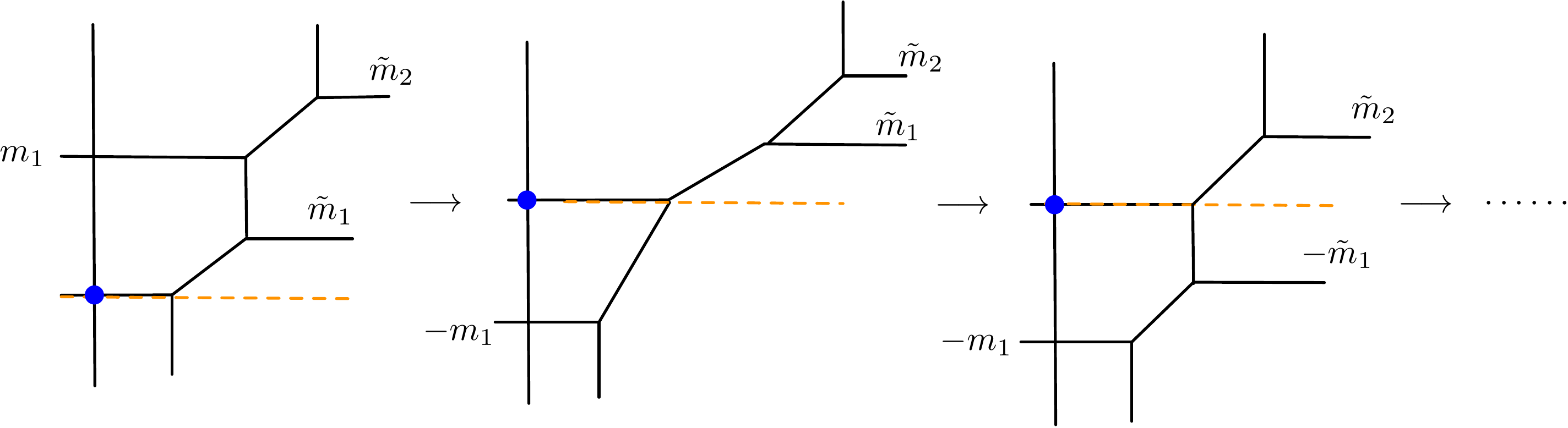}
	\caption{Flipping D5-branes lead to a chain of equivalent 3d brane webs.}
	\label{fig:flipseriesexample}
\end{figure}

The last example is $U\left(1\right)_0+ 2 \F + 2\AF$. For this theory, there are a series of equivalent brane webs as shown in Figure \ref{fig:flipseriesexample}. 
For simplicity, we do not show all of them, as it is easy to obtain them by flipping D5-branes. In total, we get eight equivalent brane webs for this theory, which are given by flips:
\begin{align}
\begin{split}
& \text{flip 0:} ~~ \{m_1, \tm_1, \tm_2 \}\,, \\
& \text{flip 1:} ~~\{-m_1, \tm_1, \tm_2 \} \,, ~~\{m_1, -\tm_1, \tm_2 \} \,, ~~ \{m_1, \tm_1, -\tm_2 \} \,, \\
& \text{flip 2:} ~~
\{-m_1, -\tm_1, \tm_2 \} \,, \{m_1, -\tm_1, -\tm_2 \}   \,,   \{-m_1, \tm_1, -\tm_2 \} \,, \\
& \text{flip 3:} ~~\{-m_1, -\tm_1, -\tm_2 \} \,.  \\
\end{split}
\end{align}


\section{Moving D5-branes}\label{movingD5}

In this section, we discuss the movement of D5-branes in 3d brane webs. The interesting point is that moving D5-branes cannot be interpreted as flipping D5-branes, so  it leads to a different class of equivalent brane webs, which are double layer webs.
For a nice review on brane constructions, see \cite{Giveon_2009}.

\subsection{Double layer webs}
It is convenient to use circles and boxes to denote gauge theories, which are well known as quiver diagrams in literature. In this notation, the circle with a number $N$ denotes gauge group $U\left(N\right)$, and the box with a number $N_{f} + N_a$ denotes the matter being $N_{f} \F$ and $N_a \AF$.

\begin{figure}[H]
	\centering
	\includegraphics[width=5.5in]{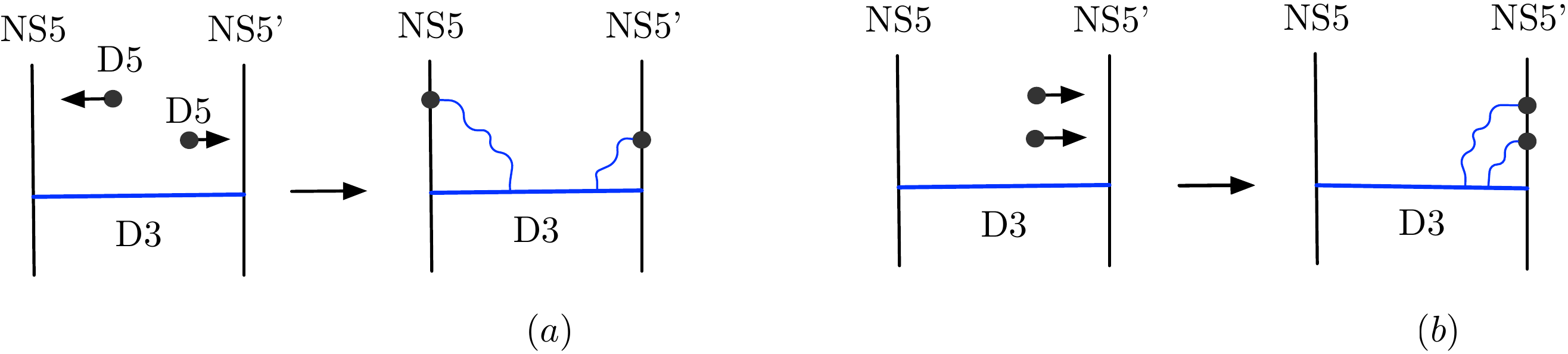}
	\caption{It is free to move flavor D5-branes (dark nodes). In brane web $\left(a\right)$, we move two D5-branes to opposite directions and attach them to NS5-brane and NS5'-brane separately. In brane web	 $\left(b\right)$, we move both flavor D5-branes to the NS5'-brane. The blue wave lines denote D3-D5 strings that give rise to chiral multiplets. }
	\label{fig:moveU12D5}
\end{figure}
Our claim is that moving D5-branes give rise to equivalent brane webs. In brane systems of 3d $\N=2$ theories, matters are given by D5-branes sandwiching between NS5-brane and NS5'-brane. Each D5-brane gives rise to one $\F$ and one $\AF$. These flavor D5-branes can move to either left or right, which leads to various brane webs. 

\begin{figure}[H]
	\centering
	\includegraphics[width=5.9in]{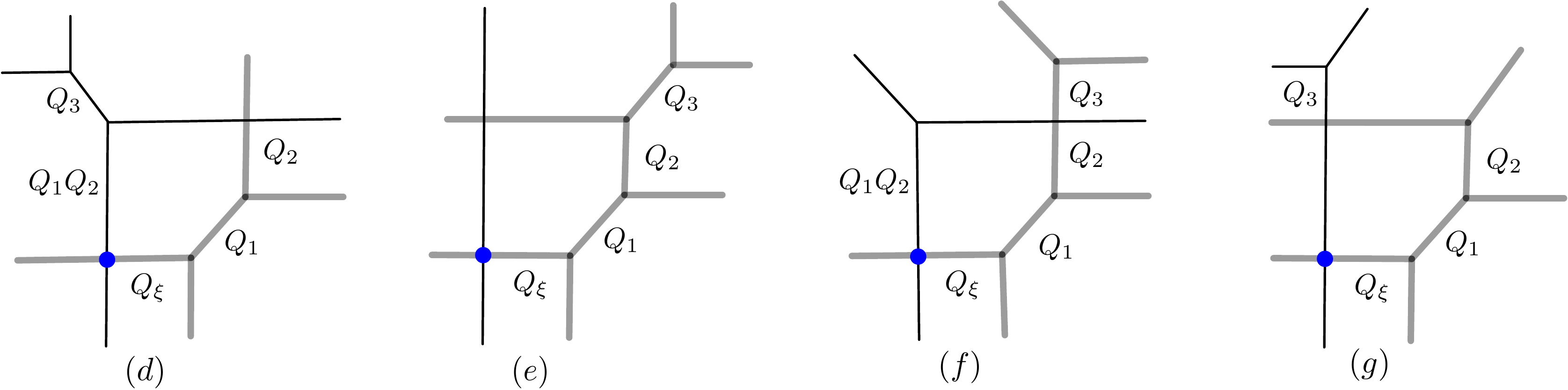}
	\caption{Some equivalent double layer brane webs for $U\left(1\right)_0 + 2 \F+2 \AF$.}
	\label{fig:twoeqlwebs}
\end{figure}

We illustrate an example in Figure \ref{fig:moveU12D5}. If we move one D5-brane to the left and another D5-brane to the right, then we get a theory denoted by the quiver diagram
$\boxed{1+1} -\mathcircled{1}_0-\boxed{1+1}$. 
On the other hand, if moving both D5-branes to the right, then we get a theory denoted by $\mathcircled{1}_0-\boxed{2+2}$. 
Since these two quiver diagrams describe the same theory $U(1)_0+2 \F+2\AF$, we have
\begin{align}
\boxed{1+1} -\mathcircled{1}_0-\boxed{1+1} ~~ =~~ \mathcircled{1}_0-\boxed{2+2}  \,,
\end{align}
which correspond to brane web $\left(a\right)$ and $\left(b\right)$ respectively in  Figure \ref{fig:moveU12D5}. The more precise brane webs are  $\left(d\right)$ and $\left(e\right)$ in Figure \ref{fig:twoeqlwebs}.  

The above configurations could be explained by open strings, as we have illustrated in Figure \ref{fig:moveU12D5}. The open strings that lead to chiral multiplets can be randomly distributed on both sides of D3-branes. If we move D5-branes, then open strings also move with them. If looking at the brane systems in Figure \ref{fig:moveU12D5} from the left direction, then one can observe some precise brane webs in Figure \ref{fig:twoeqlwebs}.

Performing Hanany-Witten transitions of D5-branes along the horizontal direction, we can get other cases. For instance, brane web $(f)$ and $(g)$ are obtained from brane web $(d)$ and $(e)$ respectively in this way.
Brane web $(f)$ has the quiver diagram $\boxed{1+0} -\mathcircled{1}_0-\boxed{1+2}$, and
brane web $(g)$ has the quiver diagram $\boxed{0+1} -\mathcircled{1}_0-\boxed{2+1}$.

Topological vertex computation shows that vortex partition functions of these brane webs are the same\footnote{One can also add one-loop part $Z^{\text{one-loop}}=\frac{ (Q_1Q_2 t/q;t)_\inf  }{ (Q_1\sqrt{t/q};t)_\inf ( Q_1Q_2Q_3 \sqrt{t/q};t )_\inf  }$.}
\begin{align}
&Z^{\text{vortex}}_{\qbarbrane} = 
 \sum_{n=0}^{\inf} \frac{ \left( Q_{\xi} \sqtq \right)^n   \left(Q_1 \sqtq ;t \right)_n \left( Q_1Q_2Q_3 \sqtq;t \right)_n }{ \left(t;t\right)_n  \left( Q_1 Q_2 \frac{t}{q};t \right)_n  }  \,,
\end{align}
where the effective Chern-Simons level is zero. Chiral multiplets come from open strings connecting D3-brane and D5-branes and can be identified from $q$-Pochhammer productions following \eqref{identifyFAF}. 

\begin{figure}[H]
	\centering
	\includegraphics[width=2.5in]{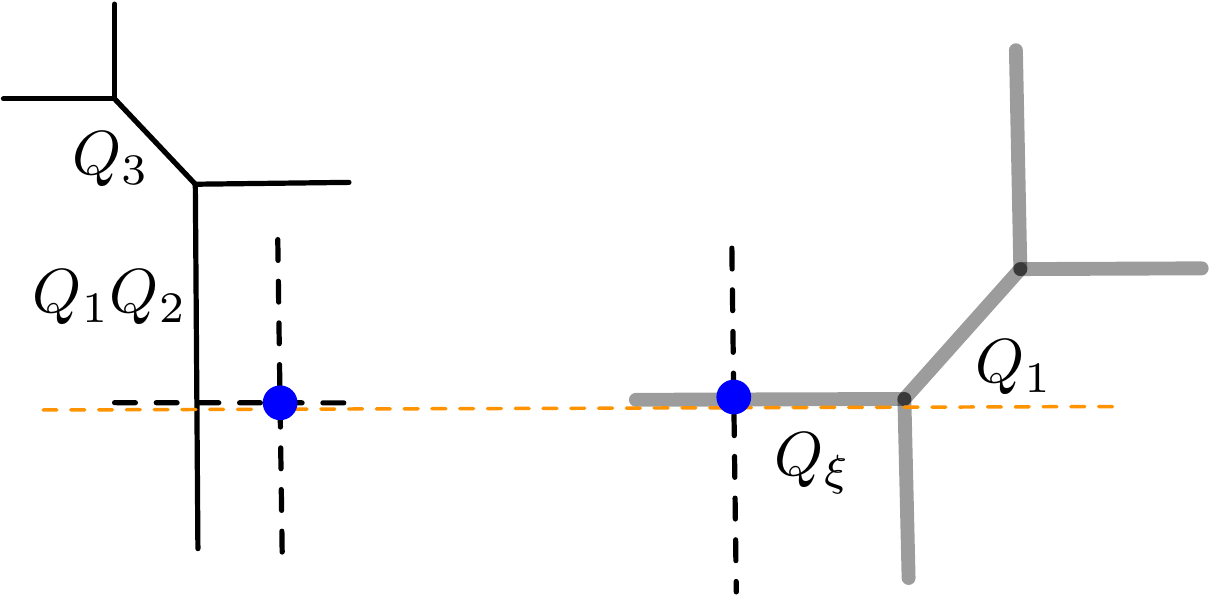}
	\caption{Splitting the brane web $(d)$ into two sub-webs. In this figure, we draw some auxiliary lines (dash lines) to let these two sub-webs look like usual 3d brane webs. }
	\label{fig:splitstrings}
\end{figure}
Let us discuss how sub-webs are related to terms in vortex partition functions.
 We can separate the brane web $(d)$ into two sub-weds, as it is shown in Figure \ref{fig:splitstrings}. 
Then the 3d vortex partition function splits into two parts corresponding to the left and the right sub-web respectively
\begin{align}\label{splitwebs}
\frac{  \left( Q_1Q_2Q_3 \sqtq;t \right)_n }{ \left( Q_1 Q_2 \frac{t}{q};t \right)_n  }  ~~  \text{and}~~
\left( Q_{\xi} \sqtq \right)^n \frac{  \left(Q_1 \sqtq ;t \right)_n  }{ \left(t;t\right)_n   } \,.
\end{align}
These two parts can be checked by topological vertex computation.

By observing \eqref{splitwebs}, we note that the left sub-web gives $1 \F + 1\AF$ and the right sub-web gives  $1 \F + 1\AF$. The only difference is that for the left sub-web the D3-brane is on the right side of the NS5-brane, so the K\"ahler parameter $Q_1Q_2Q_3$ is associated to the $\AF$ and $Q_1Q_2$ is associated to the $\F$. However, the assignment rule for mass parameters are the same as before. 

 One can see that in brane web $(d)$ the $\AF$ associated to $Q_1Q_2Q_3$ can be horizontally moved from the left sub-web to the right sub-web through the HW transition. This leads to brane web $(e)$. However, the $\F$ that is associated to $Q_1Q_2$ cannot be moved from brane web $(d)$ to brane web $(e)$ through HW transitions but can be interpreted as moving the D5-brane. 

For the theory $U(1)_k+2 \F+2\AF$ with the generic CS level $k$, we only need to rotate the relative angle between NS5-brane and NS5'-brane that are on the bottom part of the brane webs, and hence the CS level is not influenced by flavor D5-branes. This is obviously the case for the brane web $(e)$ in Figure \ref{fig:twoeqlwebs}. Similarly, in brane web $(d)$, we can straightforwardly rotate the left sub-web, since the relative angle in $k^{\eff}=\tan \theta$ is only between NS5-brane and NS5'-brane. Namely, in the left sub-web, (1,0)-branes remain the same, while the (0,1)-brane becomes ($k$,1)-brane. 

The above discussion also applies to more generic theory $ U\left(1\right)_k + N_{f} \F+N_a \AF$. Its quiver diagrams should satisfy
\begin{align}\label{splitstring}
	\boxed{N_{f}'+N_a'} -\mathcircled{1}_k-\boxed{(N_{f}-N_{f}')+(N_a-N_a')} ~~ =~~ \mathcircled{1}_k-\boxed{N_{f}+N_a}  \,.
\end{align}
When $N_f=N_a$, the above discussion is correct obviously. By decoupling some matters, one could get the theory with $N_f \neq N_a$. In addition, Hanany-Witten transitions of antifundamental chiral multiplets $\AF$ could lead to more equivalent brane webs. 
However, we still do not know 3d brane webs for all cases of \eqref{splitstring}, because we do not know how to move the D5-brane that is attached with the D3-brane. In this case, the D5-brane gives rise to the massless $\F$.

\subsection{$S$-duality}
 Let us discuss more obvious examples to illustrate that moving D5-branes is non-trivial. 
 
 \begin{figure}[H]
 	\centering
 	\includegraphics[width=2.5in]{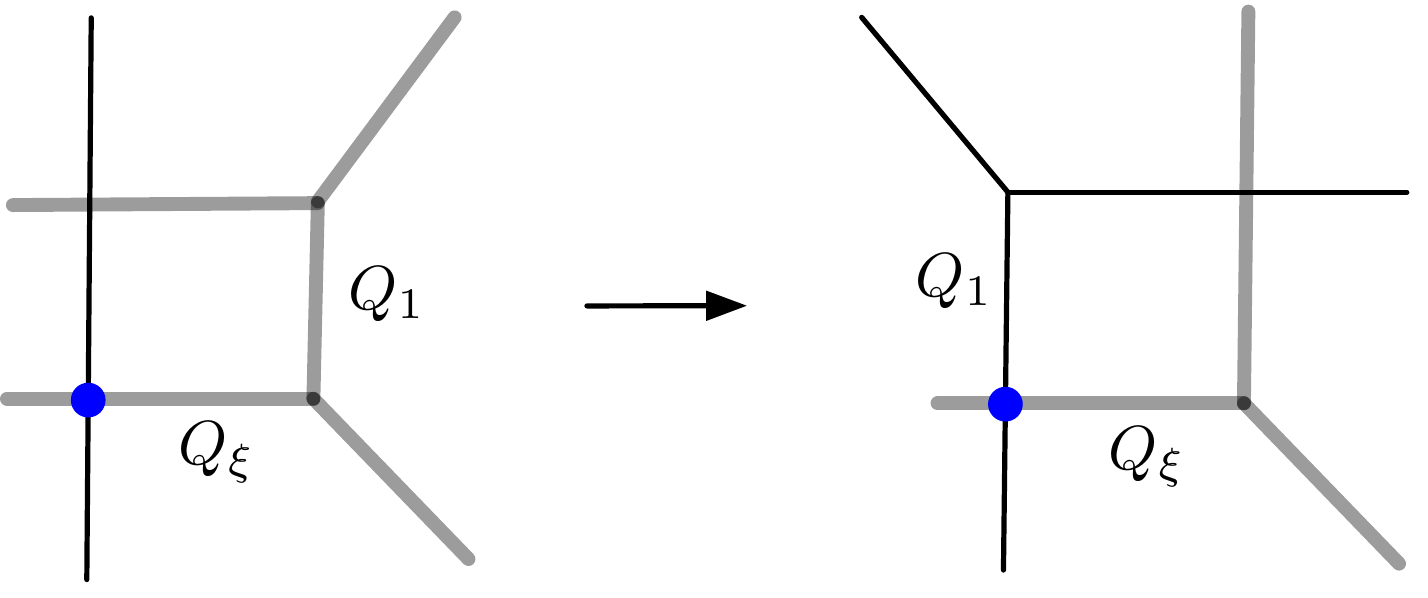}
 	\caption{These brane webs describe the theory $U(1)_0+2 \F$ with $k^{\eff} = 0 +2/2 =1$. We move the massive D5-brane to get the symmetric brane web.}
 	\label{fig:2Fkee1}
 \end{figure}
 We firstly consider the theory $U(1)_0 +2 \F$ illustrated in Figure \ref{fig:2Fkee1}. 
Its 3d partition function is \begin{align}\label{2Fkeff1}
	Z^{U(1)_0+2\F}_{\qbarbrane}(Q_\xi,Q_1) =\left(Q_1 \frac{t}{q};t\right)_\inf  \cdot \sum_{n=0}^{\inf} \frac{ \left( -\sqrt{t}\right)^{n^2} \left( \frac{\sqrt{t}}{q} Q_\xi \right)^n  }{ \left(t;t\right)_n \left(Q_1 \frac{t}{q};t\right)_n   }  \,,
\end{align}
which is just the partition function in \eqref{u12FA}. Note that if we locate the D3-brane at the massive D5-brane, then the D3-brane on the right brane web in Figure \ref{fig:2Fkee1} will locate at the other intersection.

\begin{figure}[H]
	\centering
	\includegraphics[width=2.5in]{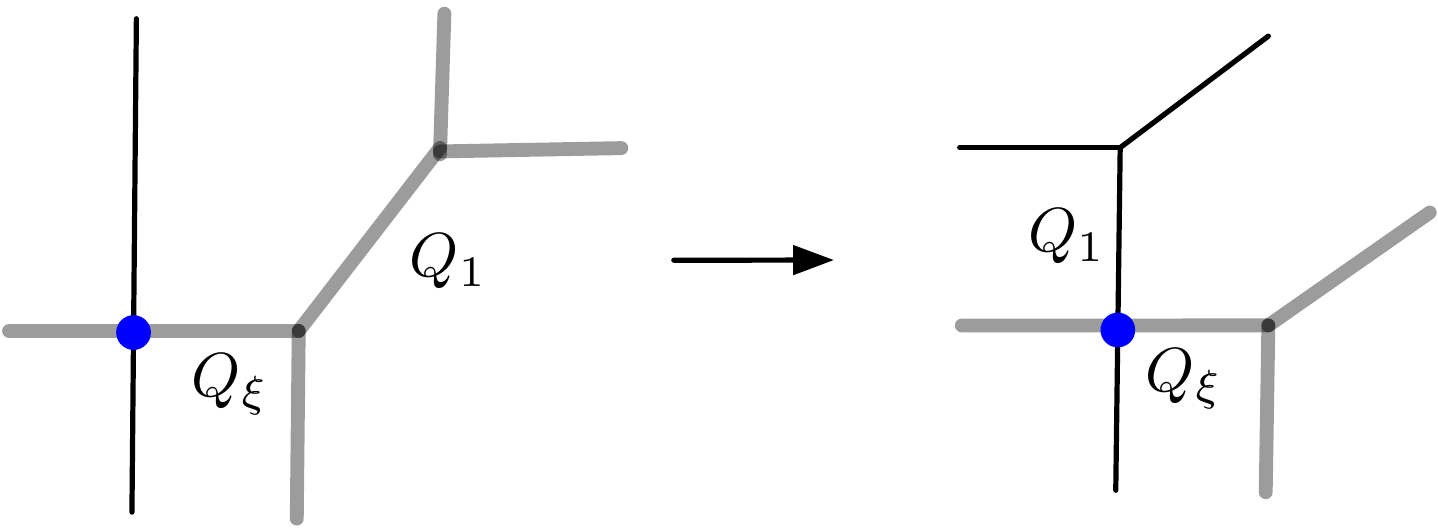}
	\caption{Brane webs describe the theory $U(1)_{0}+1 \F +1\AF$ with $k^{\eff} = 0 +(1-1)/2 =0$.}
	\label{fig:2Fkeff0}
\end{figure}
Moving $\AF$ can be viewed as the  Hanany-Witten transition of the anti-fundamental D5-brane. We consider the theory $U(1)_0 +1 \F +1 \AF$ and illustrate in Figure \ref{fig:2Fkeff0}. 
Its 3d  partition function is \begin{align}
	Z^{U(1)_0+1\F+1\AF}_{\qbarbrane}(Q_\xi ,Q_1) =\frac{1}{(Q_1 \sqtq;q)_\inf} \cdot \sum_{n=0}^{\inf} \frac{   \left(Q_{\xi} \sqtq\right)^{ n} ~   \left(Q_1 \sqtq ;t \right)_n  }{ \left(t;t\right)_n    }  \,.
\end{align}

We notice that exchanging $Q_{\xi}$ and $Q_1$ does not change partition functions:
 \begin{align}	
		\begin{split}
		Z^{U(1)_0+2\F}_{\qbarbrane}(Q_\xi,Q_1) & =			Z^{U(1)_0+2\F}_{\qbarbrane}(Q_1,Q_\xi) \,, \\
			Z^{U(1)_0+1\F +1\AF}_{\qbarbrane}(Q_\xi,Q_1) &=				Z^{U(1)_0+1\F +1\AF}_{\qbarbrane}(Q_1,Q_\xi) \,,
				\end{split}
\end{align}
which suggests that the $S$-duality (Fiber-base duality) in type-IIB string theory is manifest for the symmetric brane webs in Figure \ref{fig:2Fkee1} and Figure \ref{fig:2Fkeff0}. The $S$-duality exchanges $(p,q)$-branes to $(q, p)$-branes while D3-brane is invariant, so it is usually interpreted as the reflective symmetry of brane webs along the diagonal direction. 
One can add more chiral multiplets to brane webs in Figure \ref{fig:2Fkee1} and Figure \ref{fig:2Fkeff0} to construct more generic theories that enjoy the $S$-duality.

We need to emphasize that the $S$-duality puts the FI parameter and mass parameters on the same footing, which reflects the flavor symmetry $SU(2)$.  
Note that not all kinds of brane webs satisfy $S$-duality, since brane webs should be symmetric along the diagonal direction, which is a strong constraint on matter contents and CS levels. Hence we think that brane webs in Figure \ref{fig:2Fkee1} and Figure \ref{fig:2Fkeff0} are the only cases that enjoy $S$-duality when $N_f+N_a \leq 2$.

 We have discussed  some operations on 3d brane webs, including Hanany-Witten transitions, moving D5-branes, and fliping D5-branes. There may be more operations. In particular, one may hope that 3d
brane webs can be decomposed into many sub-webs in light of $ST$-transformations and the connection to the construction in \cite{plumb22}. We would like to leave this for future work.

\section{Nonabelian theories}\label{secnonabelian}
We can turn on the real mass deformations for nonabelian theories. The physical argument is the same as abelian theories. One can therefore get many equivalent brane webs. In this section, we discuss vortex partitions functions of nonabelian theories and find they also encode quiver matrices (mixed CS levels).  For simplicity, we ignore the one-loop parts of partition functions for simplicity, as they can be easily recovered.

\subsection{Real mass deformation}
\begin{figure}[H]
	\centering
	\includegraphics[width=4.2in]{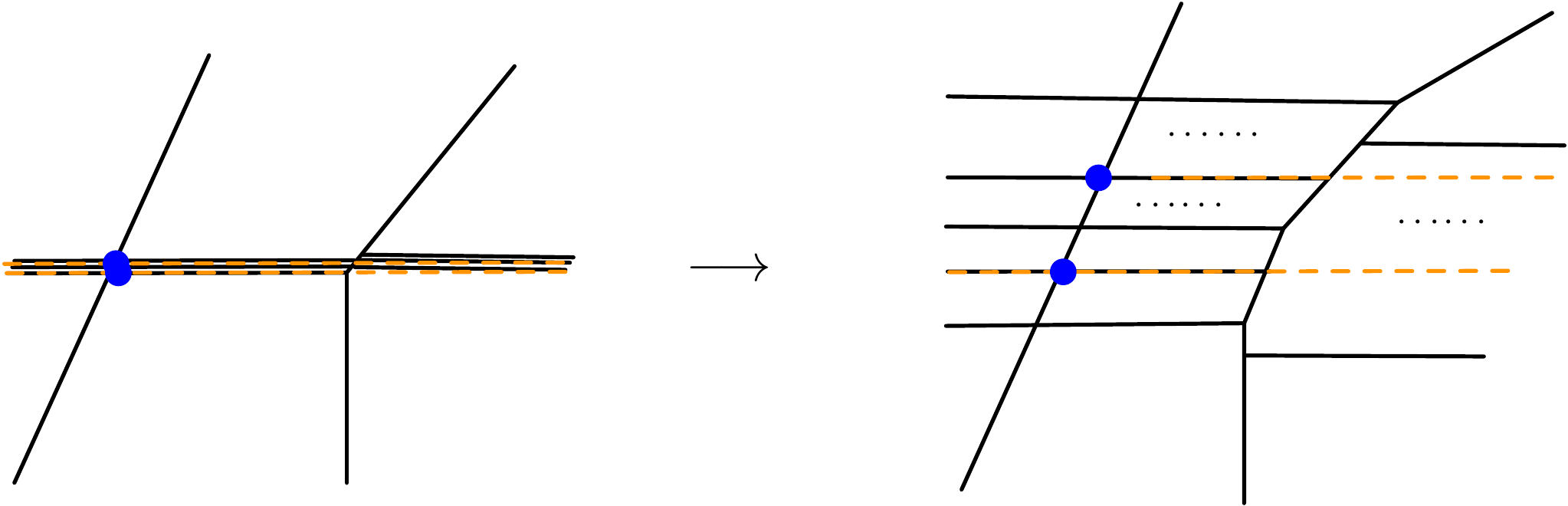}
	\caption{ This web describes the real mass deformations of the theory $U\left(2\right)_k +N_{f}\F+N_a\AF$. The mass deformations for higher rank $U\left(N\right)$ theories are similar.}
	\label{fig:turnonu2}
\end{figure}
 As illustrated in Figure \ref{fig:turnonu2}, overlapped flavor D5-branes for the theory $U\left(2\right)_k+N_{f}\F+N_a\AF$ separate from each other if one turns on real masses. Two D3-branes could locate on any pair of D5-branes, corresponding to different vacua on Higgs branch.

\subsection{Vortex partition functions}

Vortex partition functions of $U\left(N_c\right)_k+ N_{f} \F +N_a\AF$ have been computed using the factorization property of  superconformal indices in \cite{Hwang:2012jh,Taki:2013opa,Benini:2014aa}. The vortex partition function on a vacuum $\aa$  is
\begin{align}\label{vortexsigma}
&Z^{\text{vortex}} _{\aa} =\sum_{\vec{n}  =0 }^{\inf}  z^{|n|} Z_{\aa}^{\{n_j\}} \big(\tilde{M}, M, \u \big)   \,,\\
&Z_{\aa}^{\{n_j\}} \big(\tilde{M}, M, \u \big)  =\left(-1\right)^{ \left(k+\frac{N_{f}- N_a}{2}\right) \, \sum\limits_{i=1}^{N} n_i } \, e^{ i k \sum\limits_{j=1}^{N_c}  \left(  M_j n_j +\mu n_j + i \gamma n_j^2  \right)   }   \times  \\
&\hspace{1.6cm} 
 \frac{ \prod\limits_{\rho=1}^{N_a} 
	\prod\limits_{j=1}^{N_c}	\prod\limits_{k=0}^{n_j-1} 	 2 \sinh \frac{ - i \tilde{M}_\rho-i M_j -2 i \mu +2 \gamma k }{2}
  }
{  	\prod\limits_{i=1}^{N_c} 	\prod\limits_{j=1}^{N_c}	\prod\limits_{k=0}^{n_j-1} 	 2 \sinh \frac{  i {M}_i-i M_j  +2 \gamma \left(k- n_i\right) }{2}    
	\cdot
	 \prod\limits_{l=N_c+1}^{N_{f}}  
	\prod\limits_{j=1}^{N_c}  \prod\limits_{k=0}^{n_j-1}	
	2 \sinh \frac{  i M_l -i M_j + 2 \gamma\left(k+1\right) }{2}  
 }  \,,
\end{align}
where we can define $q = e^{-2 \gamma}\,, t=e^{iM}, \tilde{t} = e^{i \tilde{M}}\,, \tau=e^{i\u}$ to simplify this expression. The Higgs branch is $\mathcal{M}_H=\{ \aa \}$, and each $\aa$ is a permutation of $t_\p$.

We can rewrite the vortex partition function on vacuum $\aa$ as
\begin{align}\label{componentvortex}
Z_{\aa}^{\{n_j\}} ( \tilde{t},t,\tau ) = &
\left(-\sqrt{q}\right)^{k^{\eff} ||n||^2 
-  2 \,|n \cdot n|   }
 \prod\limits_{j=1}^{N_c}   \Bigg( t_j^{\,k^{\eff}} \cdot
 \tau^{ k-N_a  }  {q}^{ \frac{N_f+N_a-2N_c}{4}}  \prod\limits_{i=1}^{N_{f}} {t_i}^{-\frac{1}{2}}  \prod\limits_{\p=1}^{N_a}{\tilde{t}_\p}^{~-\frac{1}{2}}  \Bigg)^{ n_j}
 \nn\\
&\times 
\frac{   \prod\limits_{j=1}^{N_c}  \prod\limits_{\rho=1}^{N_a}  \left(\tau t_j \tilde{t}_\rho; ~q\right)_{n_j}    }
{   \prod\limits_{i, j=1}^{N_c} \left(  \frac{t_j}{t_i} q^{-n_i};~q \right)_{n_j}   ~\cdot~  \prod\limits_{j=1}^{N_c}
	\prod\limits_{l=N_c+1}^{N_{f}}
	\left( \frac{t_j}{t_l} q;~q\right)_{n_j}    } \,,
 \end{align}
 where we use the shorthand notation $|n| = \sum_{i=1}^{N_c} n_i$, $ |n \cdot n| =\sum_{ i ,j =1}^{N_c} 2\,n_i n_j $ and $ ||n||^2=\sum_{i=1}^{N_c} n_i^2$. Here $k$ is the bare \CS level and $k^{\eff}=k+ \frac{N_{f}-N_a}2 $ is the effective \CS level for non-abelian theories. 

We note that the vortex partition functions for non-abelian theories are themselves abelianized as \eqref{componentvortex} can be expressed in terms of $q$-Pochhammer productions, and there are mixed CS levels. Therefore, we believe there should be quiver matrices $C_{ij}$ for non-abelian theories. This is verified by examples in this subsection. We obtain quiver generating functions \eqref{quiverform1}, \eqref{qt2F2AF}, \eqref{quiverform2}, and \eqref{su22F3AF} for $U\left(2\right)_0$ theories with vanishing bare CS level for simplicity. Nonabelian theories with generic CS levels can be computed straightforwardly.

\subsection{$U\left(2\right)_0+2\F+2 \AF$ }
\begin{figure}[h!]
	\centering
	\includegraphics[width=3.5in]{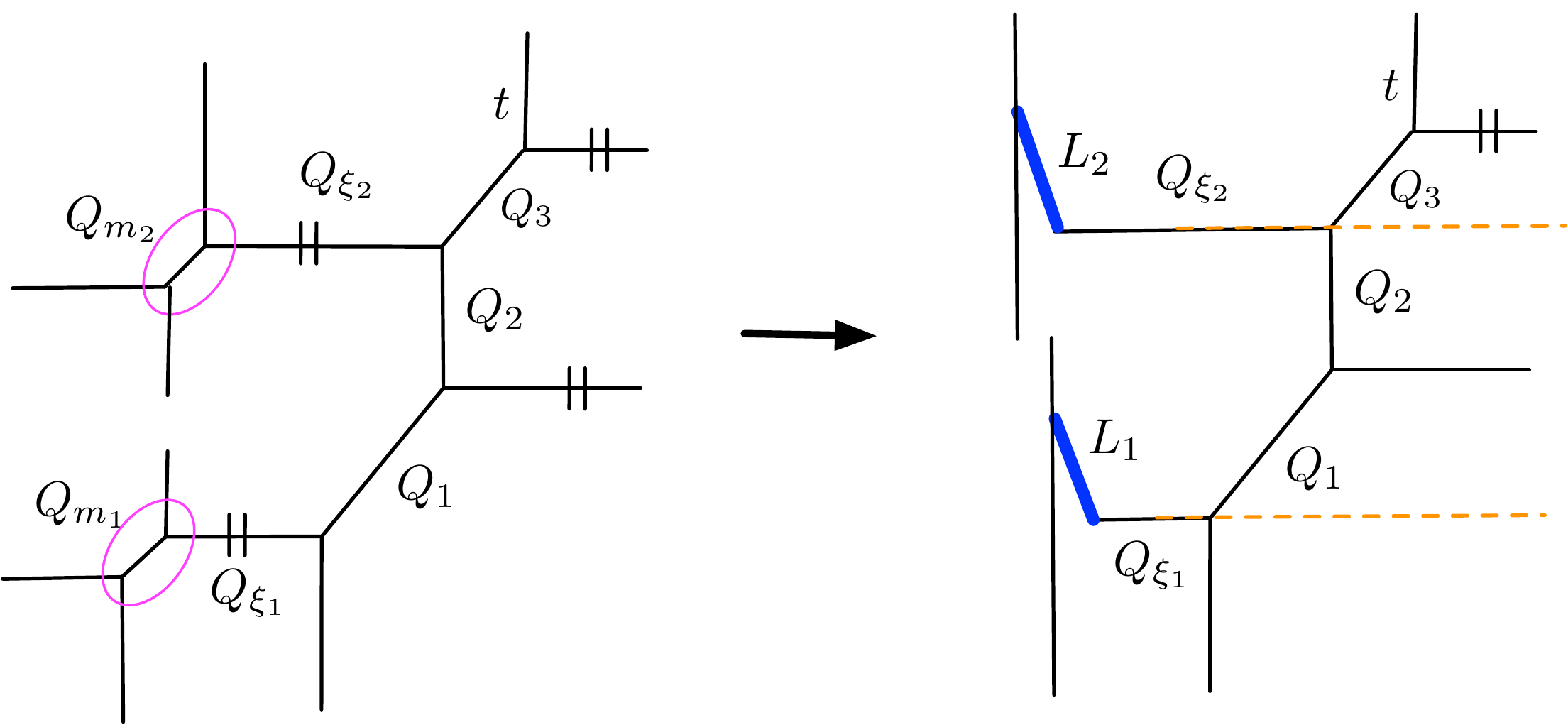}
	\caption{Implementing geometric transitions introduces D3-branes $L_1$ and $L_2$. In the topological vertex method, cutting lines means associated Young diagrams are set to be empty.}
	\label{fig:su2+2F+2AF}
\end{figure}
\begin{figure}[h!]
	\centering
	\includegraphics[width=2in]{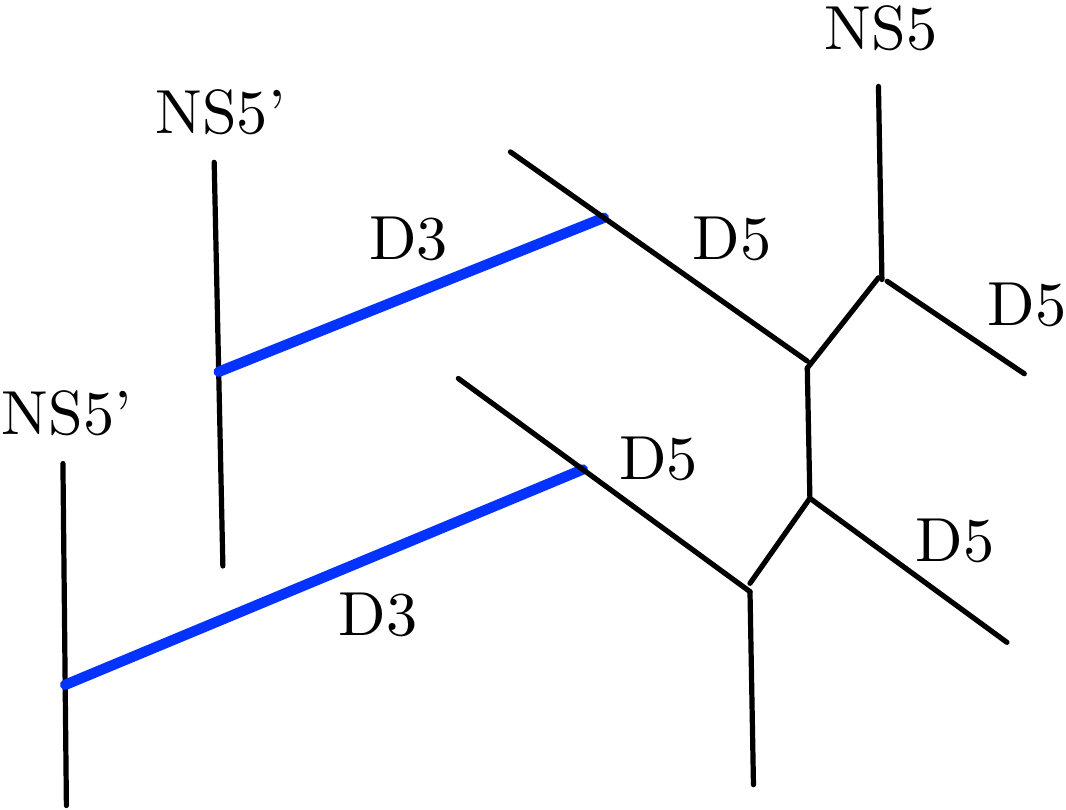}
	\caption{The new brane web matches with the vortex partition function \eqref{componentvortex}.	 }
	\label{fig:twons5}
\end{figure}

We take $U(2)_0+2\F+2 \AF$ as an example to illustrate. We implement refined topological vertex and Higgsing (geometric transition) to compute vortex partition function. Its brane web is shown in Figure \ref{fig:su2+2F+2AF}. We note that in order to match \eqref{componentvortex}, internal lines associated to NS5'-brane need to be cut. Namely, setting the corresponding Young tableaux on internal lines to be empty. This operation implies that two NS5'-branes rather than one should be introduced. We draw the more precise brane web in Figure \ref{fig:twons5}.
This brane web seems different from the brane web in e.g.\cite{Aharony:1997uv}, in which the two D3-branes connect to the same NS5'-brane. However, since we compute vortex partition functions on Higgs branch, the gauge group should be broken. We obtain this new brane web in Figure \ref{fig:su2+2F+2AF} from a practical perspective. We hope this new brane web can be understood by Higgsing mechanism. Namely, giving expectation values to scalars breaks the gauge group $U\left(1\right)\times U\left(2\right)$ to $U\left(1\right)\times U\left(1\right)$.  

In the following part of this section, we compute its refined vortex partition function for the toric diagram in Figure \ref{fig:su2+2F+2AF} using refined topological vertex. The Higgs branch for this theory only contains one vacuum. The assignment of $q$ and $t$ have been assigned on the diagram. We implement two geometric transitions by setting $Q_{m_1} = \frac{1}{q} \sqtq \,, Q_{m_2}=\frac{1}{q}\sqtq$ to obtain $\left(\bar{q}, \bar{q}\right)$-multiple brane amplitude. For more details on computation, see \cite{Cheng:2021aa}.
In this case, the Young tableaux on topological branes $L_1$ and $ L_2$ are antisymmetric, which are denoted by $\{n_1\}_V$ and $\{n_2\}_V$ respectively, and we draw the arrows for Young tableaux toward open topological branes (D3-branes). In our notation, $\{n\}_H$ denotes a symmetric Young tableau with $n$ boxes and $\{n\}_V$ denotes an anti-symmetric Young tableau with $n$ boxes; see  \eqref{HVnotation}.

Using  \eqref{antisymNn1n2} and other identities in Appendix, we express the vortex partition function as 
	\begin{small}
\begin{align}\label{refsu22F2AFpartfun}
&Z^{\left( \bar{q},\bar{q}\right)\text{-brane}} = \\
&=	
\sum_{n_1,n_2 =0}^{\inf}  
\frac{   \left({   Q_{\xi_1}\sqtq } \right)^{n_1}     \left({   Q_{\xi_2}\sqtq } \right)^{n_2}   
}
{ \left(t;t\right)_{n_1}  \left(t;t\right)_{n_2}   } \cdot
\frac{      \left(Q_1 \sqtq ;t\right)_{n_1}   \left(Q_2 \sqqt;t^{-1}\right)_{n_2}
	  \left(Q_3 \sqtq;t\right)_{n_2}  \left( Q_1Q_2Q_3 \sqtq;t\right)_{n_1}
}
{
\left(Q_1Q_2 t ^{n_1} ;t^{-1}\right)_{n_2}   \left(Q_1 Q_2 \frac{t}{q}\cdot t^{-n_2};t  \right)_{n_1}
}  \\
&{
	=  \sum_{n_1,n_2 =0}^{\inf}  
{ \left(-\sqrt{t}\right)^{-2 n_1 n_2}  \left({   Q_{\xi_1}\sqtq } \right)^{n_1}     \left(\frac{  Q_{\xi_2} }{Q_1}\right)^{n_2}    
} \cdot
\frac{      \left(Q_1 \sqtq ;t\right)_{n_1} \left( Q_1Q_2Q_3 \sqtq;t\right)_{n_1}   \left(Q_2^{-1} \sqtq;t\right)_{n_2} 
	  \left(Q_3 \sqtq;t\right)_{n_2} 
}
{ \left(t;t\right)_{n_1}  \left(t;t\right)_{n_2}  
	\left(Q_1Q_2\frac{t}{q}\cdot t^{-n_2} ;t\right)_{n_1}   \left(Q_1^{-1} Q_2^{-1} t^{-n_1};t  \right)_{n_2}
}  \,,}
\end{align}
\end{small}
which in unrefined limit $q=t$ takes form 
\begin{align}\label{unrefsu22F2AF}
&Z^{\left( \bar{q},\bar{q}\right)\text{-brane}}|_{q=t}= \nn\\
&\sum_{n_1,n_2=0}^{\inf}  \frac{ (-\sqrt{q})^{-2n_1 n_2}  Q_{\xi_1} ^{n_1}  \left(Q_1^{-1}Q_{\xi_2}\right)^{n_2}    \left(Q_1;q\right)_{n_1} \left( Q_1Q_2Q_3;q\right)_{n_1}  \left(Q_2^{-1};q\right)_{n_2}   
	\left(Q_3;q\right)_{n_2}  }
{  \left(q;q\right)_{n_1} \left(q;q\right)_{n_2}  
	\left( Q_1Q_2 q^{-n_2};q \right)_{n_1}  
	\left(Q_1^{-1} Q_2^{-1} q^{-n_1};q \right)_{n_2}     } \,,
\end{align}
which matches with \eqref{componentvortex}. For theories with generic CS levels, we only need to insert a term $ (-\sqrt{q})^{k^{\eff} ( n_1^2+ n_2^2 )}$ in \eqref{unrefsu22F2AF}.
We note that the K\"ahler parameters above the original lines are given positive power $Q_i Q_j Q_k$, and  K\"ahler parameters below the original lines are given negative power $Q_i^{-1}Q_j^{-1}Q_k^{-1}$. This assignment rule is the same as abelian theories that we have discussed in section \ref{secmassdef}.

We note that the contributions from strings connecting two separate D3-branes come from a bifundamental hypermultiplet in 5d $\N=1$ gauge theories
\begin{align}
	N_{\{n_j\}_V, \{n_i\}_V} \left( \b t q^{-1};\;t^{-1},q^{-1} \right) = \left( \b t^{n_i}, {t^{-1}}\right)_{n_j} \left( \b t q^{-1} ~q^{-n_j}, ~t \right)_{n_i}\,, 
	\end{align}
 which in our context are interpreted as $W_{\pm}$-bosons in 3d $\N=2$ nonabelian theories. 
 Each $W_\pm$-boson's contribution can split into two parts using the identity 
 \begin{align}
 \left( \a q^{-m};q \right)_n  =
 \begin{cases}
\left( \a ;q\right)_{n-m} \left(\a q^{-1} ;q^{-1} \right)_n &  \text{if} ~ n \geqslant m\\    
\left( \a q^{-1};q^{-1} \right)_n\left( \a q^{-1} ;q^{-1}  \right)_{m-n}^{-1}  & \text{if} ~ n<m
 \end{cases} \,,
 \end{align}
then the total $W_\pm$ contribution can be written as the sum of two parts
 \begin{align}
& \left(\a q^{-m};q \right)_n\left(\a^{-1} q^{-n};q\right)_m \\	
 &= \begin{cases}
  \left(\a^{-1}q^{-1};q^{-1}\right)_m \left(\a q^{-1};q^{-1}\right)_n ~ \cdot  \left(\a^{-1};q\right)_{m-n} \left(\a q^{-1};q^{-1}\right)_{m-n}^{-1}
      &  \text{if} ~ m \geqslant n\\    
 \left(\a^{-1} q^{-1};q^{-1}  \right)_m \left(  \a q^{-1};q^{-1} \right)_n \cdot \left(\a;q\right)_{n-m}\left(\a^{-1}q^{-1};q^{-1}\right)_{n-m}  & \text{if} ~ n>m
 \end{cases}  \\
&=  \begin{cases}
 \left(\a^{-1}q^{-1};q^{-1}\right)_m \left(\a q^{-1};q^{-1}\right)_n ~ \cdot  
 \frac{ \left( -\sqrt{q} \right)^{\left(m-n\right)^2}  \left( \sqrt{q}   \a \right)^{n-m}  \left( 1- \a\right) }{ 1- \a q^{n-m}   }
 &  \text{if} ~ m \geqslant n\\    
 \left(\a^{-1} q^{-1};q^{-1}  \right)_m \left(  \a q^{-1};q^{-1} \right)_n \cdot 
  \frac{ \left( -\sqrt{q}\right)^{\left(m-n\right)^2}  \left( \sqrt{q} \a^{-1}\right)^{m-n}  \left( 1- \a^{-1}\right) }{ 1- \a^{-1} q^{m-n}   }
  & \text{if} ~ n>m
 \end{cases} \,.
 \end{align}

We can also set $Q_{m_1} = t \sqtq\,, Q_{m_2} = t \sqtq$ to obtain $\left(t, t\right)$-multiple brane amplitude. We find that the exchange symmetry  $q \leftrightarrow t^{-1}$ still preserves in nonabelian theories. Therefore, there is the relation
\begin{align}
&Z^{\left( \bar{q},\bar{q}\right)\text{-brane}} \, \xlongleftrightarrow{ ~~q ~\leftrightarrow~ t^{-1} } \, Z^{\left( t, t\right)\text{-brane}} \,.
\end{align}
This exchange symmetry transposes the Young tableaux on open topological branes, so we have symmetric Young tableaux $\{n_1\}_H$ and $\{{n_2}\}_H$ for	 $\left(t, t\right)\text{-brane}$. 

Now, we compute the quiver matrices of nonabelian theories.
We note that the vortex partition function for this theory also can be written as a quiver generating function using 
\eqref{n1n2quiver}. Each term in $W$-boson contributions can be expressed as the following quiver form in a shorthand notation
\begin{align}
\frac{1}{ \left(\b q^{-n_1} ;q\right)_{n_2}} \rightarrow 
\Bigg(~~
\begin{matrix}
\\
n_1 \\
 n_2\\
d_i \\
d_j \\
\end{matrix}
\begin{split}
&\begin{array}{ccccc}
~~\,n_1&\,n_2&~\,d_i &~\,d_j
\end{array}\\
&\begin{bmatrix}
 & &-1 &-1 \\
&  &~~1 &\\
-1 &~~1 &~~1 &\\
-1 &  & &\\
\end{bmatrix}
\end{split}
\,,~
\left(\frac{\b }{ \sqrt{q}}, ~\b \right)  \Bigg)  \,.
\end{align}
We write \eqref{unrefsu22F2AF} in the form of quiver generating function\footnote{Note that these two expressions are equivalent after adding the one-loop part using \eqref{aquiverform} and \eqref{bquiverform}. In \eqref{quiverform1}, there are two $q$ parameters in $P_{C_{ij}}(q,q; x_i)$, which extends the definition in   \eqref{quivergenfun}. 
	In the following, when we consider the $(\bar{q}, t)$-brane, the refined generating function  in \eqref{quiverform3} would take form $P_{C_{ij}}(t,q^{-1}; x_i)$.}
\begin{align}\label{quiverform1}
Z^{\left( \bar{q},\bar{q}\right)\text{-brane}}|_{q=t} \simeq P_{C_{ij}} \left(q, q;\, Q_{\xi_1}, \,Q_1^{-1} Q_{\xi_2} , \frac{Q_{1,2}}{\sqrt{q}} ,Q_{1,2}, \frac{Q_{1,2}^{-1}}{\sqrt{q}},\, Q_{1,2}^{-1},\, Q_1,\, Q_{1,2,3},\, Q_2^{-1},\, Q_3
 \right)  \,,
\end{align}
where $Q_{i,j,k} := Q_i Q_j Q_k$ and the quiver matrix $C_{ij}$ is
\begin{align}
&\begin{array}{cccccccccc}
~~~n_1&~n_2&~d_1 &~d_2 &~~d_3&~d_4\,&\,d_5&\,d_6&d_7&d_8
\end{array}\\
C_{ij} =
\begin{matrix}
n_1 \\
n_2\\
d_1 \\
d_2 \\
d_3\\
d_4\\
d_5\\
d_6\\
d_7\\
d_8
\end{matrix}
	&
\left[
\begin{array}{cc|cc|cc|cccc}
& \textcolor{blue}{-1}&-1 &-1 ~& ~1& &~\,1&~\,1&~\,&~\,\\
 \textcolor{blue}{-1}&& ~~1& &-1&-1~&&&~\,1&~\,1  \\
\hline
-1&  ~~1&~~1 & &&&&&&\\
-1& & & &&&&&&\\
\hline
~~1&- 1& & &~1&&&&&\\
& -1& & &&&&&&\\
\hline
~~1& &&&&&&&&\\
~~1&&&&&&&&&\\
~~&~~1&&&&&&&&\\
~~&~~1 &&&&&&&&\\
\end{array}
\right]   \,.
\end{align}
The missing elements in $C_{ij}$ are zero. Here $n_1, n_2, n_3, d_1 ,\cdots d_{6}$ are degrees for expansion variables $x_i^{d_i}$ in the quiver generating function $P_{C_{ij}}\left(q, q; x_i\right)$.

The existence of quiver matrices implies that nonabelian theories can be abelianized into abelian theories with gauge group $U(1)\times U(1) \times \cdots \times U(1)$, and the rank of the gauge group equals to the rank of quiver matrix $C_{ij}$. Because of the equivalence \eqref{quiverform1},  we conjecture that the theory $U\left(2\right)_0 +2\F+2 \AF $ can be regarded as a theory that consists ten $U\left(1\right)+1 \F$ coupled together by effective mixed Chern-Simons level $k_{ij}^{\eff}=k_{ij} +\delta_{ij}/2$ and $C_{ij} =k_{ij}^{\eff}$. 

Similarly, one can compute the vortex partition function for $\left(\bar{q},t\right)$-brane, which takes form 
\begin{align}
	&Z^{\left( \bar{q},t\right)\text{-brane}} =
		\nn\\
		 & = \sum_{n_1,n_2 =0}^{\inf}  
		\frac{   \left({   Q_{\xi_1}\sqtq } \right)^{n_1}    \left(-\sqrt{q}\right)^{n_2^2} \left({   Q_2Q_{\xi_2} } \right)^{n_2}   
		}
		{ \left(t;t\right)_{n_1}  \left(q^{-1};q^{-1}\right)_{n_2}   }  \nn \\
	&~~~~\times 	\frac{      \left(Q_1 \sqtq ;t\right)_{n_1} \left( Q_1Q_2Q_3 \sqtq;t\right)_{n_1}  \left(Q_2^{-1} \sqtq;q^{-1}\right)_{n_2}
			\left(Q_3^{-1} \sqtq;q^{-1}\right)_{n_2}  
		}
		{
		N_{\{n_2\}_H,\{n_1\}_V } \left( Q_1 Q_2 t q^{-1};t^{-1},q^{-1}\right)
		} \\
	&=\sum_{n_1,n_2 =0}^{\inf}  
 \frac{1- Q_1 Q_2 q^{-1}  }{ 1-Q_1Q_2 q^{-1} t^{n_1}  q^{n_2}  }  \cdot
 	{   \left({   Q_{\xi_1}\sqtq } \right)^{n_1}     \left({   q^{\frac{3}{2}}Q_1^{-1}Q_{\xi_2} } \right)^{n_2}   
	}
	\label{factor1}  \nn\\
	&  ~~~~ \times 
	\frac{      \left(Q_1 \sqtq ;t\right)_{n_1} \left( Q_1Q_2Q_3 \sqtq;t\right)_{n_1}  
	}
	{\left(t;t\right)_{n_1}  
\left(Q_1 Q_2 q^{-1};t\right)_{n_1}
	}  \cdot
	\frac{      \left(Q_2^{-1} \sqtq;q^{-1}\right)_{n_2}
	\left(Q_3^{-1} \sqtq;q^{-1}\right)_{n_2}  
}
{  \left(q^{-1};q^{-1}\right)_{n_2} 
	\left( Q_1^{-1}Q_2^{-1}  q;q^{-1}\right)_{n_2}
} \,,
	\end{align}
where we use
\begin{align}
	 N_{\{{n_i}\}_{H}\,,\{{n_j}\}_{V}}\left( Q; t^{-1},q^{-1}   \right)   
	= \frac{ 1- Q t^{-1} q^{n_i} t^{n_j} }{ 1- Q t^{-1} } \cdot
	 \left(-\sqrt{q}\right)^{n_i^2} \left(Q t^{-1} q^{-\half} \right)^{n_i}
	\left(Q^{-1} t,q^{-1}\right)_{n_i}  \left(Q \;t^{-1};t  \right)_{n_j}
	 \,.
\end{align} 
An interesting point for $\left(\bar{q},t\right)$--brane is that if the mass of W-bosons $Q_1Q_2$ is much larger than one, then this vortex partition function factorizes into two independent parts, which are respectively associated to the $\bar{q}$-brane of the theory $U(1)_0+2\F+2\AF$ and the ${t}$-brane of another theory $U(1)_0+2\F+2\AF$.

We expand the denominator $(1-Q_1Q_2 q^{-1} t^{n_1} q^{n_2})^{-1} $ in \eqref{factor1} using 
$\left(1-x\right)^{-1}= \sum_{n=0}^{\inf} x^n$ and ignore the numerator $1- Q t^{-1}$ to get
\begin{align}
	&Z^{\left( \bar{q},t\right)\text{-brane}} =\nn \\
	& 
	\sum_{n_1,n_2,n_3 =0}^{\inf}  \left( -\sqrt{t}\right)^{2 n_1 n_3}  \left( -\sqrt{q^{-1}}\right)^{-2 n_2 n_3}
	{   \left({   Q_{\xi_1} \sqtq } \right)^{n_1}     \left({   \frac{q^{\frac{3}{2}} Q_{\xi_2}}{Q_1} } \right)^{n_2}   
	} \left(\frac{Q_{1,2}}{q}\right)^{n_3}  \nn\\
		&~~~~ \times 
		\frac{      \left(Q_1 \sqtq ;t\right)_{n_1} \left( Q_{1,2,3} \sqtq;t\right)_{n_1}  \left(Q_2^{-1} \sqtq;q^{-1}\right)_{n_2}
		\left(Q_3^{-1} \sqtq;q^{-1}\right)_{n_2}  
	}
	{\left(t;t\right)_{n_1} 
		\left(Q_{1,2}\, q^{-1};t\right)_{n_1}
		 \left(q^{-1};q^{-1}\right)_{n_2} 
		 	\left( Q_{1,2}^{-1}\, q;q^{-1}\right)_{n_2}
	}  \,,
	\end{align}
which also encodes a matrix $C_{ij}$, since
\begin{align}\label{quiverform3}
	Z^{\left( \bar{q},t\right)\text{-brane}} \simeq
	P_{C_{ij}} \Big(&t,q^{-1}; \;Q_{\xi_1} \sqtq,\;q^{\frac{3}{2}}Q_1^{-1}Q_{\xi_2}, \;Q_{1,2} q^{-1}, \, Q_{1,2} \frac{\sqrt{t}}{q},\, Q_1\sqtq,\,  Q_{1,2,3}\sqtq,\, \nn \\
	&Q_{1,2}^{-1}q^2, \, Q_2^{-1}\sqrt{t}, \, Q_3^{-1}\sqrt{t}
	\Big)  \,,
\end{align}
where the matrix $C_{ij}$ is
\begin{align}\label{qt2F2AF}
	&\begin{array}{ccccccccc}
		~~~n_1&~n_2&~~n_3 &~~~d_1 &~d_2 &~~d_3&~~~d_4\,&~~d_5&~~d_6
	\end{array}\\
	C_{ij} =
	\begin{matrix}
		n_1 \\
		n_2\\
		n_3\\
		d_1 \\
		d_2 \\
		d_3\\
		d_4\\
		d_5\\
		d_6
	\end{matrix}
	&
	\left[
	\begin{array}{ccc|ccc|cccc}
		& \textcolor{blue}{~~~}&~~~~1\,~~ &~~1\, ~& ~1~\,&\,~1~ ~&~&~&~\,&~\,\\
	&	\textcolor{blue}{~~1}& -1& &&&&-1&-1&\,-1\,  \\
		~1&  -1&~~ & &&&&&&\\	\hline
		~1& & &~1 &&&&&&\\
		~1&& & &&&&&&\\
		~1& & & &&&&&&\\
		\hline
	&-1 &&&&&&&&\\
		~~&-1&&&&&&&1&\\
		~~&-1&&&&&&&&1\\
	\end{array}
	\right]   \,.
\end{align}
Here $n_1, n_2, n_3, d_1 ,\cdots d_{6}$ are degrees for expansion variables $x_i^{d_i}$ in the quiver generating function $P_{C_{ij}}\left(t, q^{-1}; x_i\right)$.

Exchange symmetry still preserves, so we have
\begin{align}
	&Z^{\left( \bar{q},t\right)\text{-brane}} \, \xlongleftrightarrow{ ~~q ~\leftrightarrow ~t^{-1}  } \, Z^{\left( t,\bar{q}\right)\text{-brane}} \,.
\end{align}

\subsection{$U\left(2\right)_0+3\F+2 \AF$ }
The brane webs of this theory have three fundamental D5-branes and  two D3-branes. These D3-branes can locate at any pair of fundamental D5-branes. Then there are three vacua on Higgs branch, and we denote them by
\begin{align}
\left(\bar{q}\,, \bar{q}\,,  \emptyset\right)\,,~~\left(\bar{q}\,, \emptyset\,, \bar{q}\right)\,,~~\left(\emptyset\,, \bar{q}\,, \bar{q}\right) \,.
\end{align} 
Here the $\emptyset $ means that there is no D3-brane ending on the corresponding D5-brane, and $\bar{q}$ means that we put a $\qbarbrane$ on the D5-brane.

\begin{figure}[h!]
	\centering
	\includegraphics[width=3.5in]{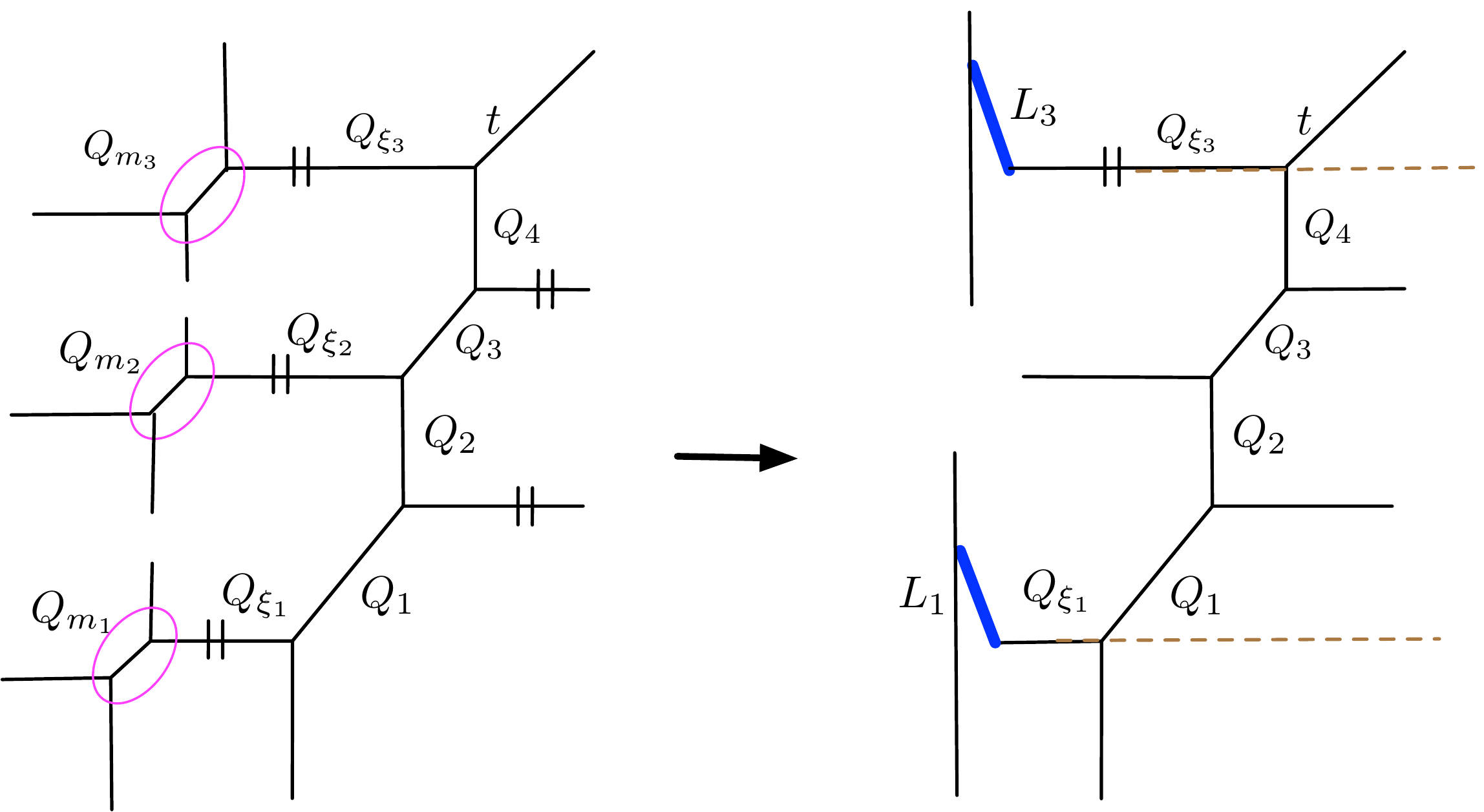}
	\caption{There are two original lines for this theory. On Higgs branch, we need to cut some internal lines. The effective \CS level for this brane web is zero.}
	\label{fig:u23F2AF}
\end{figure}

We pick up the second vacuum for computation. First, we draw the corresponding brane web in Figure \ref{fig:u23F2AF}. For this brane web, there is the freedom of flipping antifundamental flavor D5-branes up and down the original lines, 
but the effective CS levels $k^{\eff}$ remains, since flipping does not change the relative angle $\theta$ between NS5-and NS5'-brane. We set $Q_{m_1} = \frac{1}{q}\sqtq \,, Q_{m_2 }=\sqtq \,,$ and $Q_{m_3} = \frac{1}{q}\sqtq$ to obtain the vortex partition function for the $\left(\bar{q}, \emptyset, \bar{q}\right)$-brane:
\begin{align}
&Z^{  \left( \bar{q} \,, \emptyset\,,\bar{q} \right)\text{-brane}}=  \sum_{n_1,n_3 =0}^{\inf}  
\frac{ \left(-\sqrt{t}\right)^{-2 n_1 n_3}  \left({   \frac{Q_{\xi_1}}{Q_1}\sqtq } \right)^{n_1}     \left({   \frac{Q_{\xi_3}}{Q_3} \sqqt } \right)^{n_3}    
}
{ \left(t;t\right)_{n_1}  \left(t;t\right)_{n_3}   }  \nn \\
&~~~~ \times
\frac{      \left(Q_1 \sqtq ;~t \right)_{n_1} \left( Q_{1,2,3} \sqtq;~ t \right)_{n_1}   \left(Q_4^{-1} \sqtq;~t\right)_{n_3} 
	\left(Q_{2,3,4}^{-1}  \sqtq; ~t\right)_{n_3} 
}
{\left( Q_{1,2} \frac{t}{q}\,;~t  \right)_{n_1}
	\left( Q_{3,4}^{-1} \,;~t  \right)_{n_3}
	\left(Q_{1,2,3,4}\frac{t}{q}\cdot t^{-n_3} ;~ t\right)_{n_1}   \left(Q_{1,2,3,4}^{-1}t^{-n_1};~t  \right)_{n_3} 
}  \,,
\end{align}
which at unrefined limit $q=t$ is 
\begin{align}\label{U2qbarqbar}
&Z^{  \left( \bar{q} \,, \emptyset\,,\bar{q} \right)\text{-brane}} |_{q=t}=  \sum_{n_1,n_3 =0}^{\inf}  
\frac{ \left(-\sqrt{q}\right)^{-2 n_1 n_3}  \left({   \frac{Q_{\xi_1}}{Q_1} } \right)^{n_1}     \left({   \frac{Q_{\xi_3}}{Q_3}  } \right)^{n_3}    
}
{ \left(q;q\right)_{n_1}  \left(q;q\right)_{n_3}   }  \nn \\
& ~~~~ \times
\frac{      \left(Q_1  ;~q \right)_{n_1} \left( Q_{1,2,3} ;~ q \right)_{n_1}   \left(Q_4^{-1} ; ~q\right)_{n_3} 
	\left(Q_{2,3,4}^{-1}  ;~q\right)_{n_3} 
}
{\left( Q_{1,2} \,;~q  \right)_{n_1}
		\left(Q_{1,2,3,4} \, q^{-n_3} ;~q \right)_{n_1}  
	\left( Q_{3,4}^{-1} \,;~q  \right)_{n_3}
 \left(Q_{1,2,3,4}^{-1}  t^{-n_1};~q  \right)_{n_3} 
}   \,,
\end{align}
from which we can see a mixed effective CS level $ k^{\eff}_{1,3}=-\delta_{1,3}$.
Once again, \eqref{U2qbarqbar} confirms that K\"ahler parameters above original lines $L_1$ or $L_3$ are associated with positive masses, and  K\"ahler parameters below original lines are associated with negative masses.

This vortex partition function can be written in terms of the quiver generating function
\begin{align}\label{quiverform2}
	Z^{  \left( \bar{q} \,, \emptyset\,,\bar{q} \right)\text{-brane}} |_{q=t} \simeq
	 P_{C_{ij}}\Big(&q,q;\, {Q_{\xi_1}Q_1^{-1}}, {Q_{\xi_3} Q_3^{-1}}\,, \frac{Q_{1,2,3} }{\sqrt{q}}, \, Q_{1,2,3}, \frac{Q_{1,2,3,4}^{-1}}{\sqrt{q}} , \, Q_{1,2,3,4}^{-1},\,  \nn
	 \\
	 &Q_1, \, Q_{1,2,3},\, Q_4^{-1},\, Q_{2,3,4}^{-1} ,\, Q_{1,2},\, Q_{3,4}^{-1} \Big) \,,
	\end{align}
where we use shorthand notation $Q_{1,2,3,\dots}= Q_1Q_2Q_3 \cdots$. 
The quiver matrix $C_{ij}$ is 
\begin{align}
	&\begin{array}{cccccccccccc}
		~~~n_1&~n_3&~d_1 &~d_2 &~~d_3&~d_4\,&\,d_5&\,d_6&d_7&d_8&d_9&d_{10}
	\end{array}\\
	C_{ij} =
	\begin{matrix}
		n_1 \\
		n_3\\
		d_1 \\
		d_2 \\
		d_3\\
		d_4\\
		d_5\\
		d_6\\
		d_7\\
		d_8\\
		d_9\\
		d_{10}
	\end{matrix}
	&
	\left[
	\begin{array}{cc|cc|cc|cccc|cc}
		& \textcolor{blue}{-1}&-1 &-1 ~& ~1& &~\,1&~\,1&~\,&~\,&~1 &~~\\
		\textcolor{blue}{-1}&& ~~1& &-1&-1~&&&~\,1&~\,1  & &\,~1 \\
		\hline
		-1&  ~~1&~~1 & &&&&&&\\
		-1& & & &&&&&&\\
		\hline
		~~1&- 1& & &~1&&&&&\\
		& -1& & &&&&&&\\
		\hline
		~~1& &&&&&&&&\\
		~~1&&&&&&&&&\\
		~~&~~1&&&&&&&&\\
		~~&~~1 &&&&&&&&\\ \hline
			~~1& &&&&&&&&&~1\\
				~~&~~1 &&&&&&&&&&~~1\\
	\end{array}
	\right]\,.
\end{align}
Here $n_1, n_3, d_1 ,\cdots d_{10}$ are degrees for expansion variables $x_i^{d_i}$ in the quiver generating function $P_{C_{ij}}\left(q,q;x_i\right)$.

Similarly, one can computation the vortex partition function for $(\bar{q},\0,t)$-brane:
\begin{align}
	&Z^{  \left( \bar{q} \,, \emptyset\,,t \right)\text{-brane}}=  \sum_{n_1,n_3 =0}^{\inf}  
	\frac{  \left({   {Q_{\xi_1}}\sqtq } \right)^{n_1}     \left(
		{  - Q_2{Q_{\xi_3}}\sqqt } \right)^{n_3}  \left(-\sqrt{q}\right)^{n_3}  
	}
	{ \left(t;t\right)_{n_1}  \left(q^{-1};q^{-1}\right)_{n_3}   }  \nn \\
	& 
\qquad\quad \quad \times	\frac{      \left(Q_1 \sqtq ;~t \right)_{n_1} \left( Q_{1,2,3} \sqtq,~ t \right)_{n_1}   \left(Q_4^{-1} \sqtq; ~q^{-1}\right)_{n_3} 
		\left( Q_{2,3,4}^{-1}\sqtq;~q^{-1} \right)_{n_3} 
	}
	{\left( Q_{1,2} \frac{t}{q}\,;~t  \right)_{n_1}
		\left( Q_{3,4}^{-1} \,;~q^{-1} \right)_{n_3}
		N_{\{n_3\}_H,\{n_1\}_V } \left( Q_{1,2,3,4} t q^{-1};t^{-1},q^{-1}\right)
	}  \\
&= \sum_{n_1,n_3,n_4 =0}^{\inf}  
	\left(-\sqrt{t} \right)^{2 n_1 n_4} \left( -\sqrt{q^{-1}} \right)^{-2 n_3 n_4}
	  \left({   {Q_{\xi_1}}\sqtq } \right)^{n_1}     \left(
	\frac{  - Q_{1,2,2,3,4, \xi_3}  }{q \sqrt{t} } \right)^{n_3}     \left( \frac{Q_{1,2,3,4}}{q} \right)^{n_4}
\nn \\
&\times \frac{  
     \left(Q_1 \sqtq ;~t \right)_{n_1} \left( Q_{1,2,3} \sqtq;~ t \right)_{n_1}   \left(Q_4^{-1} \sqtq;~q^{-1}\right)_{n_3} 
\left( Q_{2,3,4}^{-1}\sqtq;~q^{-1} \right)_{n_3} 
}{ \left(t;t\right)_{n_1}  
\left( Q_{1,2} \frac{t}{q}\,;~t  \right)_{n_1}
 \left( Q_{1,2,3,4} q^{-1}\,; t \right)_{n_1} 
 \left(q^{-1};q^{-1}\right)_{n_3}  
 \left( Q_{3,4}^{-1} \,;~q^{-1} \right)_{n_3}  
   \left(  Q_{1,2,3,4}^{-1}q;\,q^{-1} \right)_{n_3}  }   \,,
	\end{align}
which also encodes a quiver matrix $C_{ij}$
\begin{align}
Z^{  \left( \bar{q} \,, \emptyset\,,t \right)\text{-brane}} \simeq P_{C_{ij}}	&\Big(  t,q^{-1}; 
Q_{\xi_1}\sqtq,\, -Q_{1,2,2,3,4,\xi_3}q^{-1}t^{-\half},\, Q_{1,2,3,4}q^{-1}, \, Q_1\sqtq,\, Q_{1,2,3,} \sqtq,\, \nn\\
&~\frac{Q_{1,2}}{q \sqrt{t}} \,,  Q_4^{-1}\sqrt{t},\, Q_{2,3,4}^{-1}\sqrt{t},\, Q_{3,4}^{-1} q,\, Q_{2,3,4}^{-1} q^2
\Big) \,,	\end{align}
where the matrix $C_{ij}$ is
\begin{align}\label{su22F3AF}
	&\begin{array}{ccccccccccc}
		~~~n_1&~n_3&~~n_4 &~~~d_1 &~d_2 &~~d_3&~~d_4&~~~d_5\,&~~d_6&~~d_7&~~d_8
	\end{array}\\
	C_{ij} =
	\begin{matrix}
		n_1 \\
		n_3\\
		n_4\\
		d_1 \\
		d_2 \\
		d_3\\
		d_4\\
		d_5\\
		d_6\\
		d_7\\
		d_8
	\end{matrix}
	&
	\left[
	\begin{array}{ccc|cccc|cccc}
		& \textcolor{blue}{~~~}&~~~~1\,~~ &~~1\, ~& ~1~\,&\,~1~ ~&\,~1~~&~&~\,&~\,&\\
		&	\textcolor{blue}{~~1}& -1& &&&&\,-1&~-1&\,-1\,&\,-1  \\
		~1&  -1&~~ & &&&&&&&\\	\hline
		~1& & & &&&&&&&\\
		~1&& & &&&&&&&\\
		~1& & & &&~1~&&&&&\\
		~1& &&&&&~1~&&&&\\ 		\hline
		~~&-1&&&&&&~1~&&&\\
		~~&-1&&&&&&&~1~&&\\
			~~&-1 &&&&&&&&&\\
				~~&-1 &&&&&&&&&\\
	\end{array}
	\right]   \,.
\end{align}

We can directly compute other types of branes or use the exchange symmetry between them, which leads to
\begin{align}
	&Z^{\left( \bar{q},\0,\bar{q}\right)\text{-brane}} \, \xlongleftrightarrow{ ~~q ~\leftrightarrow ~t^{-1}  } \, Z^{\left( t,\0,t\right)\text{-brane}} \,,\\
	&Z^{\left( \bar{q},\0,t\right)\text{-brane}} \, \xlongleftrightarrow{ ~~q ~\leftrightarrow ~t^{-1}  } \, Z^{\left( t,\0,\bar{q}\right)\text{-brane}} \,.
\end{align}
Moreover, since $\qbarbrane$ ($\tbarbrane$) is equivalent to $\qbrane$ ($\tbrane$) after shifting FI parameters, we do not need to compute other cases \cite{Cheng:2021aa}.

\section{Ooguri-Vafa invariants}\label{OVformula}

In \cite{Dimofte:2010tz,Cheng:2021aa,Kameyama:2017ryw}, the Ooguri-Vafa (OV) formula found in \cite{Ooguri:1999bv} was generalized to the refined version. Since there is an exchange symmetry $q\leftrightarrow t^{-1}$ relating $\qbarbrane$ and $\tbrane$, we only consider $\qbarbrane$ whose refined OV formula is
	\begin{align}\label{openGV}
\begin{split}
Z_{\qbarbrane}^{\text{vortex}} &=  \, \exp \Bigg[
\sum\limits_{\mathcal{\b} \in H_2\left(X,L, \mathbb{Z}\right)} 
\sum\limits_{s,r \in \mathbb{Z}/2} 
\sum\limits_{n=1}^{\inf}
-	\frac{  \left(-1\right)^{2 s+2 r}     t^{ -  n s   }  \left(   \frac{t}{q}   \right)^{ n \, r}   
	N_{\mathcal{\b}}^{\left(s,r\right)}
}
{  n \left( t^{ \frac{n}{2} } -   t^{ -\frac{n}{2} }   \right)   } \cdot Q_\b^n
\Bigg]   \\
&=  \, \PE  \Bigg[
\sum\limits_{{\mathcal{\b}} \in H_2\left(X,L, \mathbb{Z}\right)}
\sum\limits_{s,r \in \mathbb{Z}/2} 
-	\frac{  \left(-1\right)^{2 s+2 r}     t^{   - s   }  \left(   \frac{t}{q}   \right)^{  \, r}   
	N_{\mathcal{\b}}^{\left(s,r\right)}}
{     \left( t^{ \frac{1}{2} } -   t^{ -\frac{1}{2} }   \right)     }  \cdot Q_\b
\Bigg]  \,,
\end{split} 
\end{align}
which encodes the refined Ooguri-Vafa invariants $N_{Q_\b}^{\left(s,r\right)}$. Note that the unrefined version of \eqref{openGV} is equivalent to \eqref{DTgen}.

For abelian theories such as $U\left(1\right)_k+N_{f}\F+N_a\AF$, vortex partition functions can be written in term of quiver generating functions  \eqref{quivergenfun} and hence refined Ooguri-Vafa (OV) invariants are refined Donaldson-Thomas (DT) invariants
\begin{align}
N_{Q_\b}^{\left(s,r\right)} = \Omega^{\left(s,r\right)}_{Q_\b} \,. 
\end{align}

However, for nonabelian theories considered in section \ref{secnonabelian}, expansion variables in quiver genetrating functions are not independent, for instance \eqref{quiverform1} and \eqref{quiverform2}. Therefore, OV invariants as degeneracy numbers of vortex particles are  linear combinations of DT-invariants  $\Omega^{\left(s,r\right)}_\b$
\begin{align}
~ N_{Q_{\b}}^{\left(s,r\right)} = \sum_{\b'} ~\Omega^{\left(s',r'\right)}_{Q_{\b'}}  \,.
\end{align}
This suggests that DT-invariants compose OV invariants for nonabelian theories. 
One can use the refined formula \eqref{openGV} to extract refined DT invariants and then impose relations between variables $x_i$ to combine these DT invariants into refined OV invariants. 

As we discussed in \cite{Cheng:2021aa}, the refinement of quiver generating function is almost trivial for abelian theories $U(1)_k+N_f \F+N_a\AF$, since the refinement only shifts variables $x_i$ by some spin parameters $q^{s_i} t^{r_i}$. This shift only changes the spin indices of DT-invariants. An interesting consequence is that the index $s$ could take many values but index $r$ can only take one value. Namely for each term $Q_\b$, non-zero OV invariants are:
\begin{align}
 ~N_{Q_{\b}}^{\left(s_1, r\right)},~N_{Q_{\b}}^{\left(s_2, r\right)},~ \cdots\,,~ N_{Q_{\b}}^{\left(s_n, r\right)} \,  .
 	\end{align} 
This property also remains for nonabelian theories. Moreover, we note that there are  negative refined OV/DT-invariants from $W$-bosons in nonabelian theories. This suggests that refined OV formula \eqref{openGV} needs to  be modified, or some K\"ahler parameters need to be properly chosen for nonabelian theories.


\section{Outlook}

In this work, we discuss brane webs and Chern-Simons levels of 3d theories, in particular nonabelian theories with chiral multiplets. Many 3d brane webs are obtained by turning on real masses parameters for chiral multiplets, which separates the overlapped D5-branes. It turns out that these brane webs describe the same theory, and these equivalent brane webs are connected by a series of $ST$-transformations, based on the evidence from vortex partition functions and holomorphic blocks.  Each brane web has a corresponding quiver matrix which is interpreted as the mixed effective \CS levels of its $ST$-dual theory. In addition, we observe that these 3d brane webs are related by flipping mass parameters. Some brane webs compose the Higgs branch. Flipping mass parameters change the theory from one chamber to other chambers.  Moreover, we compute refined vortex partition functions of nonabelian theories and show that they also have quiver matrices. The homorphic blocks and effective superpotenitals are also discussed to understand $ST$-transformations.

There are some open problems. First, it deserves to further explore the movement of flavor D5-branes as well as quiver gauge theories with several gauge nodes connected by bifundamental multiplets. Quiver matrices should also be physically understood as mixed Chern-Simons levels for nonabelian theories by properly dealing with chiral multiplets in generic representations. Finding more evidence for nonabelian theories is also interesting. In addition, $ST$-transformations for abelian theories are commutative and forms a group \cite{Cheng:2020aa}, it derserves to explore if nonabelian theories also have a commutative $ST$-transformation group. Last but not the least, perhaps there are relations between intersection numbers of toric diagrams and quiver matrices; see e.g. \cite{Cabrera:2018jxt,Bourget:2020asf}.

\acknowledgments
We especially thank JHEP referee for very nice comments and suggestions to discuss holomorphic blocks and $\left(\bar{q},t\right)$-branes, which are not covered in the previous version of this note. We also especially thank Piotr Su{\l}kowski for helpful discussions and reading the manuscript carefully and correcting some mistakes.
We would like to thank Rui-Dong Zhu for hospitality when visiting Sooshow University and follow-up discussions. 
This work has been supported by the TEAM programme of the Foundation for Polish Science co-financed by the European Union under the European Regional Development Fund (POIR.04.04.00-00-5C55/17-00).

\bigskip

\appendix
\section{Nekrasov factors}
Identities for Nekrasov factors:	
\begin{align}
N_{\u \v} \left( Q; t,q \right) &:=\prod_{i,j=1}^{\inf} \frac{ 1-Q q^{\v_i-j} t^{\u_j^T-i+1} }{  1- Q q^{-j} t^{-i+1} } \,,\\	
N_{\u \v} \left( Q; t,q \right) &=N_{\v^T \u^T}\left(Q \frac{t}{q}; q;t\right)   \,,\\
N_{\u \v} \left( Q; t^{-1},q^{-1} \right) &=N_{\v^T \u^T}\left(Q \frac{q}{t}; q^{-1};t^{-1}\right)   \,, \\
 N_{\u \v} \left( Q \sqtq; t^{-1},q^{-1} \right) &=N_{\v^T \u^T}\left(Q \sqqt; q^{-1};t^{-1}\right) \,,  \\
 N_{\u \v} \left( Q^{-1} \sqtq; t^{-1},{q^{-1}}\right) &=
\left(-Q\right)^{- |\u|-|\v|  } t^{- \frac{ || \u^T||^2}{2} +\frac{ || \v^T||^2}{2} } q^{ \frac{ || \u||^2}{2} -\frac{ || \v||^2}{2} } \nn\\
&\quad \times
N_{\v^T \u^T}\left(Q \sqtq; {t^{-1}},{q^{-1}}\right)  \,,  \\
 	N_{\{n_1\}_{V},\{n_2\}_{V}}\left( Q; t^{-1},q^{-1}   \right)\,&= \left(Q ~ t^{-n_1};t\right)_{n_2}  \left(Q q t^{-1}~ t^{n_2};t^{-1}\right)_{n_1}    \,, \\
 N_{\{{n_1}\}_{H}\,,\{{n_2}\}_{H}}\left( Q; t^{-1},q^{-1}   \right)
&=\left(Q q t^{-1} q^{-n_2};q\right)_{n_1}  \left(Q~q^{n_1};q^{-1}  \right)_{n_2}    \,,  \label{antisymNn1n2}   \\
 N_{ \{ n\}_H, \{ n\}_H }\left(Q;t,q\right)&=\left(Qt,q\right)_n \left(Q q^{-n};q\right)_n  \,,\\
 N_{\{{n_1}\}_{H}\,,\{{n_2}\}_{V}}\left( Q; t^{-1},q^{-1}   \right)
 &= \frac{ 1- Q t^{-1} q^{n_1} t^{n_2} }{ 1- Q t^{-1} }
 \left(Q  t^{-1};q\right)_{n_1}  \left(Q \;t^{-1};t  \right)_{n_2}    \,,  \\
 N_{\{{n_1}\}_{V}\,,\{{n_2}\}_{H}}\left( Q; t^{-1},q^{-1}   \right) 
 &= \frac{ 1- Q q \;t^{-n_1} q^{-n_2}  }{ 1- Q q }
 \left(Q  q;t^{-1}\right)_{n_1}  \left(Q  \;q;q^{-1}  \right)_{n_2}    \,,  
\end{align}
where $\u^T$ is the transpose of the Young tableau $\u$. We use  $\{n\}_H$ and  and $\{n\}_V $ to denote symmetric and anti-symmetric Young tableaux respectively:
\begin{align}\label{HVnotation}
\{n\}_{H} :=
\tiny
\begin{ytableau}
\none & & &  & \cdots &  \\
\end{ytableau} \,,  \quad \quad
 \{n\}_{V} :=
 \tiny
 \begin{ytableau}
 \none &  \\
 \none &\\
 \none &\vdots\\
 \none &
\end{ytableau}
\,~.
\end{align}
For more details on refined topological vertex, see \cite{Cheng:2021aa,Cheng:2018ab}. The factorization of Nekrasov factors can be found in e.g. \cite{Pan:2016fbl,Kimura:2021ngu,Nieri:2017ntx}

\section{Pochhammer products}
Identities for $q$-Pochhammer products:
\begin{align}
 \left(Q;q\right)_\inf &=\prod_{k=0}^{\inf} \left(1-Q q^{k}\right) \,, \\
 \left( Q;q \right)_\inf^{-1}&=  \left( Q q^{-1};q^{-1}\right)_\inf   \,,\\
\left(Q;q\right)_n&= \left(1-Q\right)\left(1-Q q\right) \cdots \left(1-Q q^{n-1}\right) :=\prod_{k=0}^{n-1} \left(1-Q q^{k}\right) \\
\left(Q;q\right)_n &= \frac{ \left(Q;q\right)_\inf  }{ \left(Q q^n;q\right)_\inf  }\,,
\\
        \left(Q^{-1};q\right)_n &=\left(Q;q^{-1}\right)_n  ~ \left(-\sqrt{q}\right)^{n^2} \left( \sqrt{q} Q\right)^{-n}                               \,,\\
 \left(Q q^{-n};q\right)_n &=\left(Q q^{-1};q^{-1}\right)_n  \,, \\
  \left(Q;q\right)_{-n}&=\frac{1}{\left(Q~ q^{-n};q \right)_n  } =\frac{1}{ \left(Q~ q^{-1};q^{-1}\right)_n }	  = \frac{ \left( -q Q^{-1}\right)^{n}  \left(\sqrt{q}\right)^{n^2-n}   }{ \left( q Q^{-1};q\right)_n }   \,,
\\
	\left(Q;q\right)_n&= \frac{\left(Q;q\right)_\inf}{\left(Q q^n;q\right)_\inf}  =\PE\bigg[\frac{Q\left(q^n-1\right)}{1-q}  \bigg]    \,,\\
       \frac{1}{\left(Q;q\right)_\inf} &=\sum_{n=0}^\inf  \frac{Q^n}{\left(q;q \right)_n }      =\exp\bigg[  \sum_{n=1}^{\inf}  \frac{1}{n}  \frac{Q^n}{1-q^n }  \bigg] =:\PE\bigg[ \frac{Q}{1-q}\bigg]                 \,,  \\
      \left(Q;q\right)_\inf  &=      \sum_{n=0}^\inf   \frac{ \left( -\sqrt{q}\right)^{n^2} \left(\frac{Q}{\sqrt{q}} \right)^n   }   { \left(q;q\right)_n }       =\exp\bigg[ - \sum_{n=1}^{\inf}  \frac{1}{n}  \frac{Q^n}{1-q^n }  \bigg] =:\PE\bigg[ -\frac{Q}{1-q}\bigg]                  \,,\\
 \frac{\left(\a;q\right)_n}{ \left(\a;q\right)_\inf } &= \sum\limits_{d=0}^{\inf} \left(-\sqrt{q}\right)^{2 n d} \frac{\a^{d}}{\left(q;q\right)_{d} }  
 = \left(\frac{\a}{\sqrt{q}}\right)^n \sum_{d=0}^{\inf} \left(-\sqrt{q}\right)^{n^2 -2 n d+d^2} \frac{ \left(\sqrt{q} \a^{-1} \right)^d }{ \left(q;q\right)_{d} }\,,
  \label{aquiverform}\\
\frac{\left(\b;q\right)_\inf }{\left(\b;q\right)_{n}} &= \sum\limits_{d=0}^{\inf} \left(-\sqrt{q}\right)^{2 n d+d^2} \frac{\b^{d}}{\left(q;q\right)_{d} } 
= \left(\frac{ \sqrt{q}}{\b}\right)^n  \sum_{d=0}^{\inf} \left(-\sqrt{q}\right)^{ -n^2 -2 n d} \frac{\left(q \b^{-1}\right)^d }{\left(q;q\right)_d } \,,
\label{bquiverform}\\
	 \frac{\left(\a;q^{-1}\right)_n}{ \left(\a ;q^{-1}\right)_\inf   } &=
	\frac{ 1 }{\left(\a q^{-n};q^{-1}\right)_\inf   }   =
  \sum_{d=0}^{\inf}
	\left( 
	- \sqrt{q} \right)^{ d^2- 2 n d   } 
	\frac{ 
		\left(\sqrt{q} \, \a\right)^{d}  }{   \left(q;q\right)_{d}    }  \,,
	\label{cijcomefrom3}
	\\
	\frac{ \left(\b ;q^{-1}\right)_\inf   }{ \left(\b;q^{-1}\right)_n}   &={  \left(\b q^{-n};q^{-1}\right)_\inf    }
	= 
	\sum_{d=0}^{\inf}
	\left( - \sqrt{q} \right)^{-2  n d  }  
	\frac{ \left(    q\, \b \right)^{d}  }{  \left(q;q\right)_{d}     }  \,,\\
	 \frac{1}{ \left( \b q^{-n_1};q\right)_{n_2}}&= \frac{ \left(\b q^{n_2-n_1};q\right)_\inf  }{  \left( \b q^{-n_1};q  \right)_\inf }= \sum\limits_{d_i,d_j =0}^{\inf} 
	\left(-\sqrt{q}\right)^{{d_i^2} +2 \left(n_2-n_1\right) d_i -2 n_1 d_j } \frac{
		\left( \frac{\b}{\sqrt{q}}\right)^{d_i} \b^{d_j}
	}{ \left(q;q\right)_{d_i} \left(q;q\right)_{d_j} } \,.  \label{n1n2quiver}    
\end{align}

We use a shorthand notation to denote contributions from chiral multiplets to quiver matrices:
\begin{align}
	\left(\a_i;q^{\pm}\right)^{\pm}_n &=\sum_{d_i=0}^{\inf} \left(-\sqrt{q}\right)^{C_{n i}\, n\, d_i} \frac{x_i^{d_i}}{ \left(q;q\right)_{d_i} } \,,
	\\
	\left(\a_i;q^{\pm}\right)^{\pm}_n &\rightarrow \left( C_{n\, i},  x_i \right) \,,
\end{align}
where $C_{ni}$ is the component of quiver matrices
\begin{align}
	\left(\a;q^{\pm} \right)^{\pm}_n \rightarrow \Big(    
	\left[
	\begin{array}{ccc}
		a & \dots&b  \\
		\vdots&\ddots &\vdots\\
		c&\dots&d
	\end{array}
	\right] , ~x
	\Big)\,,
\end{align}	
where $x_i$ are the expansion variables, and $\cdots$ stands for contributions from other $q$-Pochhammer products.

Each $q$-Pochhammer product	 has two equivalent quiver matrices, depending on how we choose expansion variables. We summarize all cases in the following
\begin{align}
	&\left(\a;q\right)_n ~~\rightarrow~
	\big(
	\left[ 
	\begin{array}{ccc}
		~\textcolor{blue}{0}~ &\cdots  &~1~ \\
		\vdots& \ddots &\vdots\\
		~1~ &\cdots  &~0~\\
	\end{array}
	\right] \,,~ \a \big) 
	\,,  \quad \quad~~
	\big(        
	\left[ 
	\begin{array}{ccc}
		\textcolor{blue}{1} &\cdots  &-1 \\
		\vdots& \ddots &\vdots\\
		-1 &\cdots  &1\\
	\end{array}
	\right] \,, \sqrt{q}\,\a^{-1}
	\big)\,,   \label{quiverfrompoch1} &        \\ 
	&\frac{1}{ \left(\b;q\right)}_n ~~\rightarrow~\,
	\big(
	\left[ 
	\begin{array}{ccc}
		~\textcolor{blue}{0}~ &\cdots  &~1~ \\
		\vdots& \ddots &\vdots\\
		~1~ &\cdots  &~1~\\
	\end{array}
	\right] \,,~
	\frac{\b }{ \sqrt{q}}
	\big)
	\,,   \quad~~
	\big(  
	\left[ 
	\begin{array}{ccc}
		\textcolor{blue}{-1} &\cdots  &-1 \\
		\vdots& \ddots &\vdots\\
		-1 &\cdots  &0\\
	\end{array}
	\right] \,,~ q \,\b^{-1}
	\big)  \,,  \label{quiverfrompoch2}
	&            \\      
	&\left(\a;q^{-1}\right)_n \rightarrow \big(        
	\left[ 
	\begin{array}{ccc}
		\textcolor{blue}{0} &\cdots  &-1 \\
		\vdots& \ddots &\vdots\\
		-1 &\cdots  &~1\\
	\end{array}
	\right] \,, ~\sqrt{q}\,\a
	\big)\,, \quad
	\big(
	\left[ 
	\begin{array}{ccc}
		\textcolor{blue}{-1} &\cdots  &~1~ \\
		\vdots& \ddots &\vdots\\
		1 &\cdots  &0\\
	\end{array}
	\right] \,,~ \a^{-1} \big) 
	\,,    \label{quiverfrompoch3}
	&     \\
	& \frac{1}{ \left(\b;q^{-1}\right)}_n   \rightarrow  \big(  
	\left[ 
	\begin{array}{ccc}
		~\textcolor{blue}{0}~ &\cdots  &-1 \\
		\vdots& \ddots &\vdots\\
		-1 &\cdots  &0\\
	\end{array}
	\right] \,,~ q \,\b
	\big)  \,, \quad~~~
	\big(
	\left[ 
	\begin{array}{ccc}
		~\textcolor{blue}{1}~ &\cdots  &~1~ \\
		\vdots& \ddots &\vdots\\
		~1~ &\cdots  &~1~\\
	\end{array}
	\right] \,,~
	\frac{\b^{-1} }{ \sqrt{q}}
	\big)\,, \label{quiverfrompoch4}
	&  
\end{align}
Note that $\frac{1}{\left(q;q\right)_n}$ and $\frac{1}{\left(q^{-1};q^{-1}\right)_n}$ also contribute to the quiver matrix
\begin{align}
	\frac{1}{\left(q;q\right)_n} ~~\rightarrow~~
	\left[ 
	\begin{array}{ccc}
		\color{blue}{0} &\cdots  &0 \\
		\vdots& \ddots &\vdots\\
		0 &\cdots  &0\\
	\end{array}
	\right] \,,\qquad
	\frac{1}{\left(q^{-1};q^{-1}\right)_n} ~~\rightarrow~~
	\left[ 
	\begin{array}{ccc}
		\color{blue}{1} &\cdots  &0 \\
		\vdots& \ddots &\vdots\\
		0 &\cdots  &0\\
	\end{array}
	\right] \,.
\end{align}

	\section{Theta functions	}
	
	\begin{align}
		&\left(z;q\right)_\inf =\frac{M\left(z q^{-1};q \right) }{ M\left( z;q\right)} = \PE\bigg[  - \frac{z}{1-q} \bigg]\,,\\
		&\left(qz^{-1};q\right)_\inf = \frac{{M}\left(z^{-1};q \right)   }{ M\left(q z^{-1};q \right) }  = \PE\bigg[  - \frac{ q z^{-1}}{1-q} \bigg]\,,\\
		&\theta\left(-q^{-\half};q\right) =\PE\bigg[  -\frac{z+q z^{-1}}{1-q} \bigg] = \PE\bigg[ - z \cdot \frac{1+q}{1-q} \, \bigg]  \,,\\
		& \theta\left(-q^{-\half} z ;q\right) = \frac{ M\left(q^{-1} z;q \right) }{ M\left( q z;q\right)  } \,,\\
		&  \left(q z;q\right)_\inf = \frac{ M\left(z^{-1};q\right) }{ M\left(q z^{-1};q\right)  } =  \frac{ M\left(z;q\right) }{ M\left(q z;q\right)  } \,,\\
		& M\left( z;q\right) :=\PE\bigg[- \frac{q z}{\left(1-q\right)^2 }  \bigg] \,,\\
		& \frac{ M\left(z^{-1};q\right) }{ M\left( z;q\right)  } =  \frac{ M\left(q z^{-1};q\right) }{ M\left(q z;q\right)  } = \text{constant} \,.
		\end{align}


\bibliographystyle{JHEP}
\bibliography{ref}

\end{document}